\documentclass[ reprint,
 amsmath,amssymb,
 aps,
 pra,
]{revtex4-2}
\usepackage[english]{babel}

\newtheorem{Definition}{Definition}
\newtheorem{Theorem}{Theorem}

\usepackage{graphicx}\usepackage{dcolumn}\usepackage{bm}
\usepackage{xcolor}
\colorlet{red}{black}

\usepackage{multirow}
\usepackage{threeparttable}
\usepackage{makecell}

\begin{document}

\preprint{APS/123-QED}

\title{Learning Hidden Structures in Open Quantum Dynamics}

 \author{ Alexander Teretenkov} \email{taemsu@mail.ru}
\affiliation{Department of Mathematical Methods for Quantum Technologies, Steklov Mathematical Institute of Russian Academy of Sciences, Gubkina str. 8, Moscow, 119991, Russia}\author{Sergey Kuznetsov}
\email{sakuznetsov.box@gmail.com}
\affiliation{Department of Mathematical Methods for Quantum Technologies, Steklov Mathematical Institute of Russian Academy of Sciences, Gubkina str. 8, Moscow, 119991, Russia}
\author{Alexander Pechen} \email{apechen@gmail.com}
\affiliation{Department of Mathematical Methods for Quantum Technologies, Steklov Mathematical Institute of Russian Academy of Sciences, Gubkina str. 8, Moscow, 119991, Russia}

\date{\today}             
\begin{abstract}

We introduce a machine-learning approach for identifying hidden structural features of open quantum dynamics under restricted experimental access. Unlike most existing data-driven methods which focus on detection or prediction of dynamical behavior, our framework targets the inference of invariant algebraic structures underlying the effective Markovian evolution. Measurement limitations, symmetries, and superselection rules are incorporated through a $*$-algebraic description of accessible observables. The learning problem is formulated as maximum-likelihood estimation from multi-time measurement sequences, where the algebraic type of an invariant subalgebra—particularly a decoherence-free subalgebra—is treated as a discrete structural hypothesis. The feasibility of the approach is illustrated on multiple synthetic models and a waveguide quantum electrodynamics system, where nontrivial intermediate algebraic structures are identified directly from measurement data.

\end{abstract}

                              \maketitle

\section{Introduction}

Despite the growing impact of machine-learning techniques in the theory of open quantum systems, the systematic problem of uncovering hidden structural features of open quantum dynamics remains largely unresolved. Most existing approaches have primarily focused on detecting, classifying, or predicting dynamical behavior, rather than on identifying invariant algebraic or dynamical structures underlying the effective evolution. Learning control of open quantum systems based on genetic algorithms, which are widely applied in machine learning~\cite{Goldberg1989}, was developed in~\cite{Pechen_Rabitz_2006} following the approach for closed quantum systems~\cite{Judson_Rabitz_1992}. Also related developments were performed in measurement-assisted and incoherent control schemes~\cite{Dong_Rabitz_2008}. Machine learning has been extensively applied to diverse quantum control and estimation problems~\cite{Ma_Qi_Petersen_Wu_Rabitz_Dong_2025}. 
Machine learning was applied to generate autonomous
adaptive feedback schemes for quantum measurements~\cite{Hentschel_Sanders_2010}. Deep reinforcement learning was applied to simultaneously optimize the speed and fidelity of quantum computation against both leakage and stochastic control errors~\cite{Niu_Boixo_Smelyanskiy_Neven_2019}. Early machine-learning applications addressed the characterization of non-Markovian memory effects using support vector machine algorithms combined with trace-distance- and entanglement-based measures of non-Markovianity~\cite{Fanchini_Karpat_Rossatto_Norambuena_Coto_2021}. From a more formal perspective, tensor-network methods were employed to develop an efficient strategy for controlling  quantum many-body systems~\cite{Doria_Calarco_Montangero_2011}, learn representations of generic non-Markovian quantum stochastic processes~\cite{Guo_Modi_Poletti_2020}, with subsequent work analyzing bounds on the errors induced by finite-dimensional truncations of bosonic Hilbert spaces inherent to such tensor-network approaches~\cite{Vilkoviskiy_Abanin_2024}.  While these methods provide compact and numerically efficient descriptions of complex dynamics, they are not designed to explicitly reveal physically interpretable invariant structures of the underlying open-system evolution.

Related machine-learning strategies have been developed for system identification and effective modeling of open quantum dynamics, including recent approaches based on global optimization of Markovian master equations from time-series data~\cite{Popovych_Jacobs_Korpas_Marecek_Bondar_2022}. Neural-network-based models were proposed for classifying external noise sources affecting quantum evolution~\cite{Martina_Gherardini_Caruso_2023}, and neural function approximators were used to predict time-local generators of open-system dynamics for specific classes of spin systems from an underlying unitary evolution~\cite{Mazza_Zietlow_Carollo_Andergassen_Martius_Lesanovsky_2021}. Other approaches employed neural networks to directly parameterize non-Markovian open-system dynamics at the level of observables and their equations of motion~\cite{Yang_Cao_Yang_2022}, or to perform non-Markovian quantum process tomography by mapping experimental data to a process tensor represented via its Choi state~\cite{Wu_Li_Zhao_Luan_Yu_Zhang_2024}. Related data-driven approaches aim to go beyond purely black-box models by extracting interpretable features of non-Markovian dynamics~\cite{luchnikov2022probing}, as well as by constructing reduced-order descriptions of many-body quantum systems suitable for control tasks~\cite{luchnikov2024controlling}. In a complementary,  more phenomenological direction, post-Markovian master equations have been also used to model and predict non-Markovian noise in superconducting qubit systems~\cite{Zhang_Pokharel_Levenson-Falk_Lidar_2022}. 

Beyond identification and modeling, machine-learning methods have also been applied to control and optimization tasks, including deep-learning-based steering of qubit-pair states under constrained measurement settings~\cite{Wang_Ku_Lin_Chen_2024}, global optimization of quantum dynamics using AlphaZero-type algorithms~\cite{Dalgaard_2020}, supervised characterization of driven two-level systems via deep quantum exploration inspired by AlphaZero~\cite{Couturier_2023}, generation of fast and robust perfect entangling gates~\cite{Grech_Valentino_2026}. Discrete real-time learning  of the evolution of a many-body quantum state in the presence of time-dependent control fields based on the artificial neural network was developed~\cite{Gui_Ho_Rabitz_2024}. 

More broadly, machine learning has been successfully adapted for quantum control problems~\cite{Dong_Petersen_2023}, including optimal pure-state manipulation in quantum systems~\cite{Chen_Herrmann_Vamvoudakis_Vijayan_2024}. The structure of the underlying optimization landscape, which can potentially be rugged, having multiple isolated global and/or local minima, or ravine,  is essential for efficiency of machine learning and optimization. The relation between the landscape structure and the practical search effort has also been investigated for open quantum systems with Kraus-map evolution \cite{Oza_Pechen_Dominy_Beltrani_Moore_Rabitz_2009}. In this regard, for open quantum systems a ravine structure of the optimization landscapes was discovered for various commonly used optimization tasks, where
globally optimal solutions were proven to form connected sets defining the bottom of the ravine~\cite{Pechen_Prokhorenko_Wu_Rabitz_2008,Wu_Pechen_Rabitz_Hsieh_Tsou_2008} in such problems as optimization in chemistry and material properties~\cite{Moore_Pechen_Feng_Dominy_Beltrani_Rabitz_2011a,Moore_Pechen_Feng_Dominy_Beltrani_Rabitz_2011b} and in biology~\cite{Feng_Pechen_Jha_Wu_Rabitz_2012}.

Taken together, these works demonstrate the versatility of machine-learning techniques in addressing a wide range of problems in open quantum dynamics. However, the explicit inference of hidden structural features—such as invariant subalgebras, decoherence-free sectors, or algebraic constraints on the generator—from experimentally accessible data has so far received comparatively little attention, leaving an important gap between data-driven modeling and the structural theory of open quantum systems.

What structures underlying quantum dynamics are physically important? Many such structures are associated with the invariant dynamics of particular sets of observables. Examples include invariant dynamics of moments of a fixed order~\cite{barthel2022solving, zhang2022criticality, teretenkov2020dynamics, ivanov2022dynamics, nosal2020exact, linowski2022dissipative, nosal2022higher, penc2026linear}, Onsager strings~\cite{teretenkov2024exact}, and fragmentation in operator space~\cite{teretenkov2024exact, essler2020integrability, paszko2025operator}.  A related viewpoint is provided by probability representations of quantum dynamics, which generalize classical stochastic dynamics associated with invariant commutative diagonal algebras~\cite{yashin2020minimal, kulikov2024minimizing}. 
One of the natural  assumptions is that such a set of observables forms a $*$-algebra. In this case, Gorini--Kossakowski--Sudarshan--Lindblad (GKSL) generators are fully characterized in terms of the invariant $*$-algebra~\cite[Theorem 6]{hasenohrl2023generators}. In this work, however, we focus on the special case of GKSL generators with decoherence-free subalgebras~\cite{dhahri2010decoherence, dhahri2011decoherence, deschamps2016structure, agredo2022decoherence}.  In particular, this case includes decoherence-free subspaces and subsystem.

     Existing approaches to the identification of decoherence-free subspaces often
start from a reconstructed quantum map or channel, for example via quantum
process tomography, and then analyze the corresponding noise algebra or
commutant structure~\cite{choi2006method,guan2018structure}. There are also more direct protocols
for identifying decoherence-free subspaces without full process
tomography~\cite{mahler2012identification}, as well as numerical methods for
finding the DFS structure of a given noisy quantum
channel~\cite{wang2013numerical}. Our setting is different and complementary: we use multi-time measurement data
to infer the algebraic type of a hidden decoherence-free subalgebra in a
Markovian embedding, while assuming only restricted access to observables from
some fixed algebra. Thus, the target is not a DFS of an already
reconstructed map, but a latent invariant algebraic structure inferred from
multi-time statistics available under restricted observational access.

It is important to distinguish the Markovianity of the enlarged dynamics
from the Markovianity of the dynamics directly observed through a restricted
set of observables. Even if the joint dynamics on the enlarged system is
Markovian and generated by a GKSL generator, the statistics of measurements
restricted to a subalgebra of accessible observables may exhibit memory
effects. In this sense, the experimentally observed dynamics need not be
Markovian. The Markovian embedding assumption means that these non-Markovian
multi-time statistics can be represented by a Markovian evolution on a larger
finite-dimensional space.

There are well-known approaches for reconstructing the Markovian environment (or effectively Markovian via Markovian embedding approaches \cite{garraway1996cavity, garraway1997nonperturbative, dalton2001theory, tamascelli2018nonperturbative, teretenkov2019pseudomode, tamascelli2019efficient, pleasance2020generalized, kanazawa2024standard, teretenkov2025pseudomode}) of an open quantum system based on pattern recognition in statistical data obtained from system measurements \cite{luchnikov2020machine}. But they assume that one has full access to all the system degrees of freedom. Nevertheless, there are several reasons for considering a similar problem with limited access to the system degrees of freedom:
\begin{enumerate}
    \item In real experimental setup one has access only to restricted set observables. E.g. for spectroscopy \cite{mukamel1995principles, cho2009two} it is usually harder to measure the populations than coherences.
    \item For some open quantum systems model with a small parameter, not all the open quantum system observables are ''slow'' \cite{yang2002influence, seibt2017ultrafast, trushechkin2019calculation}. So the projection methods will lead to master equations for some parts of the density matrix (in the adjoint equations will be for the restricted set of observables).
    \item The dynamics for some system observables should be closed or approximately closed due to symmetries of the free dynamics \cite{accardi2002lectures, accardi2016three, fagnola2019characterization, hernandez2020stationary,  trushechkin2021unified, bolanos2025gaussian}. E.g. the dynamics of coherences and populations disentangles for usual weak coupling master equations with generic system Hamiltonian.
    \item There are some fundamental restrictions on possible states imposed by superselection rules \cite{wightman1995superselection, cisneros1998limitations, bartlett2007reference, amosov2017spectral}. 
\end{enumerate}

What set of observables should be considered as accessible? From a mathematical point of view, it is typically assumed   they should form some *-algebras, first of all von Neumann or, more generally,  $C^*$-algebras \cite{bratteli2012operator}. In this work, we consider only the finite-dimensional case for which these notions coincide with matrix *-algebras, so we restrict our discussion to them. Such algebras naturally appear when exploiting symmetries, commutant structures,
and invariant decompositions in open quantum dynamics and many-body systems,
see, e.g., \cite{moudgalya2022hilbert, grigoletto2025model, grigoletto2025exact}. 
The  paper is organized as follows. In Section~\ref{se:algebras}, we recall the structure and parameterization of finite-dimensional matrix $*$-algebras. In Section~\ref{sec:Markovian}, we summarize the Markovian open-system framework and the description of multi-time measurement statistics used in this work. In Section~\ref{se:DFA}, we review decoherence-free subalgebras and the corresponding structural form of compatible GKSL generators. In Section~\ref{se:embedding}, we formulate the generalized Markovian embedding approach and the associated likelihood-based problem of invariant algebra estimation. Section~\ref{se:numerical} presents numerical results confirming that the proposed method successfully reconstructs hidden invariant structures from measurement data. Finally, Section~\ref{se:conclusion} contains the conclusions.

\section{Algebras of observables and their parameterization}\label{se:algebras}

As we have discussed in the introduction,  matrix $*$-algebras are an essential notion both for accessible observables and for invariant ones. So let us recall their definition \cite{de2011numerical}.

\begin{Definition}
We call  $\mathcal{A} \subseteq \mathbb{C}^{n \times n} $ a matrix $*$-algebra if it is a $*$-subalgebra of $\mathbb{C}^{n \times n}$, i.e.
	\begin{enumerate}
		\item $ \forall X, Y \in \mathcal{A} , \alpha, \beta \in \mathbb{C} $ $\Rightarrow$ $ \alpha X + \beta Y \in  \mathcal{A} $
		\item $ \forall X, Y \in \mathcal{A}  $ $\Rightarrow$ $  X  Y \in  \mathcal{A} $
		\item $ \forall X \in \mathcal{A}  $ $\Rightarrow$ $  X^{\dagger} \in  \mathcal{A} $ 
	\end{enumerate}
\end{Definition}

For matrix $*$-algebras the following  Artin–Wedderburn-type theorem holds. 

\begin{Theorem}
    Any matrix $*$-algebra $\mathcal{A} \subseteq \mathbb{C}^{n \times n} $  can be represented as 
    \begin{equation}\label{eq:algebraCanonicalForm} 
			\mathcal{A} = U \left( (0 I_{n_0}) \oplus \bigoplus_{k=1}^K (\mathbb{C}^{n_k \times n_k} \otimes I_{m_k}  ) \right) U^{\dagger},
		\end{equation}
        where $n_0$ is a non-negative number (by $n_0=0$ we mean that the block $0 I_{n_0}$ is absent in the sum above), and $n_k, m_k, k=1, \ldots, K$ are  positive integer numbers, such that
        \begin{equation}\label{eq:conditionOnDimensions}
			n = n_0 + \sum_{k=1}^K n_k m_k,
		\end{equation}
        $U$ is a unitary matrix, $\mathbb{C}^{n_k \times n_k}$ are *-algebras of all complex $n_k \times n_k$ matrices, and $I_d$ is  a $d \times d$ identity matrix.
\end{Theorem}

The proof for the general case of atomic algebras can be found in \cite[Appendix]{deschamps2016structure} and the discussion of the fact that finite-dimensional *-algebras are atomic with  explicit formula \eqref{eq:algebraCanonicalForm} can be found in \cite{hasenohrl2023generators}.

    Thus, any $*$-algebra  $\mathcal{A} \subseteq \mathbb{C}^{n \times n} $ is defined by a unitary matrix $U$ and integers $n_k$ and $m_k$ under condition \eqref{eq:conditionOnDimensions}. And each specific element of this algebra  in the basis defined by unitary matrix $U $ has the block-diagonal form
    \begin{equation}\label{eq:observableFromTheAlgebra}
         X  = U
        \begin{pmatrix}
            0 I_{n_0} & 0 & \cdots & 0 & 0\\
            0 & X_1 \otimes I_{m_1} & \cdots & 0 & 0\\
             \vdots &  \vdots & \ddots & \vdots & \vdots \\
             0 & 0 & \ldots & X_{K-1} \otimes I_{m_{K-1}}& 0 \\
             0 &  0 & \ldots & 0 & X_K \otimes I_{m_K}
        \end{pmatrix}
        U^{\dagger}
    \end{equation}
    which is fully characterized by the set of matrices 
    \begin{equation}
        \{X_k \in \mathbb{C}^{n_k \times n_k}, k =1,\ldots, K\}.
    \end{equation}

In particular, for $K=1$, $n_0 =0$ and $U = I_n$, Equation \eqref{eq:observableFromTheAlgebra} takes the form
\begin{equation}\label{eq:subsystemAlgebra}
X = X_1 \otimes I_{m_1},
\end{equation}
i.e. $X$ can be considered as an observable of the subsystem of the total system in $n= n_1m_1$-dimensional Hilbert space, which is the usual open quantum systems setup \cite{breuer2002theory}.

Another widespread special case is $n_0 = 0$ and $m_k = 1$. In this case, taking into account that $X_j \otimes I_1 \simeq X_j$, we obtain
\begin{equation}
 X  = 
        \begin{pmatrix}
            X_1 & 0 & \cdots & 0 & 0\\
            0 & X_2 & \cdots & 0 & 0\\
             \vdots &  \vdots & \ddots & \vdots & \vdots \\
             0 & 0 & \ldots & X_{K-1} & 0 \\
             0 &  0 & \ldots & 0 & X_K 
        \end{pmatrix}.
\end{equation}

As an algebra of accessible observables, it arises for fast observables \cite{teretenkov2022effective} with respect to free Hamiltonian which has the following form in its eigenbasis
\begin{equation}
H_0 =
\begin{pmatrix}
\varepsilon_1 I_{n_1} & 0 & \cdots & 0 & 0\\
            0 & \varepsilon_2 I_{n_2} & \cdots & 0 & 0\\
             \vdots &  \vdots & \ddots & \vdots & \vdots \\
             0 & 0 & \ldots & \varepsilon_{K-1} I_{n_{K-1}} & 0 \\
             0 &  0 & \ldots & 0 &  \varepsilon_{K} I_{n_{K}}
\end{pmatrix},
\end{equation}
where $\varepsilon_j $ are distinct eigenvalues of $H_0$ and $n_j$ are their multiplicities. As an invariant algebra, it arises for the weak coupling limit type GKSL generators \cite{accardi2016three}.

The special case with $n_0  \neq 0$, $K=1$,  $m_k = 1$ can be represented in the form
\begin{equation}
 X  = 
\begin{pmatrix}\label{eq:restrAlgebra}
0 I_{n_0} &0\\
0 & X_1
\end{pmatrix}.
\end{equation}
As an algebra of accessible observables it arises, e.g. when one has access only to excited states, but transitions to the ground state are in very different energy range. As an invariant algebra, it occurs for GKSL generators with zero temperature,  with $ X_1$ corresponding to the observables supported on the space with restricted upper bound for the number of particles \cite{torres2014closed, teretenkov2020one}. 
The general case \eqref{eq:observableFromTheAlgebra} can be considered as a combination of these three cases and the choice of basis defined by total unitary rotation $U$.

\section{Markovian quantum dynamics and the probability of multi-time measurement outcomes}
\label{sec:Markovian}

There are two ''ingredients'' in mathematical description of Markovian dynamics of open quantum systems. The first one is the assumption that the dynamics of the density matrix is
described by a continuous one-parameter semigroup of completely positive
trace-preserving maps $\Phi_t$,
\begin{equation}\label{eq:semigroupaffectingrho}
\rho(t) = \Phi_{t-t_0}(\rho(t_0)),
\end{equation}
where $\rho(t)$ is the system density matrix at time $t$. The generators $\mathcal{L}$ of such semigroups
\begin{equation}\label{eq:semigroop}
\Phi_t = e^{\mathcal{L} t}
\end{equation}
have the GKSL form \cite{gorini1976completely, lindblad1976generators}
\begin{equation}\label{eq:GKSLgenerator}
		\mathcal{L}(\rho) = -i [H,\rho] + \sum_{j} \left(L_j \rho L_j^{\dagger} - \frac12 \{L_j^{\dagger} L_j, \rho \} \right),
\end{equation}
where $L_j  \in \mathbb{C}^{n \times n} $ are the Lindblad (jump) operators, $H = H^{\dagger} \in \mathbb{C}^{n \times n}$ is the Hamiltonian, $\rho \in \mathbb{C}^{n \times n}$.

But the density matrix can only describe the results of one-time measurements. Thus,  the second necessary ''ingredient'' is general quantum regression formulae describing multi-time measurements. The relation between such multi-time statistics and
decoherence-free structures, including a multi-time generalization of the
decoherence-free-subspace concept, was discussed in~\cite{Teretenkov2023MemoryTensor}. There are several forms of regression formulae. Usually in physical literature it is formulated in terms of some time-ordered multi-time correlation functions \cite[Section 3.2.4]{breuer2002theory}, but here we will use its equivalent form \cite[Equation (164)]{milzQuantumStochasticProcesses2020} in terms of probabilities as it is more appropriate for our further usage.

To describe a quantum measurement, we use the  notion of instrument (with finite set of outcomes). Let us recall its definition.
\begin{Definition}
    A function $\mathcal{E}_{x}$ from a finite set $\mathcal{X}$ to the set of completely positive, trace non-increasing maps is called an instrument  (with finite set of outcomes) if
    \begin{equation}
        \sum_{x \in \mathcal{X}}\mathcal{E}_{x}  
    \end{equation}
    is a completely positive trace-preserving map.
\end{Definition}
Let $ \mathcal{E}^{(j)}_{x_j}$, $j =1, \ldots, N$ be  a sequence of instruments. Then the probability of obtaining the outcomes $x_j \in \mathcal{X}_j$ at the time moments $t_j$ in the case of Markovian dynamics is given by  \cite[Equation (164)]{milzQuantumStochasticProcesses2020}
	\begin{align}
		p&(x_N, t_N; \ldots; x_0, t_0) \nonumber\\
        &= \operatorname{Tr} \mathcal{E}^{(N)}_{x_N}  \Phi_{t_N-t_{N-1}} \ldots \mathcal{E}^{(1)}_{x_1} \Phi_{t_1 - t_0} \mathcal{E}^{(0)}_{x_0} \rho(t_0). \label{eq:multiTime}
	\end{align}

There are other approaches to the definition of quantum Markovian dynamics \cite{li2018concepts}, but in this work we assume that Equations \eqref{eq:semigroop}, \eqref{eq:GKSLgenerator}, \eqref{eq:multiTime} hold.

Since Equation~\eqref{eq:multiTime} is valid for arbitrary time moments and instruments, it can be interpreted as the fact that the density matrix at time $t_N$ has the form \begin{equation}
\rho(t_N) = \frac{\mathcal{E}^{(N)}_{x_N}  \Phi_{t_N-t_{N-1}} \ldots \mathcal{E}^{(1)}_{x_1} \Phi_{t_1 - t_0} \mathcal{E}^{(0)}_{x_0} \rho(t_0)}{ \operatorname{Tr} \mathcal{E}^{(N)}_{x_N}  \Phi_{t_N-t_{N-1}} \ldots \mathcal{E}^{(1)}_{x_1} \Phi_{t_1 - t_0} \mathcal{E}^{(0)}_{x_0} \rho(t_0)}.
\end{equation}
It can be rewritten in the recursive form
\begin{equation}\label{eq:recursiveDensityMatrix}
\rho(t_N) =  \frac{\mathcal{E}^{(N)}_{x_N}  \Phi_{t_N-t_{N-1}} \rho(t_{N-1})}{ \operatorname{Tr} \mathcal{E}^{(N)}_{x_N}  \Phi_{t_N-t_{N-1}} \rho(t_{N-1})},
\end{equation}
which allows one to rewrite Equation~\eqref{eq:multiTime} in the form
\begin{align}
p&(x_N, t_N; \ldots; x_0, t_0) \nonumber \\
&= p(x_N, t_N| x_{N-1}, t_{N-1})  \ldots p(x_1, t_1| x_0, t_0), \label{eq:probabiltiyInTermsOfCond}
\end{align}
where
\begin{equation}\label{eq:condProb}
p(x_j, t_j| x_{j-1}, t_{j-1}) = \operatorname{Tr} \mathcal{E}^{(j)}_{x_j}  \Phi_{t_j-t_{j-1}} \rho(t_{j-1})
\end{equation}
for $j =1, \ldots, N$.

Semigroup \eqref{eq:semigroop} describes the quantum Markovian dynamics in the Schrödinger picture, but some notions we use above are usually introduced in the Heisenberg picture, which is described by the adjoint semigroup
\begin{equation}
\Phi_t^* = e^{\mathcal{L}^* t}.
\end{equation}
Here $*$ denotes the adjoint of a linear map on matrices (a superoperator) with respect to the Hilbert–Schmidt inner product $\operatorname{Tr} X^{\dagger} Y$ of matrices  $X , Y \in \mathbb{C}^{n \times n}$. If $\mathcal{L}$ is defined by Equation \eqref{eq:GKSLgenerator}, then the explicit form of $\mathcal{L}^*$  is 
\begin{equation}
\mathcal{L}^*(X) =  i [H,X] + \sum_{j} \left(L_j^{\dagger} X L_j - \frac12 \{L_j^{\dagger} L_j, X \} \right),
\end{equation}
where $X \in \mathbb{C}^{n \times n}$.

Equation~\eqref{eq:multiTime} can be represented in the Heisenberg picture
\begin{align*}
p&(x_N, t_N; \ldots; x_0, t_0) \\
&= \operatorname{Tr}  \rho(t_0)   (\mathcal{E}^{(0)}_{x_0})^*  \Phi_{t_1 - t_0}^*(\mathcal{E}^{(1)}_{x_1})^*  \ldots \Phi_{t_N-t_{N-1}}^* (\mathcal{E}^{(N)}_{x_N})^*(I_n).
\end{align*}

\section{Decoherence-free subalgebras}\label{se:DFA}
        
As we discussed in the introduction, many physically relevant conditions on the structure of  $\mathcal{L}^* $ arise from the fact that $\Phi_{t}^*$ leaves a matrix *-algebra $	\mathcal{N}$ invariant, i.e. $\Phi_{t}^*\mathcal{N} \subseteq \mathcal{N} $. 

One important case of such invariant matrix *-algebra is the  decoherence-free subalgebra. Usually it is defined in the following way \cite{dhahri2010decoherence, dhahri2011decoherence, deschamps2016structure}.
\begin{Definition}
    The set  of all $X \in \mathbb{C}^{n \times n}$ such that $\forall t \geqslant 0$ 
	\begin{equation}\label{eq:decohFreeDef1}
		\Phi^*_t(X^{\dagger}X) = (\Phi^*_t(X) )^{\dagger} \Phi^*_t(X)
	\end{equation}
	and
	\begin{equation}\label{eq:decohFreeDef2}
		\Phi^*_t(X X^{\dagger}) = \Phi^*_t(X)  (\Phi^*_t(X))^{\dagger}
	\end{equation}
    is called a decoherence-free  subalgebra for the completely positive trace-preserving semigroup $\Phi_t$.
\end{Definition}
	Then it can be proved that such a set is indeed a *-algebra \cite{evans1977irreducible} and that the semigroup $\Phi_{t}^* $ leaves it invariant \cite{deschamps2016structure}.

    But more physics behind such a notion is uncovered through an equivalent
maximality characterization. Here and below, by a maximal matrix $*$-algebra
with a given property we mean a matrix $*$-subalgebra that has this property
and is maximal with respect to inclusion among all matrix $*$-subalgebras
of $\mathbb C^{n\times n}$ satisfying the same property. In other words,
if $\mathcal M$ is another matrix $*$-subalgebra with this property and
$\mathcal N \subseteq \mathcal M$, then $\mathcal M=\mathcal N$.

With this convention, the decoherence-free subalgebra can be characterized
as the maximal matrix $*$-algebra $\mathcal{N}$ such that    \begin{equation}\label{eq:effectiveUnitary}
			\Phi_{t}^*\mathcal{N}  = e^{i[\tilde{H}, \; \cdot \;] t} \mathcal{N}
		\end{equation}
    for some effective Hamiltonian $\tilde{H} = \tilde{H}^{\dagger} \in \mathbb{C}^{n \times n}$ (choice of which is not unique). Thus, the dynamics is effectively unitary for the observables from the decoherence-free subalgebra and such observables are not affected by decoherence or dissipation processes.

    If the *-algebra contains the identity $I_n$  of the full matrix algebra  $\mathbb{C}^{n \times n}$, then   $n_0=0$ and Equation \eqref{eq:algebraCanonicalForm} is reduced to 
\begin{equation}\label{eq:algebraCanonicalFormWithIdentity}
    \mathcal{N} = U \left(  \bigoplus_{k=1}^K (\mathbb{C}^{n_k \times n_k} \otimes I_{m_k}  ) \right) U^{\dagger}.
\end{equation}
Since $\Phi_{t} $ is trace-preserving, the adjoint map $\Phi_{t}^* $ is unital, i.e. $\Phi_{t}^*I_n = I_n$. Thus, Equations \eqref{eq:decohFreeDef1}-\eqref{eq:decohFreeDef2} are satisfied for $I_n$.

If the canonical form \eqref{eq:algebraCanonicalFormWithIdentity} of the decoherence-free subalgebra is known, then the structure of such a GKSL generator was defined in \cite[Theorem 3.2]{deschamps2016structure}. We use this result in the following form, similar to \cite[Theorem 9]{hasenohrl2023generators}.
    \begin{Theorem}\label{th:structure}
    If the decoherence-free subalgebra $\mathcal{N}$ of semigroup \eqref{eq:semigroop} has  form \eqref{eq:algebraCanonicalFormWithIdentity}, then the matrices $H$ and  $L_j$ in \eqref{eq:GKSLgenerator} can be chosen as
        \begin{equation}\label{eq:structure}
        \begin{aligned}
            H &= \sum_{k=1}^K P_k^{\dagger} ( \kappa_k \otimes I_{m_k} +I_{n_k} \otimes \mu_k) P_k,\\
			  L_j &= \sum_{k=1}^K P_k^{\dagger} (I_{n_k} \otimes \beta_{j,k}) P_k, 
            \end{aligned}
		\end{equation}
		where $  \beta_{j,k}, \mu_k  \in \mathbb{C}^{m_k \times m_k}$, $ \mu_k = \mu_k^{\dagger} $,  $  \kappa_k  \in \mathbb{C}^{n_k \times n_k}$, $\kappa_k = \kappa_k^{\dagger}$,  and
    \begin{equation}
		P_k = P_k^{\oplus}  U^{\dagger},
	\end{equation}
        where $P_k^{\oplus} \in \mathbb{C}^{ n_k m_k \times n }$ is the projector from $\mathbb{C}^{n}$ to  $\mathbb{C}^{n_k m_k}$.
    \end{Theorem}

    The effective Hamiltonian in Equation \eqref{eq:effectiveUnitary} can be chosen as 
		\begin{equation}
			\tilde{H} = \sum_{k=1}^K P_k^{\dagger} ( \kappa_k \otimes I_{m_k}) P_k.
		\end{equation}

    In particular, in  all spaces $P_k \mathbb{C}^n \simeq \mathcal{H}_k^S \otimes  \mathcal{H}_k^B$, where $\dim \mathcal{H}_k^S = n_k$ and $\dim \mathcal{H}_k^B = m_k$, the spaces $\mathcal{H}_k^S$ support the decoherence-free subsystems \cite{ticozzi2008quantum}. And $\bigoplus_{k : m_k =1}\mathcal{H}_k^S$ is a decoherence-free subspace \cite{lidar1998decoherence, lidar2003decoherence, lidar2014review, agredo2014decoherence}.

 \section{Generalized Markovian embedding and maximum likelihood invariant algebra estimation}
 \label{se:embedding}

 Similar to \cite{luchnikov2020machine}, we assume that the observed dynamics
admits a Markovian embedding: there exists an enlarged finite-dimensional
system whose density matrix evolves according to a Markovian GKSL semigroup
in the sense of Section~\ref{sec:Markovian}. The dynamics directly accessible
in the experiment, however, is obtained only through measurements of
observables from a fixed matrix $*$-algebra $\mathcal A$. Therefore it need
not form a closed Markovian semigroup on $\mathcal A$ itself. In particular,
the conditional probabilities of future outcomes can depend on the previous
measurement history through hidden degrees of freedom of the embedding.

  In the special case, when elements of $\mathcal{A}$ have the form defined by Equation~\eqref{eq:subsystemAlgebra}, one obtains the usual Markovian embedding, so we will call our approach  generalized Markovian embedding. Moreover, we aim not only to reconstruct the environment, but also to identify the structure of the total Markovian dynamics of the system and environment. We will assume that such a structure can  be defined by the condition that this total dynamics leaves some matrix $*$-algebra of observables $\mathcal{N}$ invariant. In this work we focus on the case where $\mathcal{N}$ is decoherence-free subalgebra and use structure Theorem \ref{th:structure}, but  it can be generalized to other invariant algebras via \cite[Theorem 6]{hasenohrl2023generators}.

We will consider the special case of Equation~\eqref{eq:multiTime} assuming $\mathcal{E}^{(j)}_{x_j}$ are projective measurements
\begin{equation}
\mathcal{E}^{(j)}_{x_j} (\rho)= E_{x_j}^{(j)} \rho E_{x_j}^{(j)}, \qquad j = 1, \ldots, N,
\end{equation}
where  $E_{x_j}^{(j)}$ are orthogonal projectors from algebra $\mathcal{A}$ forming a resolution of the identity for each fixed $j$, i.e.
\begin{align*}
E_{x_j}^{(j)} E_{x_j'}^{(j)} =& \delta_{x_j, x_j'} E_{x_j}^{(j)}, \qquad E_{x_j}^{(j)} = (E_{x_j}^{(j)})^{\dagger}, \\
&\sum_{x_j \in X_j}E_{x_j}^{(j)} = I_{n-n_0}.
\end{align*}
If we know what instruments we applied at each time $\mathcal{E}^{(j)}_{x_j} $ and we have obtained the sequence of results $\{x_N, \ldots, x_1\}$ at times $\{ t_N \geqslant \ldots  \geqslant t_1 \}$, thus we effectively obtain a sequence of projectors 
\begin{equation}
E^{(j)} \equiv E_{x_j}^{(j)}
\end{equation}
belonging to the algebra $\mathcal{A}$. Similar to \cite{luchnikov2020machine}  we will consider $E^{(j)}$ as our dataset in machine learning sense. We will divide it into training and test datasets. In \cite{luchnikov2020machine} $E^{(j)} $ was assumed to be randomly chosen in the form $|\phi_j\rangle \langle \phi_j| \otimes I_B$, where $|\phi_j\rangle $ is a unit vector from the Hilbert space describing the system degrees of freedom and $I_B$ is an identity matrix in the reservoir Hilbert space whose dimension was considered as a hyperparameter.

As the initial instrument $ \mathcal{E}^{(0)}_{x_0}$ we will use either the identity superoperator, corresponding to the absence of measurement,  or  state preparation instrument $ \mathcal{E}^{(0)}\rho = \sigma \operatorname{Tr}\rho $ with fixed and known density matrix $\sigma$.

Note that while the algebra $\mathcal{A}$ is known and fixed, being defined by our experimental capabilities,  the algebra $\mathcal{N}$ is what we want to learn from the experimental data and should be considered as a hyperparameter of our model. 

We will assume  the measurements are performed at equally spaced times so that $t_j =  j \tau$. Then all $\Phi_{t_j -t_{j-1}}$ in Equation~\eqref{eq:multiTime} are equal 
\begin{equation}
\Phi_{t_j -t_{j-1}} = \Phi_{\tau},
\end{equation}
which in practice can be computed by vectorization \cite{threeapproaches}. We will assume that the time step $\tau$ is given and fixed, so we will omit the dependence of $\Phi_{\tau}$ on $\tau$. At the same time, the structure of $\Phi_{\tau}$ is defined by  the decoherence-free subalgebra $\mathcal{N}$ according to Equation~\eqref{eq:structure} and Equations~\eqref{eq:semigroop}, \eqref{eq:GKSLgenerator}. So to emphasize the dependence of $\Phi_{\tau}$ on $\mathcal{N}$ and the parameters participating in  Equation~\eqref{eq:structure} we will write
\begin{equation}
\Phi(\nu,  \lambda_{\nu}),
\end{equation}
where $\nu =( \{n_k^{\mathcal{N}}, m_k^{\mathcal{N}}\})$ defines the decoherence free subalgebra structure through Equation \eqref{eq:algebraCanonicalFormWithIdentity} and can be considered as a set of hyperparameters of the model, $\lambda_{\nu} = ( U^{\mathcal{N}}, \, \{\beta_{j,k}\}, \left\{\kappa_k\right\}, \left\{\mu_k\right\})$ defines the parameters of the subalgebra and the GKSL generator of the semigroup with such a decoherence free subalgebra through Equation \eqref{eq:structure} and can be considered as a set of model parameters.

We will use the logarithm of probability of observed data defined by  Equation~\eqref{eq:multiTime} as  likelihood function 
\begin{align}
	&F(\nu,  \lambda_{\nu}, \sigma; E)\\
    &= \ln \operatorname{Tr} \left( E^{(N)} \ldots \Phi(\nu,  \lambda_{\nu})(E^{(1)}\Phi(\nu,  \lambda_{\nu})(\sigma) E^{(1)}) \ldots E^{(N)}\right) \nonumber,
\end{align}
where $E = (E^{(1)}, \ldots, E^{(N)})$ is a sequence of measurements, $\sigma$ is an initial density matrix, which can be considered either as an additional parameter, or as a fixed known matrix.  

Using Equation \eqref{eq:probabiltiyInTermsOfCond}-\eqref{eq:condProb} the likelihood function can be rewritten in additive form
\begin{equation}
F(\nu,  \lambda_{\nu}, \sigma; E) = \sum_j f(\nu,  \lambda_{\nu}; E^{(j)}, \rho^{(j-1)}(\nu,  \lambda_{\nu}, \sigma)),
\end{equation}
where
\begin{equation}\label{eq:functionalfinal}
f(\nu,  \lambda_{\nu}; E^{(j)}, \rho) = \ln \operatorname{Tr}( E^{(j)} \Phi(\nu,  \lambda_{\nu})  \rho)
\end{equation}
and $  \rho^{(j-1)}(\nu,  \lambda_{\nu}, \sigma)$ is defined recursively by Equation~\eqref{eq:recursiveDensityMatrix}, i.e.
\begin{equation}
  \rho^{(j)}(\nu,  \lambda_{\nu}, \sigma) = \frac{ E^{(j)} (\Phi(\nu,  \lambda_{\nu}) \rho^{(j-1)}(\nu,  \lambda_{\nu}, \sigma)) E^{(j)} }{\operatorname{Tr} E^{(j)}  \Phi(\nu,  \lambda_{\nu}) \rho^{(j-1)}(\nu,  \lambda_{\nu}, \sigma)}
\end{equation}
with the initial condition
\begin{equation}\label{initial_state_sigma}
 \rho^{(0)}(\nu,  \lambda_{\nu}, \sigma) = \sigma.
\end{equation}

We also assume that algebras $\mathcal{N}$ and $\mathcal{A}$ are in a generic relative position to each other, which means that neither $\mathcal{N} \subseteq \mathcal{A}$ nor $\mathcal{A} \subseteq \mathcal{N}$, and that $\mathcal{N} \cap \mathcal{A}$ is not some invariant algebra itself. This ensures that by measurements of observables from $\mathcal{A}$ one can obtain enough information to reconstruct the environment and the fact that the  structure of  $\mathcal{N}$ is not directly presented in   $\mathcal{A}$, but can be only restored from the generalized Markovian embedding and, thus, can be considered as actually hidden.

\section{Validation and application}
\label{se:numerical}

\begin{figure*}
\centering
\includegraphics[width=0.75\linewidth]{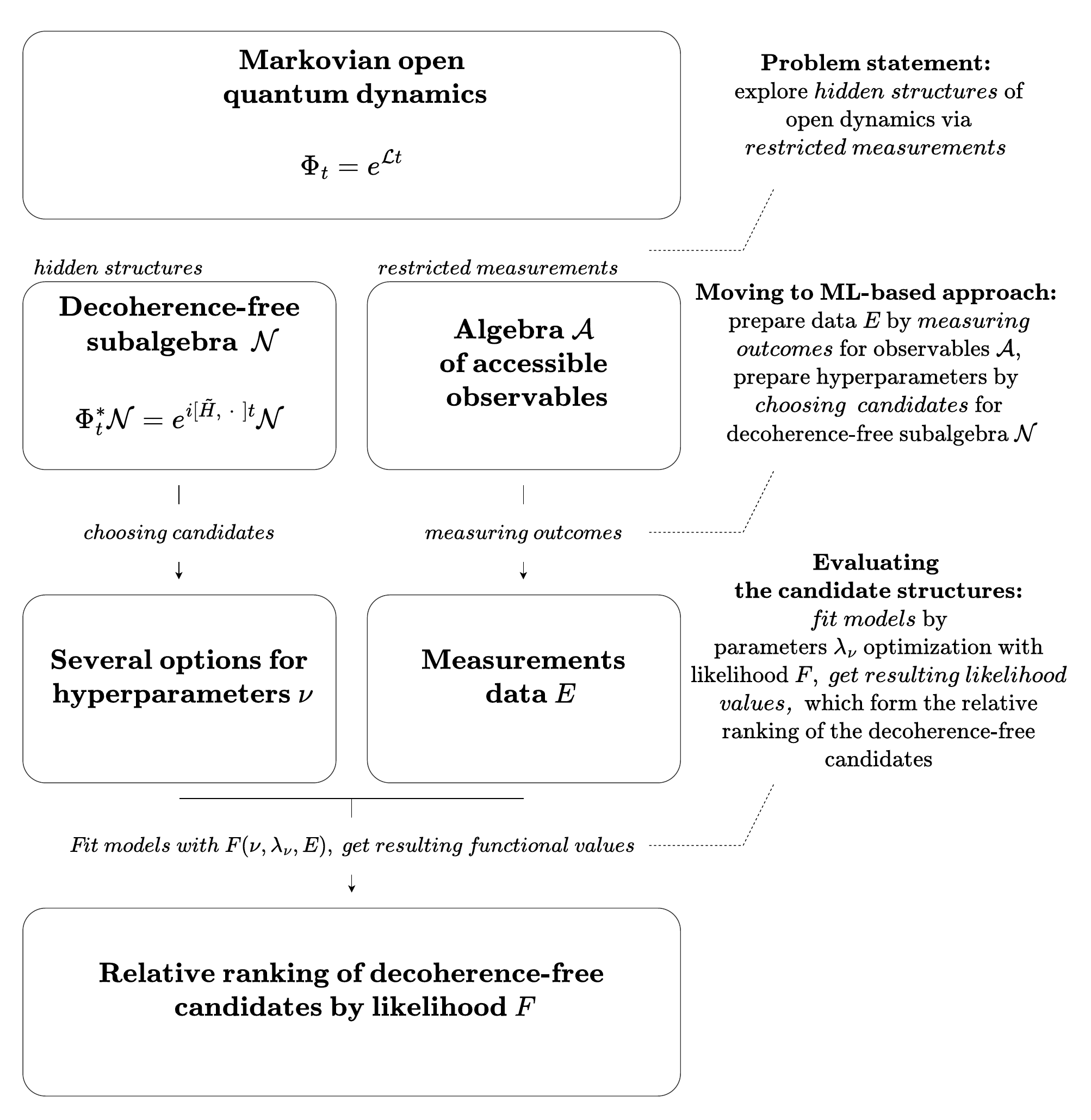}
 \caption{
\label{diag:framework}
Schematic workflow of the proposed approach for identifying hidden structural features of open quantum dynamics under restricted observational access. The accessible multi time measurement statistics are modeled through a finite-dimensional Markovian GKSL embedding. Both the algebra of accessible observables and the candidate hidden decoherence-free structures are represented by matrix $*$-algebras. Each candidate decoherence-free algebra specifies a set of structural hyperparameters, while the continuous parameters of the corresponding GKSL generator are fitted to the measurement data by maximum likelihood. The candidate structures are then ranked according to their test likelihood values.}

\end{figure*}

 In Sections \ref{se:algebras}-\ref{se:embedding}, we developed a likelihood-based machine-learning framework for identifying hidden decoherence-free structures from multi-time measurement data obtained under restricted observational access. Figure~\ref{diag:framework} provides a schematic summary of this framework. This section demonstrates that  it can efficiently identify hidden algebraic structures, is robust under variations in the training data, and yields physically meaningful effective models under observational restrictions. 

For the numerical experiments, the method was implemented in Python using the PyTorch framework \cite{PyTorch}, which provides automatic differentiation tools for solving the optimization problems arising during model training.

The numerical experiments use synthetic data generated with some chosen $ H $ and $ L_{j} $ operators in \eqref{eq:GKSLgenerator}. For each experiment, the obtained sequences of measurements $ \mathbf{E}  = \left\{ E_{k} \right\} $ form two datasets: a training dataset $ \mathbf{E}_{\text{train}} $ and a testing dataset $ \mathbf{E}_{\text{test}} $. With respect to the objective functional \eqref{eq:functionalfinal}, the first is used to solve the optimization problem for the adjustment of the model parameters $ \lambda_{\nu} $, while the second is applied to analyze the evolution of the functional. Since the conducted tests reveal neither significant differences in the amplitude of the objective functional between the training and test datasets nor considerable overfitting effects for any choice of the model hyperparameters $\nu$, we report only the values obtained for the test datasets. As the functional amplitude depends on the number of measurements in sequences, we also normalize the results by~$ N $. 

The model training procedure based on the described datasets is performed using a batch method. At each step, all measurement sequences within a batch are processed, and the value of the objective functional $ F $ is estimated by averaging its values over all elements of the batch.

Since the choice of basis in Eqs.~\eqref{eq:structure} can be absorbed into the unitary transformation $U$ in Eq.~\eqref{eq:algebraCanonicalForm}, we fix the basis so that $H$ (and hence $\kappa_k$ and $\mu_k$) is diagonal in the canonical form. The initial system state $ \sigma $ is the maximally mixed state $ I_n/n $ for all of the conducted experiments. 

Training of the models is carried out by solving an optimization problem using a gradient-based approach.  Gradient-based optimization can be sensitive to the initial starting point. In particular, presence of ravine regions or regions with suboptimal local minima or other trapping features in the optimization landscape of the objective functional may lead to a significant decrease of gradient-based optimization when starting in a vicinity of such a region, and for careful analysis of the optimization averaging over random initial parameters has to be performed~\cite{Pechen_Tannor_2012}. Following this approach to minimize such sensitivity, for the majority of the experiments below we perform several runs with different random initial values of the parameters $ \lambda_{\nu} $  and then select the best, and in some cases, average, outcomes.

The following subsections present several independent groups of numerical tests. Specifically, we use data that are generated with generators $ \mathcal{L} $ possessing certain invariant structures \eqref{eq:structure}: to demonstrate numerically how the proposed approach helps to identify these structures and to clarify the role played by their embedding hierarchy in this process \eqref{subsec:guessing}; to study how the balance between the number of used training chains and their length affects the quality of model learning  \eqref{subsec:numberlength}; to evaluate how the observables algebra structure influences information about the system contained in the resulting data in terms of the applied approach \eqref{subsec:inferring}. After that, we present results of structure learning runs for data related to the three-qubit quantum electrodynamical waveguide physical model \eqref{subsec:physmodel}. 

\subsection{Invariant structure "guessing" tests and embeddings hierarchy}\label{subsec:guessing}

In this subsection, we validate the proposed method by demonstrating that it can correctly and efficiently reconstruct the hidden algebra from synthetic measurement data. The data are generated by dynamics governed by a decoherence-free algebra that is not provided to the method, so it must infer it solely from the measurement results. So the first group of the experiments focuses on comparing the performance of the trained models for all possible structures $ \nu $, investigating whether it is feasible to find a correct algebra $ \nu_{\mathbf{E}} $ related to the considered data or not. Correctly “guessing” the right algebra is a sign of the validity of the approach and the correctness of the numerical implementation used. 

Note that the total number of structures is bounded below by the number of integer compositions of $ n $. Therefore --- even excluding structures that are identical up to permutations of the pairs $ \left\{n_{k}, \, m_{k} \right\} $ --- their total number grows at least  super-polynomially (see Appendix~\ref{app:number_structures} for details). Consequently, this complete comparison is challenging to perform for larger dimensions. For this reason, we start by considering $ n = 4 $. In this case, there are $ 18 $ different structures $ \nu $ to investigate (or $ 11 $ structures up to permutations).

\begin{figure}[h!]
\centering
\includegraphics[width=\linewidth]{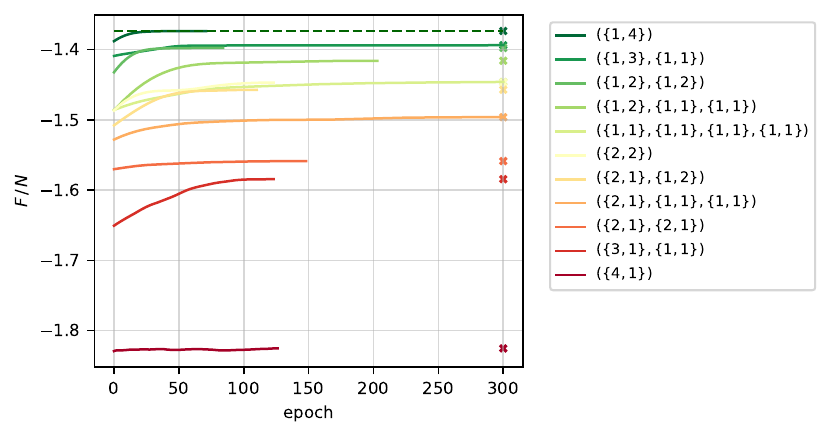}
\caption{Dynamics of the functional values on the test dataset for the models considered in the the numerical experiments from Table \ref{tab:res1A}. Each cluster of the models is presented with one model. Plots are shown up to the best epoch. Best achieved objective values are marked with the crosses at the final epoch.}
\label{fig:N4}
\end{figure}

The first dataset has been generated with the free open system dynamics structure $ \nu_{\mathbf{E}} = \left(\left\{1, \, 4\right\}\right) $. Table \ref{tab:res1A} provides a complete description of the results, sorted by the value of the objective functional, while Figure \ref{fig:N4} provides the learning plots for this case.

To analyze the results, firstly recall that the order of the blocks in the canonical form can be changed by a permutation unitary, which can be absorbed into the unitary matrix in Eq.~\eqref{eq:algebraCanonicalFormWithIdentity}. Therefore, the optimization with different orders of the pairs $\{n_k^{\mathcal{N}}, m_k^{\mathcal{N}}\}$ in the hyperparameter $ \nu $ should yield the same maximal value of the functional $ F $, but with a correspondingly permuted optimal parameter $ U $. This is consistent with the results in Table~\ref{tab:res1A}, where very close maximal values have been obtained for such structures. For this reason, in the remaining experiments, we consider algebra structures only up to permutations of $ \left\{n_{k}^{\mathcal{N}}, \, m_{k}^{\mathcal{N}} \right\}$ pairs in $ \nu $.

\begin{table}[h!]
    \centering
    \begin{tabular}{|c|c|c|c|}
        \hline
        $ \nu $ & $ F/N $ & $ \mathcal{T}_{\text{best}}$ & $ \Delta_{\mathcal{T}} F / N $ \\ \hline
        $ \left(\left\{1, 4 \right\} \right) $ & $ -1.374 $ & $ 71 $  & --- \\ \hline
         $ \left(\left\{1, 1 \right\}, \left\{1, 3\right\} \right) $ & $ -1.393 $ & $ 68 $ & --- \\ 
        $ \left(\left\{1, 3 \right\}, \left\{1, 1\right\} \right) $ & $ -1.394  $ & $ 300 $ & $ \sim 10^{-7} $ \\ 
       \hline
        $ \left(\left\{1, 2 \right\}, \left\{1, 2\right\} \right) $ & $ -1.398 $ & $ 84 $ & --- \\ \hline
        $ (\left\{1, 2 \right\}, \left\{1, 1\right\}^{2}) $ & $ -1.416 $ & $ 203 $ & --- \\
        $ \left(\left\{1, 1 \right\}, \left\{1, 2\right\}, \left\{1, 1 \right\} \right) $ & $ -1.417 $ & $ 253 $ & --- \\ 
        $ (\left\{1, 1 \right\}^{2}, \,  \left\{1, 2\right\}) $ & $ -1.417 $ & $ 73 $ & --- \\
        \hline
        $ (\left\{1, 1\right\}^{4}) $ & $ -1.446 $ & $ 300 $ & $ \sim 10^{-5} $ \\ \hline
        $ \left(\left\{2, 2 \right\} \right) $ & $ -1.447 $ & $ 123 $ & --- \\ \hline 
        $ \left(\left\{1, 2\right\}, \left\{2, 1\right\} \right) $ & $ -1.456 $ & $ 128 $ & --- \\
        $ \left(\left\{2, 1 \right\},  \left\{1, 2\right\} \right) $ & $ -1.457 $ & $ 110 $ & ---\\
        \hline
        $ \left( \left\{1, 1\right\}^{2}, \,  \left\{2, 1\right\} \right) $ & $ -1.491 $ & $ 300 $ & $ \sim 10^{-5} $ \\
        $ \left(\left\{1, 1\right\}, \left\{2, 1\right\}, \left\{1, 1\right\} \right) $ & $ -1.493 $ & $ 181 $ & --- \\
        $ (\left\{2, 1\right\},  \left\{1, 1\right\}^{2}) $ & $ -1.496 $  & $ 300 $ & $ \sim 10^{-6} $ \\ 
        \hline
        $ \left(\left\{2, 1\right\},  \left\{2, 1\right\} \right) $ & $ -1.559 $ & $ 148 $ & --- \\ \hline
        $ \left(\left\{1, 1\right\}, \left\{3, 1\right\}  \right) $ & $ -1.584 $ & $ 266 $ & --- \\
        $ \left(\left\{3, 1 \right\}, \left\{1, 1 \right\} \right) $ & $ -1.585 $ & $ 123 $ & --- \\ \hline
        $ \left(\left\{4, 1 \right\} \right) $ & $ -1.825 $ & $ 126 $ & --- \\ \hline
    \end{tabular}
    \caption{Structures exploration results for the synthetic data of $ S_{\text{train}} = S_{\text{test}} = 100 $ chains with $ N = 200 $ measurements generated by the free open dynamics structure $ \nu_{\mathbb{E}} = \left(\left\{1, \, 4\right\} \right) $. We consider all possible options for $ \nu $: from unitary evolution $ \left\{4, 1\right\} $ up to free open dynamics $ \left\{1, 4 \right\} $.  Best of 3 runs for random initial parameters values is chosen for all of the experiments; functional value for the test chains $ \mathbf{E}_{\text{test}} $ and used generative parameters $ \lambda_{\nu_{\mathbf{E}}} $ and hyperparameters $ \nu_{\mathbf{E}} $ is $ F_{\mathbf{E}} / N  = -1.373 $; total epochs number for each experiment is $ \mathcal{T} = 300 $. To verify convergence, the epoch $ \mathcal{T}_{\text{best}} $ at which the highest value of the testing objective functional has been reached is provided for each model, as well as the test functional absolute change $ \Delta_{\mathcal{T}} F / N $ at the final epoch for the models with $ \mathcal{T}_{\text{best}} \geqslant 0.95 \mathcal{T} $. As expected for the considered data, order of the structures formed by the obtained functional values completely follows the embedding structure \ref{diag:embedn4}, while the structures identical up to permutations of the pairs $ \left\{n_{k}^{\mathcal{N}}, \, m_{k}^{\mathcal{N}} \right\}$ form clusters with almost the same functional values.\\}
    \label{tab:res1A}
\end{table}
Next, note that some of the algebras can be embedded into each other up to isomorphism, which can also be absorbed by parameters $ U $. Assume that $\tilde{\mathcal N}$ is embedded into $\mathcal N$ up to isomorphism. Then each block of $\mathcal N$ decomposes into blocks of $\tilde{\mathcal N}$. More precisely, for every $k$
there exist non-negative integers $a_{k\ell}\in \mathbb{Z}_+$ such that
\begin{equation}
\mathbb{C}^{n_k}\otimes \mathbb{C}^{m_k}
\;\cong\;
\bigoplus_{\ell}
\left(
\mathbb{C}^{\tilde{n}_\ell}\otimes \mathbb{C}^{a_{k\ell}}
\right)
\otimes
\mathbb{C}^{m_k},
\label{eq:block-decomposition}
\end{equation}
then 
\begin{equation}
\bigoplus_k \mathbb{C}^{n_k}\otimes \mathbb{C}^{m_k}
\;\cong\;
\bigoplus_{\ell}
\mathbb{C}^{\tilde{n}_\ell}
\otimes \mathbb{C}^{\tilde{m}_\ell},
\end{equation}
where
\begin{equation}
\mathbb{C}^{\tilde{m}_\ell} \;\cong\;
\bigoplus_k 
\left(
\mathbb{C}^{a_{k\ell}} \otimes \mathbb{C}^{m_k}
\right).
\end{equation}
Thus,  $ \tilde{N} $ is embedded into $\mathcal N$ up to isomorphism  if and only if there exist matrix $a \in \mathbb{Z}_+^{K \times L}$  satisfying
\begin{align}
n_k &= \sum_{\ell=1}^{L} a_{k\ell}\, \tilde{n}_\ell,
\qquad \text{for all } k=1,\dots,K,
\label{eq:block-dimension-condition}
\\[6pt]
\tilde{m}_\ell &= \sum_{k=1}^{K} a_{k\ell}\, m_k,
\qquad \text{for all } \ell=1,\dots,L.
\label{eq:multiplicity-condition}
\end{align}

An embedding relation between two algebras corresponding to hyperparameters $\nu$ and $\tilde{\nu}$ is important  as it affects their relative performance. If one decoherence-free algebra is embedded into another, then the corresponding class of GKSL generators associated with the first algebra contains the class associated with the second. So, it means that the first machine learning model is more complex than the second one. In this sense, the introduced models form an embedding, which is an inversion of the decoherence-free algebras one.

\begin{figure}[h!]
\centering
\includegraphics[width=0.95\linewidth]{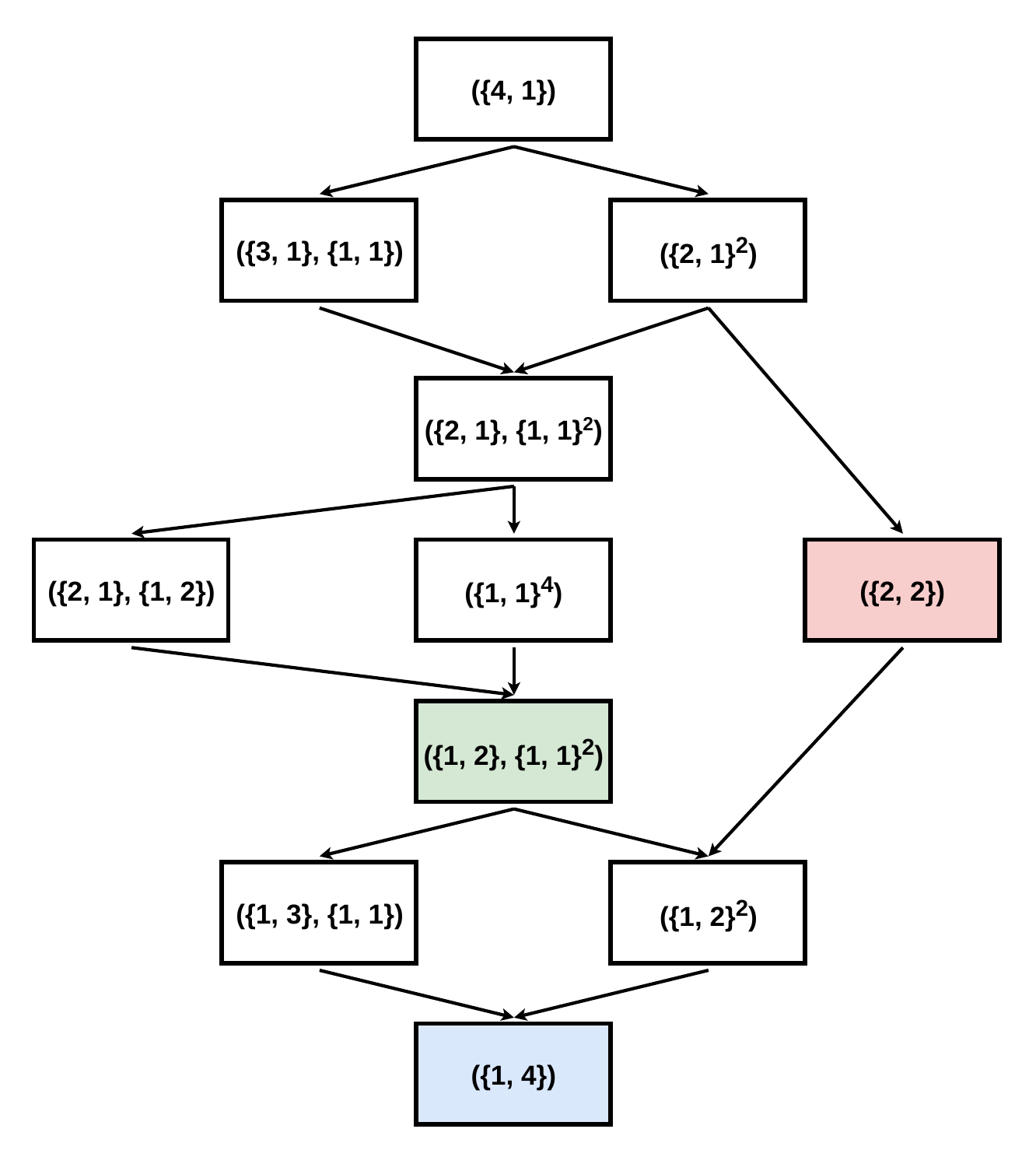}
\caption{Decoherence-free algebras and corresponding models embeddings hierarchy, arranged from unitary dynamics (the largest algebra) to free open GKSL dynamics (the smallest algebra). That is, the diagram shows transitions from the simplest possible model (top) to the most complex one (bottom). The colors indicate the structures $ \nu_{\mathbf{E}} $ being used in the conducted experiments: blue for the Table \ref{tab:res1A} experiments, red~---~Table~\ref{tab:res1B}, green --- Table~\ref{tab:res1C}.}
\label{diag:embedn4}
\end{figure}

Since the presented definition of algebras/models embedding is transitive, it is possible to build their hierarchy for a fixed value $ n $. E.g. see the complete diagram for the $ n = 4 $ case in Figure \ref{diag:embedn4}.

 It is reasonable to expect the hierarchy to have a noticeable effect on the relative performance of the models, as more complex models often perform at least as well as (and often better than) simpler ones. This can be observed in the experiments, which results are presented in Table \ref{tab:res1A}. In this table and throughout the rest of the tables, we denote $ F(\nu, \lambda_{\nu}, \mathbf{E}_{\text{test}})/N $ as $ F/N $ and, similarly, $ F(\nu_{\mathbf{E}}, \lambda_{\nu_{\mathbf{E}}}, \mathbf{E}_{\text{test}})/N $ as $ F_{\mathbf{E}}/N $ .
For this case, we consider data corresponding to the most general open Markovian dynamics described by equations \eqref{eq:semigroupaffectingrho}--\eqref{eq:GKSLgenerator}, so the algebra $ \nu_{\mathbf{E}} = \left\{1, 4\right\} $ has been used as the structure for data $ \mathbf{E} $ generation. The set of hyperparameters $ \nu $, ordered by the value of the functional, exactly corresponds to the hierarchy of embeddings: if the algebra of the first model is embedded into the algebra of the second, then the resulting value of the test objective functional $ F $ for the first model exceeds that of the second. For example, consider the structures $ \left(\left\{2,1\right\},\left\{2,1\right\}\right) $ and $ \left(\left\{2,2\right\}\right) $. 
The corresponding decoherence-free algebra of the second structure is embedded in that of the first, since the criterion \eqref{eq:block-dimension-condition}--\eqref{eq:multiplicity-condition} is satisfied for the integer matrix $ a^T=\begin{pmatrix}1 & 1\end{pmatrix} $. This embedding can be also seen from the right branch of the diagram in Figure \ref{diag:embedn4}.  Consequently, the class of GKSL generators compatible with $ \left(\left\{2,2\right\}\right) $ contains the class compatible with $ \left(\left\{2,1\right\}, \left\{2,1\right\}\right)$, and therefore its final value of $ F $ is no smaller. 

However, an important note is that this property strongly depends on the data and is not necessarily guaranteed to hold. In the previous series of experiments, all considered models are no more complex than the one used to generate the data. In contrast, the following experiments are carried out for simpler types of the dynamics, namely those which are related to the structures $ \nu_{\mathbf{E}} = (\left\{2,2 \right\}) $ (Table \ref{tab:res1B}) and $ \nu_{\mathbf{E}} = (\left\{1, 2 \right\},  \left\{1, 1 \right\}, \, \left\{1, 1 \right\})$ (Table \ref{tab:res1C}). The obtained results show that the specified property continues to hold only for models that are higher in the hierarchy (i.e. simpler models) than the one used for the data generation. At the same time, more complex models do not maintain the hierarchy order and underperform compared to the ground-truth model or even some other simpler models. For instance, note that, for both cases, models with the structure $ \nu = (\left\{1, 2 \right\}, \left\{1, 2\right\}) $ yield better results than models with $ \nu = (\left\{1, 4 \right\}) $, and all of them perform worse in comparison with $ \nu = \nu_{\mathbf{E}} $ models.

\begin{table}[h!]
    \centering
    \begin{tabular}{|c|c|c|c|}
        \hline
        $ \nu $ & $ F/N $ & $ \mathcal{T}_{\text{best}} $  &  $ \Delta_{\mathcal{T}} F / N $   \\ \hline
        $ \bm{\left(\left\{2, 2 \right\} \right)}$ & $ -1.288 $ & $ 300 $ & $ \sim 10^{-8}$ \\ \hline
        $ \left(\left\{1, 2 \right\}, \left\{1, 2\right\} \right) $ & $ -1.339 $ & $ 300 $ & $ \sim 10^{-8}$  \\ \hline
        $ (\left\{1, 2 \right\}, \left\{1, 1\right\}^{2}) $ & $ -1.342 $ & $ 300 $ & $ \sim 10^{-6}$ \\
        \hline
        $ \left(\left\{1, 4 \right\} \right) $ & $ -1.364 $ & $ 57 $ & --- \\ \hline 
        $ \left(\left\{1, 3 \right\}, \left\{1, 1\right\} \right) $ & $ -1.367 $ & $ 244 $ & --- \\ \hline
        $ \left(\left\{2, 1 \right\},  \left\{1, 2\right\} \right) $ & $ -1.371 $ & $ 300 $ & $ \sim 10^{-6}$ \\ \hline
        $ ( \left\{1, 1\right\}^{4} ) $ & $ -1.373 $ & $ 300 $ & $ \sim 10^{-6} $ \\ \hline
        $ ( \left\{2, 1\right\},  \left\{1, 1\right\}^{2}) $ & $ -1.377 $ & $ 300 $ & $ \sim 10^{-5} $ \\ \hline 
        $ \bm{\left(\left\{2, 1\right\},  \left\{2, 1\right\} \right)} $ & $ -1.402 $ & $ 300 $ & $ \sim 10^{-5} $ \\ \hline
        $ \bm{\left(\left\{3, 1 \right\}, \left\{1, 1 \right\} \right)} $ & $ -1.491 $ & $ 168 $ & --- \\ \hline
        $ \bm{\left(\left\{4, 1 \right\} \right)} $ & $ -1.653 $ & $ 249 $ & --- \\ \hline
    \end{tabular}
    \caption{Results of the experiments similar to those presented in Table \ref{tab:res1A} for the data generated with the $ \nu_{\mathbf{E}} = (\left\{2, 2 \right\}) $ setup. All runs settings are identical to what have been used previously, $ F_{\mathbf{E}} / N = -1.288 $. For each class of models that are equal up to permutations of the $ \left\{n_{k}, \, m_{k} \right\}$ pairs, only one structure is considered. The results for the structures (in bold) that are not lower in the embedding hierarchy than the one used to generate the data are consistent with this hierarchy, whereas this property does not hold for more complex models.}
    \label{tab:res1B}
\end{table}

\begin{table}[h!]
    \centering
    \begin{tabular}{|c|c|c|c|}
        \hline
        $ \nu $ & $ F/N $ & $ \mathcal{T}_{\text{best}} $ &  $ \Delta_{\mathcal{T}} F / N $ \\ \hline
        $ \bm{(\left\{1, 2 \right\}, \left\{1, 1\right\}^{2})} $ & $ -1.345 $ & $ 142 $ & --- \\
        \hline
        $ \left(\left\{1, 2 \right\}, \left\{1, 2\right\} \right) $ & $ -1.365 $ & $ 300 $ & $ \sim 10^{-7} $  \\ \hline
        $ \left(\left\{1, 3 \right\}, \left\{1, 1\right\} \right) $ & $ -1.367 $ & $ 84 $ & --- \\ \hline
        $ \bm{( \left\{1, 1\right\}^{4} )} $ & $ -1.369 $ & $ 282 $ & $ \sim 10^{-6} $ \\ \hline 
        $ \bm{\left(\left\{2, 1 \right\},  \left\{1, 2\right\} \right)} $ & $ -1.372 $ & $ 300 $ & $ \sim 10^{-5} $ \\ \hline
        $ \left(\left\{1, 4 \right\} \right) $ & $ -1.386 $ & $ 46 $ & --- \\ \hline 
        $ \bm{( \left\{2, 1\right\},  \left\{1, 1\right\}^{2})} $ & $ -1.396 $ & $ 30 $ & $ \sim 10^{-6} $ \\ \hline
        $ \left(\left\{2, 2 \right\} \right) $ & $ -1.414 $ & $ 300 $ & $ \sim 10^{-6} $ \\ \hline
        $ \bm{\left(\left\{2, 1\right\},  \left\{2, 1\right\} \right)} $ & $ -1.458 $ & $ 146 $ & --- \\ \hline
        $ \bm{\left(\left\{3, 1 \right\}, \left\{1, 1 \right\} \right)} $ & $ -1.497 $ & $ 267 $ & --- \\ \hline
        $ \bm{\left(\left\{4, 1 \right\} \right)} $ & $ -1.715 $ & $ 253 $ & --- \\ \hline
    \end{tabular}
    \caption{Results of the analogous to Table \ref{tab:res1A} testing runs for $ \nu_{\mathbf{E}} = (\left\{1, 2 \right\}, \left\{1, 1 \right\},  \left\{1, 1\right\})$. $ F_{\mathbf{E}} / N = -1.345 $. The same hierarchy pattern as in Table \ref{tab:res1B} is met for current $ \nu_{\mathbf{E}} $: the "correct" order holds only up to the structure by which the data has been generated, but no further.}
    \label{tab:res1C}
\end{table}

This phenomenon can be explained by the following idea: when the complexity of a model exceeds the complexity of  the dataset (as, in our examples, considerably more parameters are used to be trained by the dataset than to generate it), the landscape of the optimization problem may turn out to have more traps and suboptimal local minima. As a result, for datasets of relatively low complexity, an excessively sophisticated model may tend to perform worse than a simpler one. This is a crucial fact for identifying the underlying structure when working with some experimental data, not necessarily related to a generator $ \mathcal{L} $ of this specific structure. If the models order ranked according to the values of the test functional $ F $ violates the order determined  by the embedding hierarchy, then the data under consideration may, to a certain approximation, correspond to a simpler
\begin{table}[h!]
    \centering
    \begin{tabular}{|c|c|c|c|c|c|}
        \hline
    $ \nu_{\mathbf{E}}$ & $ F_{\mathbf{E}}/N $ & $ \nu $ & $ F/N  $ & $ \mathcal{T}_{\text{best}} $ & $  \Delta_{\mathcal{T}} F / N $ \\
    \hline

    \multirow{3}{*}{ $ \left(\left\{2,3 \right\}\right) $} & \multirow{3}{*}{ $ -1.749 $ } & $ \left(\left\{2, 3 \right\}\right) $ & $ -1.750 $ & $ 500 $ & $ \sim 10^{-8} $ \\ \cline{3-6}
            & & $ \left(\left\{1, 5 \right\}, \left\{1, 1 \right\}\right) $   & $ -1.790 $ & $ 115 $ &  --- \\  \cline{3-6}
        & & $ \left(\left\{3, 2 \right\}\right) $  & $ -1.826 $ &  $ 428 $ & --- \\
    \hline

    \multirow{3}{*}{$ \left(\left\{3, 2 \right\}\right)$} & \multirow{3}{*}{$ -1.688 $} & $ \left(\left\{3, 2 \right\}\right) $ & $ -1.689 $ & $ 500 $ & $ \sim 10^{-5} $ \\ \cline{3-6}
    
            & & $ \left(\left\{2, 3 \right\}\right) $ & $ -1.769 $ & $ 316 $ & --- \\ \cline{3-6}
    & & $ \left(\left\{1, 5 \right\}, \left\{1, 1 \right\}\right) $ & $ -1.780 $ & $ 130 $ & --- \\ 
                                \hline
    & \multirow{3}{*}{ $ -1.779 $ } &  $ \left(\left\{1, 5 \right\}, \left\{1, 1 \right\}\right) $ & $ -1.779 $ & $ 206 $ & --- \\ \cline{3-6}
    $ (\left\{1, 5 \right\}, $ & & $ \left(\left\{2, 3 \right\}\right) $ & $ -1.819 $ & $ 252 $ & --- \\ \cline{3-6}
         $ \left\{1, 1 \right\}) $ & &  $ \left(\left\{3, 2 \right\}\right) $ & $ -1.870 $ & $ 334 $ & --- \\
    \hline
    
    \end{tabular}
    \caption{Results of the experiments on "guessing" the structure via several relatively embedding-free options for the corresponding datasets $ \mathbf{E} $ with $ n = 6 $, $ S_{\text{train}} = S_{\text{test}} = 100 $, $ N = 200 $. The format of Tables \ref{tab:res1A}-\ref{tab:res1C} is followed, with adding information on the data original structure $ \nu_{\mathbf{E}} $ and functional $ F_{\mathbf{E}}/N$. Total epochs number is $\mathcal{T} = 500 $, best of 3 runs is chosen. All the structures are correctly identified.}
    \label{tab:res2}
\end{table}
\begin{figure}[h!]
    \centering
    \includegraphics[width=\linewidth]{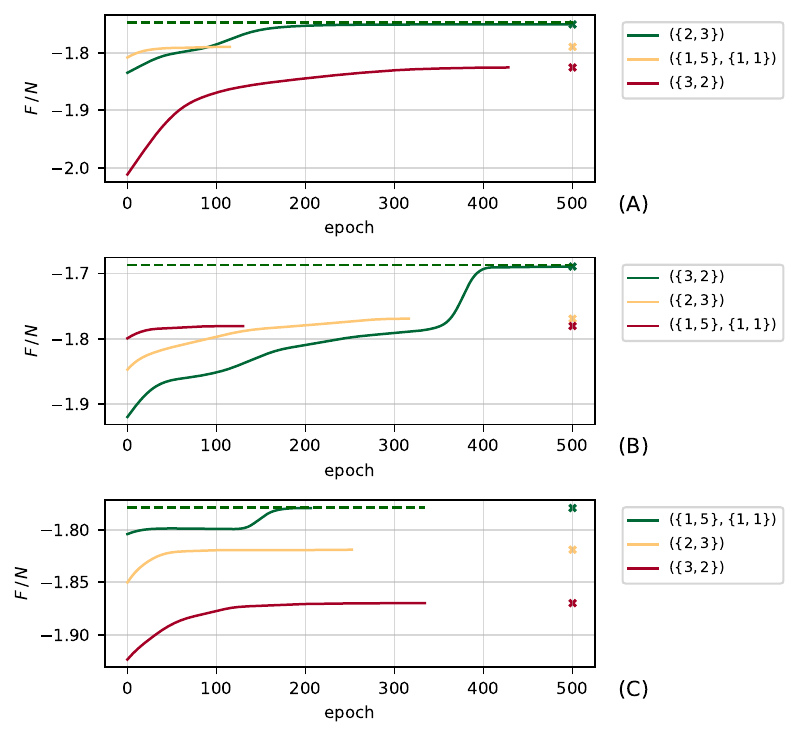}
    \caption{Test functional evolutions for the structure investigation experiments from Table \ref{tab:res2}: (A) for $ \nu_{\mathbf{E}} = (\left\{2, \, 3\right\}) $ data; (B) $ \nu_{\mathbf{E}} = (\left\{3, \, 2\right\}) $ data; (C) $ \nu_{\mathbf{E}} = (\left\{1, \, 5\right\}, \left\{1, \, 1\right\} )$ data; same hyperparameters $ \nu $ are used for the trained models, and the correct structures are successfully found.} 
    \label{fig:N6}
\end{figure} 
\noindent underlying structure, related to the last model for which the hierarchy order is still respected. This idea largely underlies the mechanism of structure exploration.

Choosing a simpler structure is also reasonable in situations when several models of different complexity produce approximately equal results, even if the correct hierarchical order is preserved. In this case, the consistency of the results indicates that the more complex models have not stuck in local minima or other traps and have also converged to optimal values. More generally, the smaller the gap between the simple and the complex models, the more reasonable it is to choose the simpler structure. A situation such as the one described in the previous paragraph can be referred to as the most clear indicator of presence of a specific structure in the data.

One more series of numerical tests explored whether it is possible to find the correct structure for higher-dimensional problems among several significantly different alternatives. This means that none of the considered optional structures $ \nu $ is an embedding of another in the above-mentioned sense (in other words, these structures are pairwise not on the same branch of the embedding hierarchy). Specifically, we consider three datasets of $ n = 6 $, generated with $ \nu_{\mathbf{E}} = \left(\left\{2, \, 3\right\}\right), \, \left(\left\{3, \, 2\right\}\right) $ and $ \left(\left\{1, \, 5\right\}, \, \left\{1, \, 1\right\}\right) $. For the models being trained, the same true options $ \nu $ are considered.

In all cases, the models with the correct structures achieved, up to a small numerical error, likelihood values corresponding to those recorded for $\lambda_{\mathbf{E}}$ and significantly outperformed models with other structures $ \nu $.
Results of these tests are presented in Table~\ref{tab:res2} and Fig.~\ref{fig:N6}. {Table~\ref{tab:res2} shows that, for each of the three datasets generated with mutually non-embedded structures, the model corresponding to the data-generating structure attains the largest test likelihood. Moreover, its final value of $F/N$ agrees, up to a small numerical discrepancy, with the reference value $F_{\mathbf{E}}/N$ obtained from the generative parameters, whereas the competing structures yield noticeably lower likelihoods. Figure~\ref{fig:N6} complements this comparison by showing the evolution of the test functional during training: in all three cases, the model with the correct structure converges to the highest objective-functional value and remains separated from the alternative candidates. Thus, the method consistently recovers the prescribed hidden structure from the measurement data, while the agreement with the generative likelihood also verifies that the numerical optimization and the implementation of the likelihood calculation reproduce the expected model when the correct structural hypothesis is used.}

 The exhaustive comparison of candidate algebraic structures used in the present numerical experiments is feasible only for relatively small Hilbert-space dimensions. The number of admissible unital matrix $*$-algebra structures grows rapidly with the system dimension, as quantified in Appendix~\ref{app:number_structures}. For larger systems, exhaustive
enumeration may be replaced by a local exploration of the embedding hierarchy. A possible greedy split--merge search strategy is outlined in Appendix~\ref{app:heuristic_search}.

 \subsection{Robustness under noisy conditions}

The results presented in Tables  \ref{tab:res1A}–\ref{tab:res2} demonstrate the feasibility of identifying the underlying structures from noise-free synthetic data. In realistic experimental settings, however, data distortions are unavoidable. It is therefore important to assess the robustness of the proposed method against possible noise effects. To this end, in this subsection, we consider experiments conducted on data affected by two different noise models. 

The first noise model perturbs the original generator by mixing it with an auxiliary random generator $\mathcal{L}_{\operatorname{rand}} $ that does not possess a structure associated with any nontrivial decoherence-free subalgebra. In this way, the dynamics is modified while remaining close to the original one, with an additional component that partially destroys its underlying structure. The strength of this perturbation is controlled by the weight parameter $ \varepsilon $. Thus, we consider
\begin{equation}
    \mathcal{L}_{\varepsilon} = (1 - \varepsilon)  \mathcal{L} + \varepsilon \mathcal{L}_{\operatorname{rand}},
\end{equation}
where $ 0 < \varepsilon < 1 $, $ \mathcal{L} $ is an original generator possessing the decoherence-free structure, while $ \mathcal{L}_{\operatorname{rand}} $ is a fixed random generator \eqref{eq:GKSLgenerator}. Since the set of GKSL generators forms a convex cone, the resulting $\mathcal{L}_{\varepsilon} $ also defines a valid open Markovian quantum dynamics.

The second noise model accounts for measurement errors while leaving the underlying open dynamics unchanged. Specifically, with probability $ p $ , the projector corresponding to the actual measurement outcome $ E^{(k)} $ is replaced by an incorrect projector $ E^{(k)}_{\operatorname{error}} $ associated with a different outcome, "flipping" the result
\begin{equation}
    \text{With probability } p: \quad E^{(k)} \to E^{(k)}_{\operatorname{error}}.
\end{equation}
As in the previous case, where $ \varepsilon $ controls the strength of the dynamical perturbation, the parameter $ p $ determines the noise level in the measurement data.

For these tests, we use an approach similar to that employed in the experiments reported in Table~\ref{tab:res2}. Specifically, we consider several mutually non-embedded decoherence-free structures and attempt to identify the structure hidden in the data with respect to various noise levels. We set $ n = 4 $, $ \nu_{\mathbf{E}} = \left(\{2,1\},\{1,2\}\right) $ and $ \nu = \left(\{2,1\},\{1,2\}\right), \, \left(\{1,1\}^{4}\right), \, \left(\{2,2\}\right) $. As can be seen from the hierarchy presented in Figure \ref{diag:embedn4}, none of the structures $ \nu $ is embedded into another. Consequently, they can be readily distinguished in the absence of noise. Table \ref{tab:noise} shows how the results change as noise of varying intensities, characterized by $ \varepsilon $ and $ p $, arises in the data, relative to the noise-free case. 
\begin{table}[h!]
         \centering
    {
    \scriptsize
    \begin{tabular}{|c|c|c|c|c|c|c|c|}
        \hline
        \multicolumn{2}{|c|}{Noise type}
        & $ F_{\mathbf{E}}/N $
        & $ \nu $
        & $ F / N $ 
        & Gap
        & $ \mathcal{T}_{\text{best}} $
        &  $ \Delta_{\mathcal{T}} F / N $ \\
        \hline
        \multicolumn{2}{|c|}{\multirow{3}{*}{No noise}}
        & \multirow{3}{*}{$ -1.296 $}
        & $\left(\{2,1\},\{1,2\}\right)$
        & $ -1.296 $
        & ---
        & $ 300 $
        & $ \sim 10^{-8} $
        \\
        \cline{4-8}

        \multicolumn{2}{|c|}{}
        &
        & $\left(\{1,1\}^{4}\right)$
        & $ -1.322 $
        & $ 0.026 $
        & $ 300 $
        & $ \sim 10^{-5} $
        \\
        \cline{4-8}

        \multicolumn{2}{|c|}{}
        &
        & $\left(\{2,2\}\right)$
        & $ -1.385$
        & $ 0.089 $
        & $ 300 $
        & $ \sim 10^{-6} $
        \\
        \hline

        \multirow{9}{*}{            \shortstack{
                $ \mathcal{L}_{\varepsilon}$\\
                noise
            }
        }
        & \multirow{3}{*}{$\varepsilon= 0.05 $}
        & \multirow{3}{*}{$ -1.334 $}
        & $\left(\{2,1\},\{1,2\}\right)$
        & $ -1.333 $
        & ---
        & $ 223 $
        & ---
        \\
        \cline{4-8}

        &
        &
        &  $\left(\{1,1\}^{4}\right)$
        & $ -1.355 $
        & $ 0.022 $
        & $ 300 $
        & $ \sim 10^{-5} $
        \\
        \cline{4-8}

        &
        & 
        & $\left(\{2,2\}\right)$
        & $ -1.395 $
        & $ 0.062 $
        & $ 157 $
        & ---
        \\
        \cline{2-8}

        & \multirow{3}{*}{$ \varepsilon = 0.10 $}
        & \multirow{3}{*}{$ -1.365 $}
        & $\left(\{2,1\},\{1,2\}\right)$
        & $ -1.360 $
        & ---
        & $ 291 $
        & $ \sim 10^{-6} $
        \\
        \cline{4-8}

        &
        & 
        & $\left(\{1,1\}^{4}\right)$
        & $ -1.377 $
        & $ 0.017 $
        & $ 300 $
        & $ \sim 10^{-6} $
        \\
        \cline{4-8}

        &
        & 
        & $\left(\{2,2\}\right)$
        & $ -1.409 $
        & $ 0.049 $
        & $ 215 $
        & $ \sim 10^{-6} $
        \\
        \cline{2-8}
        
        &
        \multirow{3}{*}{$\varepsilon= 0.15 $}
        & \multirow{3}{*}{$ -1.396 $}
        & $\left(\{2,1\},\{1,2\}\right)$
        & $ -1.385 $
        & ---
        & $ 191 $
        & ---
        \\
        \cline{4-8}

        &
        & 
        & $\left(\{1,1\}^{4}\right)$
        & $ -1.393 $
        & $ 0.008 $
        & $ 150 $
        & ---
        \\
        \cline{4-8}

        &
        & 
        & $\left(\{2,2\}\right)$
        & $ -1.418 $
        & $ 0.033 $
        & $ 300 $
        & $  \sim 10^{-5} $
        \\
        \hline

                \multirow{9}{*}{            \shortstack{
                Flip\\
                noise
            }
        }
        & \multirow{3}{*}{$ p = 0.05 $}
        & \multirow{3}{*}{$ -1.324 $}
        & $\left(\{2,1\},\{1,2\}\right)$
        & $ -1.325 $
        & ---
        & $ 300 $
        & $  \sim 10^{-8} $
        \\
        \cline{4-8}

        &
        & 
        & $\left(\{1,1\}^{4}\right)$
        & $ -1.347 $
        & $ 0.022 $
        & $ 300 $
        & $ \sim 10^{-5} $
        \\
        \cline{4-8}

        &
        & 
        & $\left(\{2,2\}\right)$
        & $ -1.392 $
        & $ 0.067 $
        & $ 300 $
        & $ \sim 10^{-5}$
        \\
        \cline{2-8}

        & \multirow{3}{*}{$ p = 0.15 $}
        & \multirow{3}{*}{$ -1.364 $}
        & $\left(\{2,1\},\{1,2\}\right)$
        & $ -1.364 $
        & ---
        & $ 224 $
        & ---
        \\
        \cline{4-8}

        &
        & 
        & $\left(\{1,1\}^{4}\right)$
        & $ -1.380 $
        & $ 0.016 $
        & $ 300 $
        & $ \sim 10^{-6} $
        \\
        \cline{4-8}

        &
        & 
        & $\left(\{2,2\}\right)$
        & $ -1.409 $
        & $ 0.045 $
        & $ 300 $
        & $  \sim 10^{-6} $
        \\
        \cline{2-8}

        & \multirow{3}{*}{$ p = 0.25 $}
        & \multirow{3}{*}{$ -1.399 $}
        & $\left(\{2,1\},\{1,2\}\right)$
        & $ -1.401 $
        & ---
        & $ 251 $
        & ---
        \\
        \cline{4-8}

        &
        & 
        & $\left(\{1,1\}^{4}\right)$
        & $ -1.407 $
        & $ 0.006 $
        & $ 300 $
        & $  \sim 10^{-6} $
        \\
        \cline{4-8}

        &
        & 
        & $ \left(\{2,2\}\right) $
        & $ -1.423 $
        & $ 0.022 $
        & $ 117 $
        & ---
        \\
        \hline
    \end{tabular}
    }
    \caption{Comparison of models trained on data generated for $ \nu_{\mathbf{E}}=\left(\{2,1\}, \, \{1,2\}\right) $ under different noise models (as before, we have $ S_{\text{train}} = S_{\text{test}} = 100 $, $ N = 200 $). The “Gap” column reports the difference between the value of the objective function $ F/N $ obtained for the model with $ \nu=\nu_{\mathbf{E}} $ and that obtained for the model with the alternative structure under consideration. For each model, the best result among three runs is reported, total epochs number is $ \mathcal{T}=300 $. For both noise models, increasing the noise intensity reduces the gap between the correct structure and the alternative candidates, while also leading to an overall deterioration in the achieved objective-function value.
}
    \label{tab:noise}
\end{table}

The conducted experiments demonstrate how the proposed method for identifying hidden structures performs in the presence of noise. As the noise intensity parameters $ p $ and $ \varepsilon $ increase, the correct decoherence-free structure becomes more difficult to distinguish: the gap between the objective-function values obtained for the correct and incorrect structures decreases, while the overall quality of the attained values also falls. Nevertheless, the results indicate that the method is robust, since this degradation occurs gradually for both of the noise models. When the dataset is only weakly corrupted and still retains sufficient information about the underlying dynamics, the method remains capable of identifying the correct structure, although with a smaller margin over the alternative candidates.

Although robustness is observed for both noise models, the experiments reveal an important distinction between them in terms of the relationship between the objective-function value $F_{\mathbf{E}}/N$, evaluated on the noisy data using the original generator, and the value $ F/N $ obtained for the trained model with the correct structure. In the presence of measurement-flip errors, the dataset, although corrupted, still reflects the dynamics generated by the original generator $\mathcal{L}$. By contrast, the first noise model perturbs the underlying dynamics itself. Accordingly, under measurement-flip noise, one observes that
\begin{equation}
\frac{F_{\mathbf{E}}}{N}\geqslant \frac{F}{N},
\end{equation}
with both quantities decreasing as the noise level increases. This inequality does not generally hold for the first noise model, since the trained model attempts to approximate the perturbed generator $\mathcal{L}_{\varepsilon}$ rather than the unperturbed generator $\mathcal{L}$.

\subsection{Repetitive measurement -- preparation tradeoff}\label{subsec:numberlength}

In this subsection we show that the proposed method demonstrates robust performance with respect to variations in the training data structure. Specifically, we discuss the tradeoff between the data chain length $ N_{\text{train}} $ and the number of measurement chains $ S_{\text{train}} $ in the training dataset $ \mathbf{E}_{\text{train}}$.

From a  physical perspective,  the setups with fixed $N_{\text{train}} S_{\text{train}}$ but different values of $N_{\text{train}}$ and $S_{\text{train}}$ are essentially different. At the beginning of each measurement chain, the system must be prepared in a given initial state; thus, $S_{\text{train}}$ represents the number of such state preparations. A large $N_{\text{train}}$ corresponds to a passive regime of experimental interrogation, where one monitors the system dynamics without intervention. In contrast, a large $S_{\text{train}}$ corresponds to an active regime, where the system state is frequently reset by external interactions or even reinitialized as a new instance.

In machine learning terms, this topic is important since the balance between these two properties of the data. Suppose the total number of measurements $ N_\text{train} S_\text{train} $ is fixed. In that case, training with a larger number of shorter chains might be considerably more effective computationally, since they can be processed jointly in a batch-based way (which works significantly faster when performing computations on GPU). However, it is essential to verify whether it does not lead to loss of training robustness or degradation of the objective functional value.
\begin{table}[h!]
    \centering
    \begin{tabular}{|c|c|c|c|c|}
        \hline
        $ N_{\text{train}} $ & $ S_{\text{train}} $ & $ F/N $ & $ \mathcal{T}_{\text{best}} $ & $  \Delta_{\mathcal{T}} F / N $ \\ \hline
         10 & 6000 & $ -1.3461 $ & $ 164	$ & --- \\ \hline
         20 & 3000 & $ -1.3461 $ & $ 700 $ & $ \sim 10^{-9} $ \\ \hline
         30 & 2000 & $ -1.3459 $ & $ 29 $ & --- \\ \hline
         40 & 1500 & $ -1.3461 $ & $ 54 $ & --- \\ \hline
         50 & 1200 & $ -1.3460 $ & $ 59 $ & --- \\ \hline
         100 & 600 & $ -1.3460 $ & $ 69 $ & --- \\ \hline
         200 & 300 & $ -1.3458 $ & $ 700 $ & $ \sim 10^{-7} $ \\ \hline
         300 & 200 & $ -1.3459 $ & $ 112 $ & ---  \\ \hline
         400 & 150 & $ -1.3459 $ & $ 700 $ & $ \sim 10^{-9} $ \\ \hline
         500 & 120 & $ -1.3459 $ & $ 273 $ & --- \\ \hline
    \end{tabular}
    \caption{Results of experiments on exploring the tradeoff between the length of training data chains $ N_{\text{train}} $ and their quantity $ S_{\text{train}} $. We consider various combinations of these settings with the invariant $ N_{\text{train}}S_{\text{train}} = 6 \cdot 10^{4} $ of total training measurements for the $ \nu_\mathbf{E} = (\left\{1, 2 \right\}, \left\{1, 2 \right\}) $ structure. Epochs number is $ \mathcal{T} = 700 $. As a testing dataset we use independent $ S_{\text{test}} = 100 $ measurement chains of $ N_{\text{test}} = 500 $ length, $ F_{\mathbf{E}} / N = -1.3458 $. Experiments indicate no considerable difference between choosing long or short chains in terms of the objective functional value if maintaining the same number of measurements.}
    \label{tab:res4}
\end{table}

A number of experimental runs has been conducted to address the problem of $ N_{\text{train}} $ and $ S_{\text{train}} $ tradeoff, their results are presented in Table \ref{tab:res4}. 

They demonstrate that for the same total number of measurements $ N_{\text{train}} S_{\text{train}} $ no significant difference is observed in the final value of the objective functional for the considered data. From these runs, it can be concluded that, given a sufficiently large total number of measurements, the tradeoff between data chains length and quantity does not affect the amount of information contained in the analyzed data, allowing the model to be trained to approximately the same test objective values.

The obtained results also indicate that, provided the total number of measurements is sufficiently large, the method does not exhibit a substantial dependence on the initial state $ \sigma $ introduced in \eqref{initial_state_sigma}. Otherwise, longer measurement sequences, which are expected to be less sensitive to the initial state, would likely produce results that differ from those obtained using shorter sequences. No systematic difference of this kind was observed in the experiments.

In addition, these outcomes provide evidence of robustness with respect to the randomness inherent in the generation of synthetic datasets. In both respects, the total number of measurements appears to play a key role: when it is sufficiently large, the method demonstrates stable behavior and the results appear to be highly reproducible across different realizations of the data.

\subsection{Inferring effective physics from restricted observations} \label{subsec:inferring}

In this subsection we show that the proposed method is effective for constructing reduced models of dynamics from restricted observational data.

A natural and essential question is what constraints the structure of the algebra of observables $ \mathcal{A} $ imposes on data. Algebra $ \mathcal{A} $ serves as a "bottleneck" that connects the internal structure of the system with the information that is encoded in the measurement data we have access to. In the carried out experiments, we focus on investigating the effect of structural degradation of $ \mathcal{A} $ as the value of $ n_{0} $ increases. Specifically, we have taken into consideration measurements for the structures $ ( \{n_{1}^{\mathcal{A}}, \, m_{1}^{\mathcal{A}} \}) $ with $ n = n_{0}^{\mathcal{A}} + n_{1}^{\mathcal{A}} m_{1}^{\mathcal{A}} $ and $ m_{1}^{\mathcal{A}} = 1 $.  This algebra fits the physical setup described below Eq.~\eqref{eq:restrAlgebra} and naturally arises, e.g., in spectroscopy \cite{cho2009two}, in cases where only part of the energy levels can be excited by the external field. Moreover, this algebra models limited access to experimental observations in the most drastic way, which is why we have chosen it for our experiments.

We have conducted two distinct series of experiments. In each series, models have been trained on two families of datasets ($ n = 5 $): for the free open GKSL dynamics $ \nu_{\mathbf{E}} = \left(\{1, 5\}\right) $ and dynamics with the substantially large decoherence-free algebra $  \nu_{\mathbf{E}} = \left(\{2, 2\}, \{1, 1\}\right) $. For both series, training datasets are considered for previously introduced types of algebra $ \mathcal{A} $. In the first series though, test datasets $ \mathbf{E}_{\text{test}} $ remain corresponding to the value $ n_{0} = 0 $ (Table \ref{tab:res3a}), while in the second series they follow the same $ \mathcal{A} $ structures as for the $ \mathbf{E}_{\text{train}}$ (Table \ref{tab:res3b}). This approach provides an opportunity to investigate from two different perspectives how the training proceeds: from the standpoint of the most general information about the system, and from the standpoint of only that dynamic information which is available to us under the given observation constraints.

The roles of Tables~\ref{tab:res3a} and~\ref{tab:res3b} are complementary,
and the two tables should be read together. In Table~\ref{tab:res3a}, the
models are trained on data obtained with increasingly restricted observable
algebras, while they are evaluated on the same full-access test dataset
corresponding to $n_0^{\mathcal A}=0$. Since the test dataset is common to all
rows, the resulting values of $F/N$ can be compared directly. Their
deterioration as $n_0^{\mathcal A}$ increases shows that the restricted
measurement records do not contain sufficient information to reconstruct a
model capable of predicting general measurements outside the accessible
algebra.

Table~\ref{tab:res3b} addresses the complementary question. In this case, the training and test datasets are generated using the same restricted observable algebra. For each fixed observable algebra, the fitted likelihood $F/N$ can therefore be compared with the corresponding reference value $F_{\mathbf E}/N$ obtained from the original data-generating dynamics under the same observational restrictions.

As shown in Table~\ref{tab:res3b}, the fitted effective models attain empirical
test-likelihood values comparable to or higher than the corresponding reference
values $F_{\mathbf E}/N$. This indicates that, once observational access is
restricted, a model fitted directly to the accessible measurement records can
provide a higher-likelihood effective description of these records than the
original microscopic model evaluated under the same restrictions. This result
should not be interpreted as a more accurate reconstruction of the full
underlying dynamics. Indeed, Table~\ref{tab:res3a} shows that models trained on
increasingly restricted measurement records generally perform worse when they
are evaluated on the common full-access test dataset.

This measurement-induced bottleneck is also an identifiability limitation.
For a fixed accessible algebra $\mathcal A$ and a fixed measurement protocol,
two models $(\nu,\lambda_\nu)$ and $(\nu',\lambda_{\nu'})$ are observationally
indistinguishable if
\begin{equation*}
p_{\nu,\lambda_\nu}
\left(E^{(N)},\ldots,E^{(1)}\right)
=
p_{\nu',\lambda_{\nu'}}
\left(E^{(N)},\ldots,E^{(1)}\right)
\end{equation*}
for every measurement sequence allowed by the protocol, with
$E^{(j)}\in\mathcal A$. Such observationally indistinguishable models may
possess different invariant algebraic structures. In this case, neither the
maximum-likelihood procedure nor any other inference method based on the same
restricted records can distinguish them. For finite datasets, models whose
accessible probability distributions are only approximately equal may
similarly yield statistically indistinguishable likelihood values.

Taken together, Tables~\ref{tab:res3a} and~\ref{tab:res3b} show that the
information lost under restricted access may be essential for reconstructing
the full microscopic dynamics, while being unnecessary for constructing a
high-likelihood effective description of the multi-time statistics that remain
experimentally accessible. Thus, restricted observations may support effective
models adapted to all observations available under the same restrictions, even
though these models do not provide unique reconstructions of the complete
underlying dynamics. The fitted model and its inferred algebraic structure
should consequently be interpreted relative to the accessible algebra
$\mathcal A$ and the chosen measurement protocol.

Restricted access does not, however, necessarily imply non-identifiability.
If competing algebraic structures induce distinct probability distributions
on the accessible measurement records, they can in principle be distinguished
from sufficiently informative data. The successful structure recovery in the
preceding numerical examples demonstrates such practical identifiability for
the particular candidate sets and measurement protocols considered here,
rather than a general uniqueness result for arbitrary restricted
observations. This distinction is important, for example, in spectroscopic
settings, where only a restricted set of transitions or coherences may be
experimentally accessible.

\begin{table}[h!]
    \centering
    \begin{tabular}{|c|c|c|c|c|c|c|}
        \hline
        $ \nu_{\mathbf{E}}, \, \nu $ & $ F_{\mathbf{E}}/N $ & $ n_{0}^{\mathcal{A}} $ & $ \left( \left\{n_{1}^{\mathcal{A}}, \, m_{1}^{\mathcal{A}} \right\} \right) $ & $ F/N $ & $ \mathcal{T}_{\text{best}} $ &  $ \Delta_{\mathcal{T}} F / N $ \\ \hline
        \multirow{4}{*}{$\left(\left\{1, 5\right\} \right)$}& \multirow{4}{*}{$ -1.607 $} & 0 & $ \left(\left\{5, 1\right\}\right) $ & $ -1.607 $ & $ 74 $ & --- \\ \cline{3-7}
        & & 1 & $ \left(\left\{4, 1\right\}\right) $ & $ -1.612 $ & $ 3 $ & --- \\ \cline{3-7}
        & & 2 & $ \left(\left\{3, 1\right\}\right) $ & $ -1.612 $ & $ 2 $ & --- \\ \cline{3-7}
        & & 3 & $ \left(\left\{2, 1\right\}\right) $ & $ -1.612 $ & $ 1 $ & --- \\ \hline
        & \multirow{4}{*}{$ -1.533 $} & 0 & $ \left(\left\{5, 1\right\}\right) $ & $ -1.536 $ & $ 400 $ & $ \sim 10^{-5} $ \\ \cline{3-7}
        $(\left\{2, 2\right\}, $ & & 1 & $ \left(\left\{4, 1\right\}\right) $ & $ -1.681 $ & $ 14 $ & --- \\ \cline{3-7}
        $ \left\{1, 1\right\}) $ & & 2 & $ \left(\left\{3, 1\right\}\right) $ & $ -1.681 $ & $ 149 $ & --- \\ \cline{3-7}
        & & 3 & $ \left(\left\{2, 1\right\}\right) $ & $ -1.684 $ & $ 283 $ & --- \\ \hline
    \end{tabular}
    \caption{Experimental results for training models on data $ E_{\text{train} } $ gained by using the $*$-matrix algebras $ \mathcal{A} $ of observables with various $ n_{0}$, complemented to $ n = 5 $ by normal block of $ \left\{n_{1}, \, m_{1} \right\} $ having $ m_{1} = 1 $. Datasets with two different $ \nu_{\mathbf{E}} $ are considered. For this runs, $ E_{\text{test}} $ is taken for $ n_{0} = 0 $ case and considered all the same for all the experiments in order to assess training degradation when moving to more degenerate data. The results drop drastically when using to $ n > 0 $: the models start to demonstrate significantly weaker training performance. Total epochs number is $ \mathcal{T} = 400 $.}      \label{tab:res3a}
\end{table}

\begin{table}[h!]
    \centering
    \begin{tabular}{|c|c|c|c|c|c|c|}
        \hline
        $ \nu_{\mathbf{E}}, \, \nu $ & $ n_{0}^{\mathcal{A}} $ & $ \left( \left\{n_{1}^{\mathcal{A}}, \, m_{1}^{\mathcal{A}} \right\} \right) $ & $ F_{\mathbf{E}}/N $ & $ F/N $ & $ \mathcal{T}_{\text{best}} $ &  $ \Delta_{\mathcal{T}} F / N $  \\ \hline
        \multirow{4}{*}{$\left(\left\{1, 5\right\} \right)$}& 0 & $ \left(\left\{5, 1\right\}\right) $ & $ -1.607 $ & $ -1.607 $ & $ 74 $ & --- \\ \cline{2-7}
        & 1 & $ \left(\left\{4, 1\right\}\right) $ & $ -1.629 $ & $ -1.384 $ & $ 400 $ & $ \sim 10^{-6}$ \\ \cline{2-7}
        & 2 & $ \left(\left\{3, 1\right\}\right) $ & $ -1.596 $ & $ -1.097 $ & $ 400 $ & $ \sim 10^{-6} $ \\ \cline{2-7}
        & 3 & $ \left(\left\{2, 1\right\}\right) $ & $ -1.593 $ & $ -0.692 $ & $ 400 $ & $ \sim 10^{-7}$\\ \hline
        & 0 & $ \left(\left\{5, 1\right\}\right) $ & $ -1.533 $ & $ -1.536 $ & $ 400 $ & $ \sim 10^{-5} $ \\ \cline{2-7}
        $ (\left\{2, 2\right\}, $ & 1 & $ \left(\left\{4, 1\right\}\right) $ & $ -1.577 $ & $ -1.427 $ & $ 400 $ & $ \sim 10^{-5} $ \\ \cline{2-7}
        $ \left\{1, 1\right\}) $  & 2 & $ \left(\left\{3, 1\right\}\right) $ & $ -1.578 $ & $ -1.098 $ & $ 400 $ & $ \sim 10^{-7} $\\ \cline{2-7}
        & 3 & $ \left(\left\{2, 1\right\}\right) $ & $ -1.542 $ & $ -0.677 $ & $ 291 $ & --- \\ \hline
    \end{tabular}
    \caption{Results of the analogues of the numerical runs from Table \ref{tab:res3a} on the $ \mathbf{E}_{\text{test}} $ dataset, generated for the same observable algebra structure as $ \mathbf{E}_{\text{train}} $. The values of the functional drop significantly as $ n_{0} $ increases and do not match the values for the true Hamiltonians and Lindbladians of the system on the considered test chains. It reveals a measurement-induced “bottleneck” effect: due to the loss of information in the data caused by the degradation of the observable algebra $ \mathcal{A} $, the models learn a more primitive dynamics actually encoded in the data rather than the true dynamics of the system. Total epochs number is $ \mathcal{T} =  400 $.}  
    \label{tab:res3b}
\end{table}

\subsection{Genuinely hidden structures in waveguide quantum electrodynamics}\label{subsec:physmodel}

In this subsection we demonstrate that the proposed method can reveal genuinely hidden algebraic structures in physically relevant models. We consider the waveguide quantum electrodynamics model with parametric gain from the work~\cite{karnieli2025decoherence}. 
Namely, an array  of two-level emitters coupled to $\chi^{(2)}$  nonlinear waveguide with coupling strength $\gamma>0$, the total   squeeze parameter $r \in \mathbb{R}$ and squeezing phase $\theta \in [0, 2 \pi)$. 

In~\cite{karnieli2025decoherence}, it was shown that this model exhibits approximate (for small $r$) decoherence-free subspaces for $\geqslant 4$ atoms. According to~\cite{deschamps2016structure}, this implies the existence of a decoherence-free algebra with some of the multiplicities $m_j$ equal to one. Therefore, it is natural to analyze the structure of this decoherence-free algebra numerically. But what is more important by our approach we can try to find the analogous structures for $3$ atoms, when this model is considered to be completely dissipative. So we focus on three atom case, because for this case the  decoherence-free type structure is genuinely hidden in this sense.

The Hamiltonian and Lindblad operators of this model are given by
\begin{equation}
\begin{split}
    &H=\frac{\gamma}{2} \sum_{i, j=1}^3 \sinh \!\left(\frac{r|i-j|}{2}\right)
\!\left(e^{-i \theta} \sigma_{+, j} \sigma_{+, i}+e^{i \theta} \sigma_{-, j} \sigma_{-, i}\right), \\
    &L_s=\sqrt{\gamma} \sum_{j=1}^3\!\left(\cosh r_{s, j} \sigma_{-, j}-i e^{-i \theta} \sinh r_{s, j} \sigma_{+, j}\right). \\
\end{split}
\end{equation}
 Here, $s\in\{\rightarrow,\leftarrow\}$ labels the two counter-propagating waveguide modes, corresponding to the right- and left-moving channels, respectively. The position-dependent squeezing parameters entering the corresponding Lindblad operators are defined as
\begin{equation}
r_{\rightarrow, j}=r\frac{j-1}{2}, \qquad  r_{\leftarrow, j}=r\frac{3-j}{2}.
\end{equation}
 Thus, $L_{\rightarrow}$ and $L_{\leftarrow}$ describe the dissipative coupling of the emitters to the right- and left-propagating waveguide channels, respectively.

\begin{figure}[h!]
\centering
\includegraphics[width=0.95\linewidth]{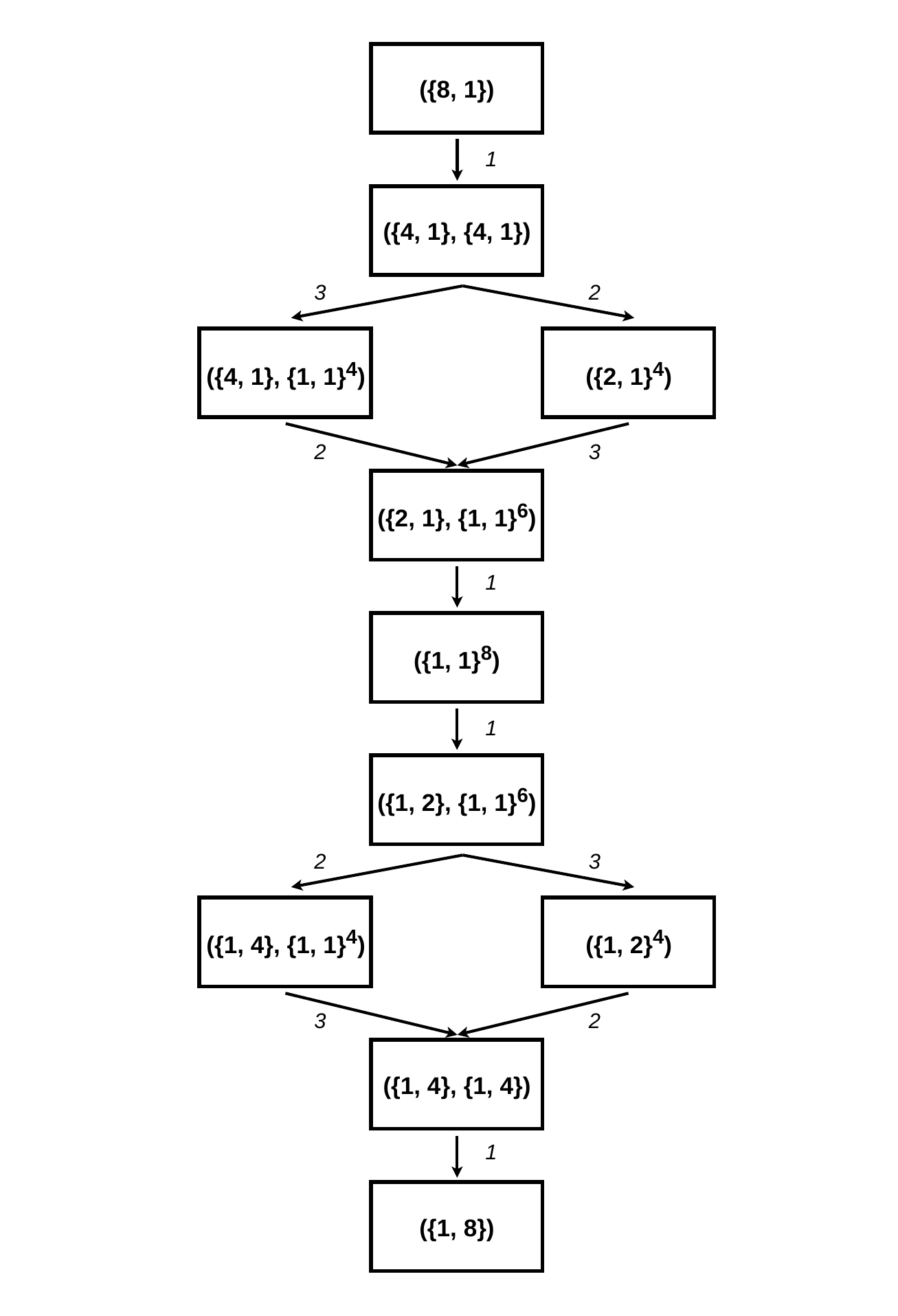}
\caption{The embedding hierarchy of decoherence-free algebra structures presented in Table~\ref{tab:res5} for the three-qubit model. Each node corresponds to a structure $\nu$, and arrows indicate embeddings between the corresponding algebras. The structures are arranged from simpler (top) to more complex (bottom) models in terms of the associated class of GKSL generators. The length of an arrow reflects the number of intermediate structures in the hierarchy that are not explicitly shown here.}
\label{diag:embedn8}
\end{figure}

\begin{table}[h!]
    \centering
    {\scriptsize
         \begin{tabular}{|c|c|c|c|c|c|c|}
        \hline
        \multirow{2}{*}{$ \nu $} & \multicolumn{3}{c|}{$ F / N $} & \multirow{2}{*}{Gap} & $ \multirow{2}{*}{$ \mathcal{T}_{\text{best}} $} $ & \multirow{2}{*}{$ \Delta_{\mathcal{T}} F / N $} \\ \cline{2-4}
        & Mean & Var & \textbf{Best} & & &\\
        \hline
        $ ( \left\{1, 8\right\}) $ & $ -2.0556 $ & $ \sim 10^{-8} $ & $ \mathbf{-2.0555} $ & --- & $ 76 $ & --- \\ \hline
        $ ( \left\{1, 4\right\}^{2}) $ & $ -2.0624 $ & $ \sim 10^{-6} $ & $ \mathbf{-2.0586} $ & $ 0.0031 $ & $ 82 $ & --- \\ \hline
        $ ( \left\{1, 2\right\}^{4}) $ & $ -2.0685 $ & \multirow{7}{*}{$ \sim 10^{-5} $} & $ \mathbf{-2.0632} $ & $ 0.0046 $ & $ 1490 $ & $ \sim 10^{-6} $  \\ \cline{1-2} \cline{4-7}
        $ ( \left\{1, 4\right\}, \left\{1, 1\right\}^{4}) $ & $ -2.0724 $ & & $ \mathbf{-2.0688} $ & $ 0.0102 $ & $ 287 $ & --- \\ \cline{1-2} \cline{4-7}
         \multirow{2}{*}{$ ( \left\{1, 2\right\}, \left\{1, 1\right\}^{6}) $}  & \multirow{2}{*}{$ -2.0815 $} & & \multirow{2}{*}{$ \mathbf{-2.0751} $}  & $ 0.0119 $ & \multirow{2}{*}{$ 691 $} & \multirow{2}{*}{---} \\
         & & & & $ 0.0063 $ & &\\ \cline{1-2} \cline{4-7}
        $ (\left\{1, 1\right\}^{8}) $ & $ -2.0878 $ & & $ \mathbf{-2.0829} $ &  $ 0.0078 $ & $ 646 $ & --- \\ \cline{1-2} \cline{4-7}
        $ ( \left\{2, 1\right\}, \left\{1, 1\right\}^{6}) $ & $ -2.1008 $ &  & $ \mathbf{-2.0934} $ & $ 0.0105 $ & $ 424 $ & ---  \\ \cline{1-2} \cline{4-7}
        $ \mathbf{( \left\{2, 1\right\}^{4})} $ & $ -2.1455 $ &  & $ \mathbf{-2.1367} $ & $ 0.0433 $ & $ 980 $ & ---  \\ \cline{1-2} \cline{4-7}
        $ ( \left\{4, 1\right\}, \left\{1, 1\right\}^{4}) $ & $ -2.1612 $ &  & $ \mathbf{-2.1511} $ & $ 0.0577 $ & $ 623 $ & ---  \\ \cline{1-2} \cline{4-7}
        \multirow{2}{*}{$ \mathbf{( \left\{4, 1\right\}^{2})} $} & \multirow{2}{*}{$ -2.2628 $} &  & \multirow{2}{*}{ $ \mathbf{-2.2490} $ } & $ 0.1123 $ & \multirow{2}{*}{ $ 781 $ } & \multirow{2}{*}{ --- }  \\
        & & & & $ 0.0979 $ & &\\ \cline{1-2} \cline{4-7}
        \hline
        $ \mathbf{( \left\{8, 1\right\})}  $ & $ -2.5312 $ & $ \sim 10^{-4} $ & $ \mathbf{-2.4988} $ & $ 0.2498 $ & $ 45 $ & --- \\ \hline
    \end{tabular}
    
    }
    \caption{Training results for the considered three-qubit model ($ n = 8 $). We explore various $ \nu $ structures presented in Figure \ref{diag:embedn8} for the dataset $ \mathbf{E} $ with $ F_{\mathbf{E}} / N =  -2.033     2 $, $ S_{\text{train}} = S_{\text{test}} = 100 $, $ N_{\text{train}} S_{\text{train}} = 10^{4} $. The rows of the table are ordered in accordance with the hierarchy and the       best value of the test functional $ F/N $      the subcolumn in bold. The “Gap” column of the table contains information on how much the objective functional value for the given structure is lower than that for the previous one in the hierarchy. The information in the remaining columns is of same type as in the previously presented tables. Total epochs number is $ \mathcal{T} =  1500 $. Best of      twenty runs is taken,       the mean and variance of the results are reported in the corresponding subcolumns of the  $ F/N $ column. The structure highlighted in bold correspond to the main transition in the hierarchy discussed in the text and presented in      Figure \ref{fig:N8bars}, where a noticeable change in the model performance occurs.
    }
    \label{tab:res5}
\end{table}
\begin{figure}[h!]
\centering
\includegraphics[width=\linewidth]{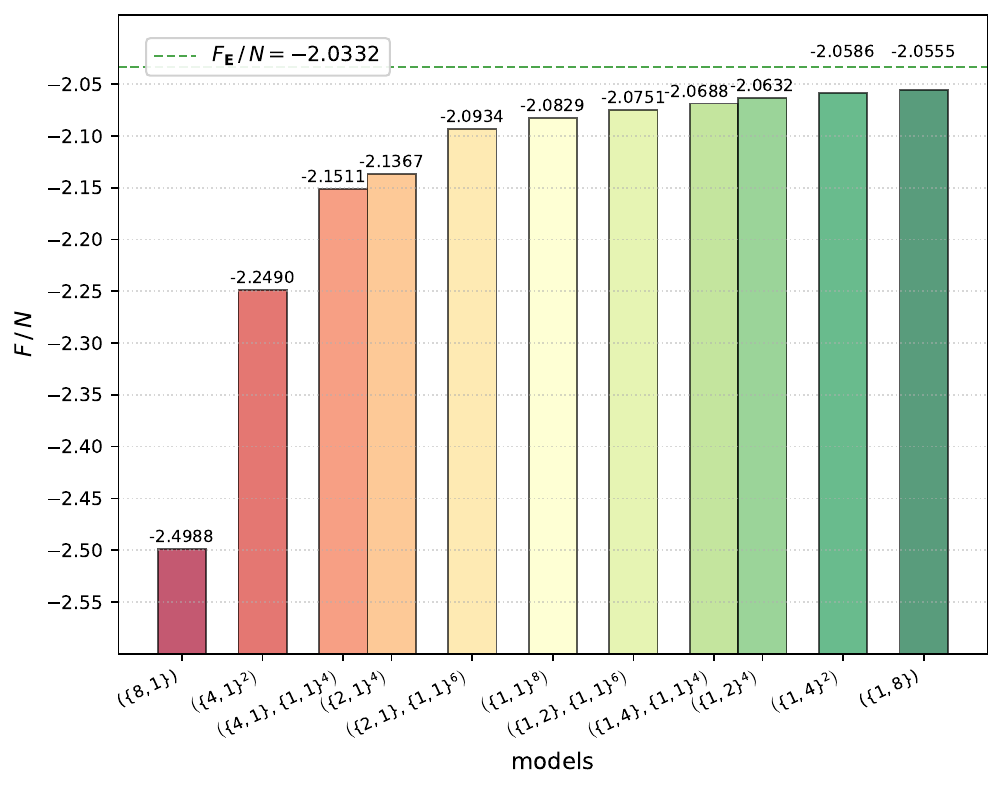}
\caption{The bar chart of the test functional values $ F/N $ 
 presented in Table \ref{tab:res5}. Columns which are related to structures located in different branches of the embeddings hierarchy (see Figure \ref{diag:embedn8}) are grouped together.}
\label{fig:N8bars}
\end{figure} 
The results presented in Table~\ref{tab:res5}, together with the embedding
hierarchy shown in Figure~\ref{diag:embedn8}, reveal a clear structure in the
model performance. In the upper part of the hierarchy, models such as
$(\{1,8\})$, $(\{1,4\}^{2})$, and $(\{1,2\}^{4})$ achieve the largest
values of the objective functional, and transitions between these structures
lead only to very small changes in $F/N$, indicating a stable performance
plateau.

A first noticeable degradation appears when passing to more fragmented
structures, starting from $(\{1,4\},\{1,1\}^{4})$ and continuing through
$(\{1,2\},\{1,1\}^{6})$ and $(\{1,1\}^{8})$. In this regime, the decrease
in $F/N$ becomes systematic and more pronounced, showing that excessive
fragmentation of the algebra into small blocks deteriorates the quality of the
model.

An even sharper deterioration is observed when moving to the structure
$(\{2,1\}^{4})$, where the objective functional drops significantly.
This marks a clear separation between models that reproduce the observed
multi-time statistics relatively well and models whose performance is
substantially poorer.

Further enlargement of the blocks, as in $(\{4,1\}^{2})$ and $(\{8,1\})$,
leads to an even stronger degradation.

Altogether, the nonuniform changes in the test likelihood suggest a
characteristic structural scale at which additional algebraic constraints
begin to cause a substantial loss of model performance, indicating a
pronounced structural crossover along the embedding hierarchy.

\section{Conclusion}
\label{se:conclusion}

 We have proposed a likelihood-based framework for reconstructing hidden invariant algebraic structures of open quantum dynamics from multi-time measurement data obtained under restricted observational access. We emphasize that the method does not assume that the dynamics observed through the restricted algebra of accessible observables is Markovian by itself. Rather, it assumes the existence of a Markovian embedding: an enlarged finite-dimensional system whose dynamics is governed by a GKSL semigroup. The key idea is to fix the algebra of accessible observables $\mathcal{A}$ by the measurement setup and to treat the invariant algebra $\mathcal{N}$ of the total embedded GKSL dynamics as a latent structural variable to be inferred from data.
In the present work, we focused on the case when $\mathcal{N}$ is a decoherence-free subalgebra, which allows one to combine the canonical parametrization of matrix $*$-algebras with the known structural form of compatible GKSL generators.

 This leads to a generalized Markovian embedding scheme, extending the standard subsystem-based embedding setting to arbitrary observable algebras.
Within this scheme, multi-time projective measurement statistics define a natural likelihood functional, whose maximization makes it possible to fit both dynamical parameters and the hidden algebraic structure. In this way, the reconstruction of effective open-system dynamics becomes a structure-learning problem.

The numerical results support the validity of the approach. For synthetic datasets, the method correctly identifies the generating invariant structure and reveals a clear relation between model performance and the embedding hierarchy of decoherence-free algebras. In particular, the breakdown of this hierarchy for more complex models appears to be a useful practical indicator that the data are already well described by a simpler effective structure, while further increase in model complexity mainly introduces optimization difficulties.

We also found that, for a fixed total number of measurements, the balance between the length of measurement chains and their number has only a weak effect on the final likelihood values. By contrast, the structure of the accessible observable algebra $\mathcal{A}$ plays a crucial role. Restricted observational access may significantly reduce the information contained in the data and prevent reconstruction of the full underlying generator. However, this should not be viewed solely as a limitation: the restriction imposed by $\mathcal{A}$ naturally induces an effective description of the dynamics adapted to the accessible observables. In this sense, $\mathcal{A}$ acts as an information bottleneck that not only limits, but also organizes the inferred physics, selecting those dynamical features that are operationally relevant and allowing for the emergence of simpler effective models consistent with the available data.

Finally, the waveguide QED example shows that the method can detect nontrivial hidden effective structures even in situations where standard analysis does not predict exact decoherence-free sectors. In particular, for the three-emitter model, the analysis of the likelihood landscape across the embedding hierarchy reveals a high-performance plateau followed by successive degradation regimes. Models with moderately sized blocks yield nearly identical values of the objective functional, while increasing fragmentation leads to a gradual decline, and a more pronounced drop occurs for structurally incompatible block configurations; further enlargement of irreducible blocks results in an additional strong degradation. Altogether, this identifies a preferred structural scale of the effective decoherence-free algebra: both overly fine-grained decompositions and excessively large blocks degrade performance, indicating that the effective algebra exhibits a nontrivial intermediate matrix-block organization emerging directly from the data.

Overall, the proposed approach provides a bridge between the operator-algebraic structure of quantum Markov semigroups and data-driven learning of effective physical models. It opens a route toward identifying hidden symmetry and decoherence-free patterns directly from experimentally accessible multi-time statistics. Importantly, such patterns should not be viewed merely as convenient features of a particular model representation: they often correspond to physically meaningful structures, such as emergent subsystems, conserved quantities, or collective degrees of freedom, and thus reflect elements of the underlying physical ontology revealed through the accessible observations.

\begin{acknowledgments}
This work was supported by the Ministry of Science and Higher Education of the Russian Federation (Grant No. 075-15-2024-529).

 The authors are grateful to A.~S.~Holevo, I.~A.~Lopatin, V.~Umanit\`a, and V.~I.~Yashin for fruitful discussions on the questions considered in this work. The authors also thank the referees for their valuable comments and suggestions.

\end{acknowledgments}

\appendix

\section{Counting the number of algebras}
\label{app:number_structures}
If the order of the Wedderburn blocks is retained, the corresponding
structures are counted by colored compositions of $n$, where a part of size
$q$ has $d(q)$ colors corresponding to the factorizations $q=ab$. Hence the
ordinary generating function is
\begin{equation}
B(z)=\frac{1}{1-\sum_{q\ge 1}d(q)z^q}.
\end{equation}
Consequently $B_n$ grows exponentially, $B_n\sim C\rho^{-n}$, where
$\rho$ is the unique positive solution of $\sum_{q\ge 1}d(q)\rho^q=1$.

\begin{equation*}
\rho \approx 0.406148,
\qquad
\rho^{-1} \approx 2.462.
\end{equation*}

The number of unital matrix $*$-algebra types in dimension $n$, up to
permutations of the Wedderburn blocks, is the sequence OEIS A006171. Equivalently,
it is the number of partitions of $n$ with $d(a)$ copies of each part $a$,
whose generating function is
\[
\sum_{n\ge 0} A_n z^n
=
\prod_{a\ge 1}(1-z^a)^{-d(a)}.
\]
This sequence and its partition interpretation were studied in
Ref.~\cite{agarwal1988partitions}; its interpretation as the number of unital
$*$-subalgebras of $M_n(\mathbb C)$ up to unitary similarity is also recorded
in Ref.~\cite{erickson2025demazure}.

It satisfies
\begin{equation*}
\log A_n \sim \pi\sqrt{\frac{n\log n}{3}}
\end{equation*}
as $n\to\infty$~\cite{brigham1950general}.
\begin{table}[h!]

\centering
\begin{tabular}{|c|c|c|}
\hline
$n$ & $A_n$ up to block permutations & $B_n$ ordered blocks \\
\hline
1  & 1   & 1    \\
2  & 3   & 3    \\
3  & 5   & 7    \\
4  & 11  & 18   \\
5  & 17  & 43   \\
6  & 34  & 108  \\
7  & 52  & 263  \\
8  & 94  & 651  \\
9  & 145 & 1599 \\
10 & 244 & 3942 \\
\hline
\end{tabular}
\caption{Number of candidate unital matrix $*$-algebra structures in dimension $n$.
Here $A_n$ counts structures up to permutations of Wedderburn blocks, while
$B_n$ counts ordered block decompositions.}
\label{tab:algebra-structure-counts}
\end{table}

\section{Heuristic exploration of algebraic structures}
\label{app:heuristic_search}

For larger Hilbert-space dimensions, the exhaustive enumeration of all
structures $\nu$ becomes impractical. A possible heuristic alternative is the
following greedy split--merge search on the embedding graph of algebraic
structures.

\begin{enumerate}
    \item \textbf{Initialization.}
    Choose an initial structure $\nu^{(0)}$. This structure may be selected
    as the simplest admissible model, or using prior physical information such
    as known symmetries, conservation laws, locality constraints, or accessible
    excitation sectors. Set $r=0$.

    \item \textbf{Local neighborhood construction.}
    For the current structure $\nu^{(r)}$, construct a local neighborhood
    $\mathcal{N}_{\rm loc}(\nu^{(r)})$ in the embedding graph. This neighborhood
    consists of structures obtained from $\nu^{(r)}$ by elementary splitting or
    merging of Wedderburn blocks.

    \item \textbf{Training of local candidates.}
    For each candidate structure $\nu'\in\mathcal{N}_{\rm loc}(\nu^{(r)})$,
    train the corresponding model by maximizing the likelihood over the
    continuous parameters $\lambda_{\nu'}$:
    \begin{equation*}
        \widehat{\lambda}_{\nu'}
        =
        \operatorname*{arg\,max}_{\lambda_{\nu'}}
        F(\nu',\lambda_{\nu'},\sigma;\mathbf E_{\rm train}).
    \end{equation*}
    Whenever possible, initialize the optimization from the parameters already
    found for the current structure $\nu^{(r)}$. This warm-start procedure is
    natural when $\nu'$ is a more flexible refinement of $\nu^{(r)}$.

    \item \textbf{Validation scoring.}
    Compute the validation score
    \begin{equation*}
        S(\nu')
        =
        \frac{1}{N_{\rm val}}
        F(\nu',\widehat{\lambda}_{\nu'},\sigma;\mathbf E_{\rm val})
    \end{equation*}
    for all $\nu'\in\mathcal{N}_{\rm loc}(\nu^{(r)})$, and also for the current
    structure $\nu^{(r)}$.

    \item \textbf{Model-complexity check.}
    Let $\delta_{\rm val}>0$ be a tolerance level describing the statistical
    uncertainty of validation scores. This tolerance may be fixed directly,
    estimated from the variability over random initializations, or obtained
    from a bootstrap analysis over independent measurement chains.

    \item \textbf{Stopping rule.}
    If no more flexible candidate $\nu'\in\mathcal{N}_{\rm loc}(\nu^{(r)})$
    improves the validation score by more than $\delta_{\rm val}$, i.e.
    \begin{equation*}
        S(\nu') - S(\nu^{(r)}) \leq \delta_{\rm val}
    \end{equation*}
    for all more flexible local candidates $\nu'$, stop the search and select
    $\nu^{(r)}$ as the effective algebraic structure supported by the data.

    \item \textbf{Greedy update.}
    If at least one more flexible candidate improves the validation score by
    more than $\delta_{\rm val}$, choose the best such candidate,
    \begin{equation*}
        \nu^{(r+1)}
        =
        \operatorname*{arg\,max}_{\nu'\in\mathcal{N}_{\rm loc}(\nu^{(r)})}
        S(\nu'),
    \end{equation*}
    set $r\leftarrow r+1$, and return to step 2.

    \item \textbf{Tie-breaking by simplicity.}
    If several candidate structures have validation scores that differ by less
    than $\delta_{\rm val}$, choose the simplest among them, i.e. the structure
    higher in the embedding hierarchy. This prevents interpreting small
    likelihood fluctuations as evidence for a more complex hidden algebra.
\end{enumerate}

\nocite{*}

\bibliography{refphysrev2.bib}

\begin{thebibliography}{106}%
\makeatletter
\providecommand \@ifxundefined [1]{%
 \@ifx{#1\undefined}
}%
\providecommand \@ifnum [1]{%
 \ifnum #1\expandafter \@firstoftwo
 \else \expandafter \@secondoftwo
 \fi
}%
\providecommand \@ifx [1]{%
 \ifx #1\expandafter \@firstoftwo
 \else \expandafter \@secondoftwo
 \fi
}%
\providecommand \natexlab [1]{#1}%
\providecommand \enquote  [1]{``#1''}%
\providecommand \bibnamefont  [1]{#1}%
\providecommand \bibfnamefont [1]{#1}%
\providecommand \citenamefont [1]{#1}%
\providecommand \href@noop [0]{\@secondoftwo}%
\providecommand \href [0]{\begingroup \@sanitize@url \@href}%
\providecommand \@href[1]{\@@startlink{#1}\@@href}%
\providecommand \@@href[1]{\endgroup#1\@@endlink}%
\providecommand \@sanitize@url [0]{\catcode `\\12\catcode `\$12\catcode `\&12\catcode `\#12\catcode `\^12\catcode `\_12\catcode `\%12\relax}%
\providecommand \@@startlink[1]{}%
\providecommand \@@endlink[0]{}%
\providecommand \url  [0]{\begingroup\@sanitize@url \@url }%
\providecommand \@url [1]{\endgroup\@href {#1}{\urlprefix }}%
\providecommand \urlprefix  [0]{URL }%
\providecommand \Eprint [0]{\href }%
\providecommand \doibase [0]{https://doi.org/}%
\providecommand \selectlanguage [0]{\@gobble}%
\providecommand \bibinfo  [0]{\@secondoftwo}%
\providecommand \bibfield  [0]{\@secondoftwo}%
\providecommand \translation [1]{[#1]}%
\providecommand \BibitemOpen [0]{}%
\providecommand \bibitemStop [0]{}%
\providecommand \bibitemNoStop [0]{.\EOS\space}%
\providecommand \EOS [0]{\spacefactor3000\relax}%
\providecommand \BibitemShut  [1]{\csname bibitem#1\endcsname}%
\let\auto@bib@innerbib\@empty
\bibitem [{\citenamefont {Goldberg}(1989)}]{Goldberg1989}%
  \BibitemOpen
  \bibfield  {author} {\bibinfo {author} {\bibfnamefont {D.~E.}\ \bibnamefont {Goldberg}},\ }\href@noop {} {\emph {\bibinfo {title} {Genetic Algorithms in Search, Optimization and Machine Learning}}}\ (\bibinfo  {publisher} {Addison-Wesley},\ \bibinfo {address} {Reading},\ \bibinfo {year} {1989})\BibitemShut {NoStop}%
\bibitem [{\citenamefont {Pechen}\ and\ \citenamefont {Rabitz}(2006)}]{Pechen_Rabitz_2006}%
  \BibitemOpen
  \bibfield  {author} {\bibinfo {author} {\bibfnamefont {A.}~\bibnamefont {Pechen}}\ and\ \bibinfo {author} {\bibfnamefont {H.}~\bibnamefont {Rabitz}},\ }\href {https://doi.org/10.1103/PhysRevA.73.062102} {\bibfield  {journal} {\bibinfo  {journal} {Phys. Rev. A}\ }\textbf {\bibinfo {volume} {73}},\ \bibinfo {pages} {062102} (\bibinfo {year} {2006})}\BibitemShut {NoStop}%
\bibitem [{\citenamefont {Judson}\ and\ \citenamefont {Rabitz}(1992)}]{Judson_Rabitz_1992}%
  \BibitemOpen
  \bibfield  {author} {\bibinfo {author} {\bibfnamefont {R.~S.}\ \bibnamefont {Judson}}\ and\ \bibinfo {author} {\bibfnamefont {H.}~\bibnamefont {Rabitz}},\ }\href {https://doi.org/10.1103/PhysRevLett.68.1500} {\bibfield  {journal} {\bibinfo  {journal} {Phys. Rev. Lett.}\ }\textbf {\bibinfo {volume} {68}},\ \bibinfo {pages} {1500} (\bibinfo {year} {1992})}\BibitemShut {NoStop}%
\bibitem [{\citenamefont {Dong}\ \emph {et~al.}(2008)\citenamefont {Dong}, \citenamefont {Chen}, \citenamefont {Tarn}, \citenamefont {Pechen},\ and\ \citenamefont {Rabitz}}]{Dong_Rabitz_2008}%
  \BibitemOpen
  \bibfield  {author} {\bibinfo {author} {\bibfnamefont {D.}~\bibnamefont {Dong}}, \bibinfo {author} {\bibfnamefont {C.}~\bibnamefont {Chen}}, \bibinfo {author} {\bibfnamefont {T.-J.}\ \bibnamefont {Tarn}}, \bibinfo {author} {\bibfnamefont {A.}~\bibnamefont {Pechen}},\ and\ \bibinfo {author} {\bibfnamefont {H.}~\bibnamefont {Rabitz}},\ }\href {https://doi.org/10.1109/TSMCB.2008.926603} {\bibfield  {journal} {\bibinfo  {journal} {IEEE Trans. Syst. Man Cybern. B}\ }\textbf {\bibinfo {volume} {38}},\ \bibinfo {pages} {957} (\bibinfo {year} {2008})}\BibitemShut {NoStop}%
\bibitem [{\citenamefont {Ma}\ \emph {et~al.}(2025)\citenamefont {Ma}, \citenamefont {Qi}, \citenamefont {Petersen}, \citenamefont {Wu}, \citenamefont {Rabitz},\ and\ \citenamefont {Dong}}]{Ma_Qi_Petersen_Wu_Rabitz_Dong_2025}%
  \BibitemOpen
  \bibfield  {author} {\bibinfo {author} {\bibfnamefont {H.}~\bibnamefont {Ma}}, \bibinfo {author} {\bibfnamefont {B.}~\bibnamefont {Qi}}, \bibinfo {author} {\bibfnamefont {I.~R.}\ \bibnamefont {Petersen}}, \bibinfo {author} {\bibfnamefont {R.-B.}\ \bibnamefont {Wu}}, \bibinfo {author} {\bibfnamefont {H.}~\bibnamefont {Rabitz}},\ and\ \bibinfo {author} {\bibfnamefont {D.}~\bibnamefont {Dong}},\ }\href {https://doi.org/10.1093/nsr/nwaf269} {\bibfield  {journal} {\bibinfo  {journal} {Natl. Sci. Rev.}\ }\textbf {\bibinfo {volume} {12}},\ \bibinfo {pages} {nwaf269} (\bibinfo {year} {2025})}\BibitemShut {NoStop}%
\bibitem [{\citenamefont {Hentschel}\ and\ \citenamefont {Sanders}(2010)}]{Hentschel_Sanders_2010}%
  \BibitemOpen
  \bibfield  {author} {\bibinfo {author} {\bibfnamefont {A.}~\bibnamefont {Hentschel}}\ and\ \bibinfo {author} {\bibfnamefont {B.~C.}\ \bibnamefont {Sanders}},\ }\href {https://doi.org/10.1103/PhysRevLett.104.063603} {\bibfield  {journal} {\bibinfo  {journal} {Phys. Rev. Lett.}\ }\textbf {\bibinfo {volume} {104}},\ \bibinfo {pages} {063603} (\bibinfo {year} {2010})}\BibitemShut {NoStop}%
\bibitem [{\citenamefont {Niu}\ \emph {et~al.}(2019)\citenamefont {Niu}, \citenamefont {Boixo}, \citenamefont {Smelyanskiy},\ and\ \citenamefont {Neven}}]{Niu_Boixo_Smelyanskiy_Neven_2019}%
  \BibitemOpen
  \bibfield  {author} {\bibinfo {author} {\bibfnamefont {M.~Y.}\ \bibnamefont {Niu}}, \bibinfo {author} {\bibfnamefont {S.}~\bibnamefont {Boixo}}, \bibinfo {author} {\bibfnamefont {V.~N.}\ \bibnamefont {Smelyanskiy}},\ and\ \bibinfo {author} {\bibfnamefont {H.}~\bibnamefont {Neven}},\ }\href {https://doi.org/10.1038/s41534-019-0141-3} {\bibfield  {journal} {\bibinfo  {journal} {npj Quantum Inf.}\ }\textbf {\bibinfo {volume} {5}},\ \bibinfo {pages} {33} (\bibinfo {year} {2019})}\BibitemShut {NoStop}%
\bibitem [{\citenamefont {Fanchini}\ \emph {et~al.}(2021)\citenamefont {Fanchini}, \citenamefont {Karpat}, \citenamefont {Rossatto}, \citenamefont {Norambuena},\ and\ \citenamefont {Coto}}]{Fanchini_Karpat_Rossatto_Norambuena_Coto_2021}%
  \BibitemOpen
  \bibfield  {author} {\bibinfo {author} {\bibfnamefont {F.~F.}\ \bibnamefont {Fanchini}}, \bibinfo {author} {\bibfnamefont {G.}~\bibnamefont {Karpat}}, \bibinfo {author} {\bibfnamefont {D.~Z.}\ \bibnamefont {Rossatto}}, \bibinfo {author} {\bibfnamefont {A.}~\bibnamefont {Norambuena}},\ and\ \bibinfo {author} {\bibfnamefont {R.}~\bibnamefont {Coto}},\ }\href {https://doi.org/10.1103/PhysRevA.103.022425} {\bibfield  {journal} {\bibinfo  {journal} {Phys. Rev. A}\ }\textbf {\bibinfo {volume} {103}},\ \bibinfo {pages} {022425} (\bibinfo {year} {2021})}\BibitemShut {NoStop}%
\bibitem [{\citenamefont {Doria}\ \emph {et~al.}(2011)\citenamefont {Doria}, \citenamefont {Calarco},\ and\ \citenamefont {Montangero}}]{Doria_Calarco_Montangero_2011}%
  \BibitemOpen
  \bibfield  {author} {\bibinfo {author} {\bibfnamefont {P.}~\bibnamefont {Doria}}, \bibinfo {author} {\bibfnamefont {T.}~\bibnamefont {Calarco}},\ and\ \bibinfo {author} {\bibfnamefont {S.}~\bibnamefont {Montangero}},\ }\href {https://doi.org/10.1103/PhysRevLett.106.190501} {\bibfield  {journal} {\bibinfo  {journal} {Phys. Rev. Lett.}\ }\textbf {\bibinfo {volume} {106}},\ \bibinfo {pages} {190501} (\bibinfo {year} {2011})}\BibitemShut {NoStop}%
\bibitem [{\citenamefont {Guo}\ \emph {et~al.}(2020)\citenamefont {Guo}, \citenamefont {Modi},\ and\ \citenamefont {Poletti}}]{Guo_Modi_Poletti_2020}%
  \BibitemOpen
  \bibfield  {author} {\bibinfo {author} {\bibfnamefont {C.}~\bibnamefont {Guo}}, \bibinfo {author} {\bibfnamefont {K.}~\bibnamefont {Modi}},\ and\ \bibinfo {author} {\bibfnamefont {D.}~\bibnamefont {Poletti}},\ }\href {https://doi.org/10.1103/PhysRevA.102.062414} {\bibfield  {journal} {\bibinfo  {journal} {Phys. Rev. A}\ }\textbf {\bibinfo {volume} {102}},\ \bibinfo {pages} {062414} (\bibinfo {year} {2020})}\BibitemShut {NoStop}%
\bibitem [{\citenamefont {Vilkoviskiy}\ and\ \citenamefont {Abanin}(2024)}]{Vilkoviskiy_Abanin_2024}%
  \BibitemOpen
  \bibfield  {author} {\bibinfo {author} {\bibfnamefont {I.}~\bibnamefont {Vilkoviskiy}}\ and\ \bibinfo {author} {\bibfnamefont {D.~A.}\ \bibnamefont {Abanin}},\ }\href {https://doi.org/10.1103/PhysRevB.109.205126} {\bibfield  {journal} {\bibinfo  {journal} {Phys. Rev. B}\ }\textbf {\bibinfo {volume} {109}},\ \bibinfo {pages} {205126} (\bibinfo {year} {2024})}\BibitemShut {NoStop}%
\bibitem [{\citenamefont {Popovych}\ \emph {et~al.}()\citenamefont {Popovych}, \citenamefont {Jacobs}, \citenamefont {Korpas}, \citenamefont {Marecek},\ and\ \citenamefont {Bondar}}]{Popovych_Jacobs_Korpas_Marecek_Bondar_2022}%
  \BibitemOpen
  \bibfield  {author} {\bibinfo {author} {\bibfnamefont {Z.}~\bibnamefont {Popovych}}, \bibinfo {author} {\bibfnamefont {K.}~\bibnamefont {Jacobs}}, \bibinfo {author} {\bibfnamefont {G.}~\bibnamefont {Korpas}}, \bibinfo {author} {\bibfnamefont {J.}~\bibnamefont {Marecek}},\ and\ \bibinfo {author} {\bibfnamefont {D.~I.}\ \bibnamefont {Bondar}},\ }\href@noop {} {}\Eprint {https://arxiv.org/abs/2203.17164} {arXiv:2203.17164} \BibitemShut {NoStop}%
\bibitem [{\citenamefont {Martina}\ \emph {et~al.}(2023)\citenamefont {Martina}, \citenamefont {Gherardini},\ and\ \citenamefont {Caruso}}]{Martina_Gherardini_Caruso_2023}%
  \BibitemOpen
  \bibfield  {author} {\bibinfo {author} {\bibfnamefont {S.}~\bibnamefont {Martina}}, \bibinfo {author} {\bibfnamefont {S.}~\bibnamefont {Gherardini}},\ and\ \bibinfo {author} {\bibfnamefont {F.}~\bibnamefont {Caruso}},\ }\href {https://doi.org/10.1088/1402-4896/acb39b} {\bibfield  {journal} {\bibinfo  {journal} {Phys. Scr.}\ }\textbf {\bibinfo {volume} {98}},\ \bibinfo {pages} {035104} (\bibinfo {year} {2023})}\BibitemShut {NoStop}%
\bibitem [{\citenamefont {Mazza}\ \emph {et~al.}(2021)\citenamefont {Mazza}, \citenamefont {Zietlow}, \citenamefont {Carollo}, \citenamefont {Andergassen}, \citenamefont {Martius},\ and\ \citenamefont {Lesanovsky}}]{Mazza_Zietlow_Carollo_Andergassen_Martius_Lesanovsky_2021}%
  \BibitemOpen
  \bibfield  {author} {\bibinfo {author} {\bibfnamefont {P.~P.}\ \bibnamefont {Mazza}}, \bibinfo {author} {\bibfnamefont {D.}~\bibnamefont {Zietlow}}, \bibinfo {author} {\bibfnamefont {F.}~\bibnamefont {Carollo}}, \bibinfo {author} {\bibfnamefont {S.}~\bibnamefont {Andergassen}}, \bibinfo {author} {\bibfnamefont {G.}~\bibnamefont {Martius}},\ and\ \bibinfo {author} {\bibfnamefont {I.}~\bibnamefont {Lesanovsky}},\ }\href {https://doi.org/10.1103/PhysRevResearch.3.023084} {\bibfield  {journal} {\bibinfo  {journal} {Phys. Rev. Res.}\ }\textbf {\bibinfo {volume} {3}},\ \bibinfo {pages} {023084} (\bibinfo {year} {2021})}\BibitemShut {NoStop}%
\bibitem [{\citenamefont {Yang}\ \emph {et~al.}(2022)\citenamefont {Yang}, \citenamefont {Cao},\ and\ \citenamefont {Yang}}]{Yang_Cao_Yang_2022}%
  \BibitemOpen
  \bibfield  {author} {\bibinfo {author} {\bibfnamefont {J.}~\bibnamefont {Yang}}, \bibinfo {author} {\bibfnamefont {J.}~\bibnamefont {Cao}},\ and\ \bibinfo {author} {\bibfnamefont {W.-L.}\ \bibnamefont {Yang}},\ }\href {https://doi.org/10.1088/1674-1056/ac2490} {\bibfield  {journal} {\bibinfo  {journal} {Chin. Phys. B}\ }\textbf {\bibinfo {volume} {31}},\ \bibinfo {pages} {010314} (\bibinfo {year} {2022})}\BibitemShut {NoStop}%
\bibitem [{\citenamefont {Wu}\ \emph {et~al.}(2024)\citenamefont {Wu}, \citenamefont {Li}, \citenamefont {Zhao}, \citenamefont {Luan}, \citenamefont {Yu},\ and\ \citenamefont {Zhang}}]{Wu_Li_Zhao_Luan_Yu_Zhang_2024}%
  \BibitemOpen
  \bibfield  {author} {\bibinfo {author} {\bibfnamefont {Y.}~\bibnamefont {Wu}}, \bibinfo {author} {\bibfnamefont {Z.}~\bibnamefont {Li}}, \bibinfo {author} {\bibfnamefont {D.}~\bibnamefont {Zhao}}, \bibinfo {author} {\bibfnamefont {T.}~\bibnamefont {Luan}}, \bibinfo {author} {\bibfnamefont {X.}~\bibnamefont {Yu}},\ and\ \bibinfo {author} {\bibfnamefont {Z.}~\bibnamefont {Zhang}},\ }in\ \href {https://doi.org/10.1109/ICTC61510.2024.10602090} {\emph {\bibinfo {booktitle} {2024 5th Information Communication Technologies Conference (ICTC)}}}\ (\bibinfo  {publisher} {IEEE},\ \bibinfo {address} {Nanjing},\ \bibinfo {year} {2024})\ pp.\ \bibinfo {pages} {54--58}\BibitemShut {NoStop}%
\bibitem [{\citenamefont {Luchnikov}\ \emph {et~al.}(2022)\citenamefont {Luchnikov}, \citenamefont {Kiktenko}, \citenamefont {Gavreev}, \citenamefont {Ouerdane}, \citenamefont {Filippov},\ and\ \citenamefont {Fedorov}}]{luchnikov2022probing}%
  \BibitemOpen
  \bibfield  {author} {\bibinfo {author} {\bibfnamefont {I.~A.}\ \bibnamefont {Luchnikov}}, \bibinfo {author} {\bibfnamefont {E.~O.}\ \bibnamefont {Kiktenko}}, \bibinfo {author} {\bibfnamefont {M.~A.}\ \bibnamefont {Gavreev}}, \bibinfo {author} {\bibfnamefont {H.}~\bibnamefont {Ouerdane}}, \bibinfo {author} {\bibfnamefont {S.~N.}\ \bibnamefont {Filippov}},\ and\ \bibinfo {author} {\bibfnamefont {A.~K.}\ \bibnamefont {Fedorov}},\ }\href {https://doi.org/10.1103/PhysRevResearch.4.043002} {\bibfield  {journal} {\bibinfo  {journal} {Phys. Rev. Res.}\ }\textbf {\bibinfo {volume} {4}},\ \bibinfo {pages} {043002} (\bibinfo {year} {2022})}\BibitemShut {NoStop}%
\bibitem [{\citenamefont {Luchnikov}\ \emph {et~al.}(2024)\citenamefont {Luchnikov}, \citenamefont {Gavreev},\ and\ \citenamefont {Fedorov}}]{luchnikov2024controlling}%
  \BibitemOpen
  \bibfield  {author} {\bibinfo {author} {\bibfnamefont {I.~A.}\ \bibnamefont {Luchnikov}}, \bibinfo {author} {\bibfnamefont {M.~A.}\ \bibnamefont {Gavreev}},\ and\ \bibinfo {author} {\bibfnamefont {A.~K.}\ \bibnamefont {Fedorov}},\ }\href {https://doi.org/10.1103/PhysRevResearch.6.013161} {\bibfield  {journal} {\bibinfo  {journal} {Phys. Rev. Res.}\ }\textbf {\bibinfo {volume} {6}},\ \bibinfo {pages} {013161} (\bibinfo {year} {2024})}\BibitemShut {NoStop}%
\bibitem [{\citenamefont {Zhang}\ \emph {et~al.}(2022)\citenamefont {Zhang}, \citenamefont {Pokharel}, \citenamefont {Levenson-Falk},\ and\ \citenamefont {Lidar}}]{Zhang_Pokharel_Levenson-Falk_Lidar_2022}%
  \BibitemOpen
  \bibfield  {author} {\bibinfo {author} {\bibfnamefont {H.}~\bibnamefont {Zhang}}, \bibinfo {author} {\bibfnamefont {B.}~\bibnamefont {Pokharel}}, \bibinfo {author} {\bibfnamefont {E.}~\bibnamefont {Levenson-Falk}},\ and\ \bibinfo {author} {\bibfnamefont {D.}~\bibnamefont {Lidar}},\ }\href {https://doi.org/10.1103/PhysRevApplied.17.054018} {\bibfield  {journal} {\bibinfo  {journal} {Phys. Rev. Appl.}\ }\textbf {\bibinfo {volume} {17}},\ \bibinfo {pages} {054018} (\bibinfo {year} {2022})}\BibitemShut {NoStop}%
\bibitem [{\citenamefont {Wang}\ \emph {et~al.}(2024)\citenamefont {Wang}, \citenamefont {Ku}, \citenamefont {Lin},\ and\ \citenamefont {Chen}}]{Wang_Ku_Lin_Chen_2024}%
  \BibitemOpen
  \bibfield  {author} {\bibinfo {author} {\bibfnamefont {H.-M.}\ \bibnamefont {Wang}}, \bibinfo {author} {\bibfnamefont {H.-Y.}\ \bibnamefont {Ku}}, \bibinfo {author} {\bibfnamefont {J.-Y.}\ \bibnamefont {Lin}},\ and\ \bibinfo {author} {\bibfnamefont {H.-B.}\ \bibnamefont {Chen}},\ }\href {https://doi.org/10.1038/s42005-024-01563-3} {\bibfield  {journal} {\bibinfo  {journal} {Commun. Phys.}\ }\textbf {\bibinfo {volume} {7}},\ \bibinfo {pages} {72} (\bibinfo {year} {2024})}\BibitemShut {NoStop}%
\bibitem [{\citenamefont {Dalgaard}\ \emph {et~al.}(2020)\citenamefont {Dalgaard}, \citenamefont {Motzoi}, \citenamefont {Sørensen},\ and\ \citenamefont {Sherson}}]{Dalgaard_2020}%
  \BibitemOpen
  \bibfield  {author} {\bibinfo {author} {\bibfnamefont {M.}~\bibnamefont {Dalgaard}}, \bibinfo {author} {\bibfnamefont {F.}~\bibnamefont {Motzoi}}, \bibinfo {author} {\bibfnamefont {J.~J.}\ \bibnamefont {Sørensen}},\ and\ \bibinfo {author} {\bibfnamefont {J.}~\bibnamefont {Sherson}},\ }\href {https://doi.org/10.1038/s41534-019-0241-0} {\bibfield  {journal} {\bibinfo  {journal} {npj Quantum Inf.}\ }\textbf {\bibinfo {volume} {6}},\ \bibinfo {pages} {6} (\bibinfo {year} {2020})}\BibitemShut {NoStop}%
\bibitem [{\citenamefont {Couturier}\ \emph {et~al.}(2023)\citenamefont {Couturier}, \citenamefont {Dionis}, \citenamefont {Guérin}, \citenamefont {Guyeux},\ and\ \citenamefont {Sugny}}]{Couturier_2023}%
  \BibitemOpen
  \bibfield  {author} {\bibinfo {author} {\bibfnamefont {R.}~\bibnamefont {Couturier}}, \bibinfo {author} {\bibfnamefont {E.}~\bibnamefont {Dionis}}, \bibinfo {author} {\bibfnamefont {S.}~\bibnamefont {Guérin}}, \bibinfo {author} {\bibfnamefont {C.}~\bibnamefont {Guyeux}},\ and\ \bibinfo {author} {\bibfnamefont {D.}~\bibnamefont {Sugny}},\ }\href {https://doi.org/10.3390/e25030446} {\bibfield  {journal} {\bibinfo  {journal} {Entropy}\ }\textbf {\bibinfo {volume} {25}},\ \bibinfo {pages} {446} (\bibinfo {year} {2023})}\BibitemShut {NoStop}%
\bibitem [{\citenamefont {Grech}\ \emph {et~al.}(2026)\citenamefont {Grech}, \citenamefont {Krauss}, \citenamefont {Consiglio}, \citenamefont {Apollaro}, \citenamefont {Koch}, \citenamefont {Hirlaender},\ and\ \citenamefont {Valentino}}]{Grech_Valentino_2026}%
  \BibitemOpen
  \bibfield  {author} {\bibinfo {author} {\bibfnamefont {L.}~\bibnamefont {Grech}}, \bibinfo {author} {\bibfnamefont {M.~G.}\ \bibnamefont {Krauss}}, \bibinfo {author} {\bibfnamefont {M.}~\bibnamefont {Consiglio}}, \bibinfo {author} {\bibfnamefont {T.~J.~G.}\ \bibnamefont {Apollaro}}, \bibinfo {author} {\bibfnamefont {C.~P.}\ \bibnamefont {Koch}}, \bibinfo {author} {\bibfnamefont {S.}~\bibnamefont {Hirlaender}},\ and\ \bibinfo {author} {\bibfnamefont {G.}~\bibnamefont {Valentino}},\ }\href {https://doi.org/10.1088/2058-9565/ae2c16} {\bibfield  {journal} {\bibinfo  {journal} {Quantum Sci. Technol.}\ }\textbf {\bibinfo {volume} {11}},\ \bibinfo {pages} {015030} (\bibinfo {year} {2026})}\BibitemShut {NoStop}%
\bibitem [{\citenamefont {Gui}\ \emph {et~al.}(2024)\citenamefont {Gui}, \citenamefont {Ho},\ and\ \citenamefont {Rabitz}}]{Gui_Ho_Rabitz_2024}%
  \BibitemOpen
  \bibfield  {author} {\bibinfo {author} {\bibfnamefont {S.}~\bibnamefont {Gui}}, \bibinfo {author} {\bibfnamefont {T.-S.}\ \bibnamefont {Ho}},\ and\ \bibinfo {author} {\bibfnamefont {H.}~\bibnamefont {Rabitz}},\ }\href {https://doi.org/10.1103/PhysRevA.110.052412} {\bibfield  {journal} {\bibinfo  {journal} {Phys. Rev. A}\ }\textbf {\bibinfo {volume} {110}},\ \bibinfo {pages} {052412} (\bibinfo {year} {2024})}\BibitemShut {NoStop}%
\bibitem [{\citenamefont {Dong}\ and\ \citenamefont {Petersen}(2023)}]{Dong_Petersen_2023}%
  \BibitemOpen
  \bibfield  {author} {\bibinfo {author} {\bibfnamefont {D.}~\bibnamefont {Dong}}\ and\ \bibinfo {author} {\bibfnamefont {I.~R.}\ \bibnamefont {Petersen}},\ }\bibinfo {title} {Machine learning for quantum control},\ in\ \href {https://doi.org/10.1007/978-3-031-20245-2_5} {\emph {\bibinfo {booktitle} {Learning and Robust Control in Quantum Technology}}},\ \bibinfo {series and number} {Communications and Control Engineering}\ (\bibinfo  {publisher} {Springer International Publishing},\ \bibinfo {address} {Cham},\ \bibinfo {year} {2023})\ pp.\ \bibinfo {pages} {93--140}\BibitemShut {NoStop}%
\bibitem [{\citenamefont {Chen}\ \emph {et~al.}(2024)\citenamefont {Chen}, \citenamefont {Herrmann}, \citenamefont {Vamvoudakis},\ and\ \citenamefont {Vijayan}}]{Chen_Herrmann_Vamvoudakis_Vijayan_2024}%
  \BibitemOpen
  \bibfield  {author} {\bibinfo {author} {\bibfnamefont {A.~S.}\ \bibnamefont {Chen}}, \bibinfo {author} {\bibfnamefont {G.}~\bibnamefont {Herrmann}}, \bibinfo {author} {\bibfnamefont {K.~G.}\ \bibnamefont {Vamvoudakis}},\ and\ \bibinfo {author} {\bibfnamefont {J.}~\bibnamefont {Vijayan}},\ }\href {https://doi.org/10.1109/LCSYS.2024.3409671} {\bibfield  {journal} {\bibinfo  {journal} {IEEE Control Syst. Lett.}\ }\textbf {\bibinfo {volume} {8}},\ \bibinfo {pages} {1319} (\bibinfo {year} {2024})}\BibitemShut {NoStop}%
\bibitem [{\citenamefont {Oza}\ \emph {et~al.}(2009)\citenamefont {Oza}, \citenamefont {Pechen}, \citenamefont {Dominy}, \citenamefont {Beltrani}, \citenamefont {Moore},\ and\ \citenamefont {Rabitz}}]{Oza_Pechen_Dominy_Beltrani_Moore_Rabitz_2009}%
  \BibitemOpen
  \bibfield  {author} {\bibinfo {author} {\bibfnamefont {A.}~\bibnamefont {Oza}}, \bibinfo {author} {\bibfnamefont {A.}~\bibnamefont {Pechen}}, \bibinfo {author} {\bibfnamefont {J.}~\bibnamefont {Dominy}}, \bibinfo {author} {\bibfnamefont {V.}~\bibnamefont {Beltrani}}, \bibinfo {author} {\bibfnamefont {K.}~\bibnamefont {Moore}},\ and\ \bibinfo {author} {\bibfnamefont {H.}~\bibnamefont {Rabitz}},\ }\href {https://doi.org/10.1088/1751-8113/42/20/205305} {\bibfield  {journal} {\bibinfo  {journal} {J. Phys. A: Math. Theor.}\ }\textbf {\bibinfo {volume} {42}},\ \bibinfo {pages} {205305} (\bibinfo {year} {2009})}\BibitemShut {NoStop}%
\bibitem [{\citenamefont {Pechen}\ \emph {et~al.}(2008)\citenamefont {Pechen}, \citenamefont {Prokhorenko}, \citenamefont {Wu},\ and\ \citenamefont {Rabitz}}]{Pechen_Prokhorenko_Wu_Rabitz_2008}%
  \BibitemOpen
  \bibfield  {author} {\bibinfo {author} {\bibfnamefont {A.}~\bibnamefont {Pechen}}, \bibinfo {author} {\bibfnamefont {D.}~\bibnamefont {Prokhorenko}}, \bibinfo {author} {\bibfnamefont {R.}~\bibnamefont {Wu}},\ and\ \bibinfo {author} {\bibfnamefont {H.}~\bibnamefont {Rabitz}},\ }\href {https://doi.org/10.1088/1751-8113/41/4/045205} {\bibfield  {journal} {\bibinfo  {journal} {J. Phys. A: Math. Theor.}\ }\textbf {\bibinfo {volume} {41}},\ \bibinfo {pages} {045205} (\bibinfo {year} {2008})}\BibitemShut {NoStop}%
\bibitem [{\citenamefont {Wu}\ \emph {et~al.}(2008)\citenamefont {Wu}, \citenamefont {Pechen}, \citenamefont {Rabitz}, \citenamefont {Hsieh},\ and\ \citenamefont {Tsou}}]{Wu_Pechen_Rabitz_Hsieh_Tsou_2008}%
  \BibitemOpen
  \bibfield  {author} {\bibinfo {author} {\bibfnamefont {R.}~\bibnamefont {Wu}}, \bibinfo {author} {\bibfnamefont {A.}~\bibnamefont {Pechen}}, \bibinfo {author} {\bibfnamefont {H.}~\bibnamefont {Rabitz}}, \bibinfo {author} {\bibfnamefont {M.}~\bibnamefont {Hsieh}},\ and\ \bibinfo {author} {\bibfnamefont {B.}~\bibnamefont {Tsou}},\ }\href {https://doi.org/10.1063/1.2883738} {\bibfield  {journal} {\bibinfo  {journal} {J. Math. Phys.}\ }\textbf {\bibinfo {volume} {49}},\ \bibinfo {pages} {022108} (\bibinfo {year} {2008})}\BibitemShut {NoStop}%
\bibitem [{\citenamefont {Moore}\ \emph {et~al.}(2011{\natexlab{a}})\citenamefont {Moore}, \citenamefont {Pechen}, \citenamefont {Feng}, \citenamefont {Dominy}, \citenamefont {Beltrani},\ and\ \citenamefont {Rabitz}}]{Moore_Pechen_Feng_Dominy_Beltrani_Rabitz_2011a}%
  \BibitemOpen
  \bibfield  {author} {\bibinfo {author} {\bibfnamefont {K.~W.}\ \bibnamefont {Moore}}, \bibinfo {author} {\bibfnamefont {A.}~\bibnamefont {Pechen}}, \bibinfo {author} {\bibfnamefont {X.-J.}\ \bibnamefont {Feng}}, \bibinfo {author} {\bibfnamefont {J.}~\bibnamefont {Dominy}}, \bibinfo {author} {\bibfnamefont {V.}~\bibnamefont {Beltrani}},\ and\ \bibinfo {author} {\bibfnamefont {H.}~\bibnamefont {Rabitz}},\ }\href {https://doi.org/10.1039/C0SC00425A} {\bibfield  {journal} {\bibinfo  {journal} {Chem. Sci.}\ }\textbf {\bibinfo {volume} {2}},\ \bibinfo {pages} {417} (\bibinfo {year} {2011}{\natexlab{a}})}\BibitemShut {NoStop}%
\bibitem [{\citenamefont {Moore}\ \emph {et~al.}(2011{\natexlab{b}})\citenamefont {Moore}, \citenamefont {Pechen}, \citenamefont {Feng}, \citenamefont {Dominy}, \citenamefont {Beltrani},\ and\ \citenamefont {Rabitz}}]{Moore_Pechen_Feng_Dominy_Beltrani_Rabitz_2011b}%
  \BibitemOpen
  \bibfield  {author} {\bibinfo {author} {\bibfnamefont {K.~W.}\ \bibnamefont {Moore}}, \bibinfo {author} {\bibfnamefont {A.}~\bibnamefont {Pechen}}, \bibinfo {author} {\bibfnamefont {X.-J.}\ \bibnamefont {Feng}}, \bibinfo {author} {\bibfnamefont {J.}~\bibnamefont {Dominy}}, \bibinfo {author} {\bibfnamefont {V.~J.}\ \bibnamefont {Beltrani}},\ and\ \bibinfo {author} {\bibfnamefont {H.}~\bibnamefont {Rabitz}},\ }\href {https://doi.org/10.1039/C1CP20353C} {\bibfield  {journal} {\bibinfo  {journal} {Phys. Chem. Chem. Phys.}\ }\textbf {\bibinfo {volume} {13}},\ \bibinfo {pages} {10048} (\bibinfo {year} {2011}{\natexlab{b}})}\BibitemShut {NoStop}%
\bibitem [{\citenamefont {Feng}\ \emph {et~al.}(2012)\citenamefont {Feng}, \citenamefont {Pechen}, \citenamefont {Jha}, \citenamefont {Wu},\ and\ \citenamefont {Rabitz}}]{Feng_Pechen_Jha_Wu_Rabitz_2012}%
  \BibitemOpen
  \bibfield  {author} {\bibinfo {author} {\bibfnamefont {X.}~\bibnamefont {Feng}}, \bibinfo {author} {\bibfnamefont {A.}~\bibnamefont {Pechen}}, \bibinfo {author} {\bibfnamefont {A.}~\bibnamefont {Jha}}, \bibinfo {author} {\bibfnamefont {R.}~\bibnamefont {Wu}},\ and\ \bibinfo {author} {\bibfnamefont {H.}~\bibnamefont {Rabitz}},\ }\href {https://doi.org/10.1039/C1SC00648G} {\bibfield  {journal} {\bibinfo  {journal} {Chem. Sci.}\ }\textbf {\bibinfo {volume} {3}},\ \bibinfo {pages} {900} (\bibinfo {year} {2012})}\BibitemShut {NoStop}%
\bibitem [{\citenamefont {Barthel}\ and\ \citenamefont {Zhang}(2022)}]{barthel2022solving}%
  \BibitemOpen
  \bibfield  {author} {\bibinfo {author} {\bibfnamefont {T.}~\bibnamefont {Barthel}}\ and\ \bibinfo {author} {\bibfnamefont {Y.}~\bibnamefont {Zhang}},\ }\href {https://doi.org/10.1088/1742-5468/ac8e5c} {\bibfield  {journal} {\bibinfo  {journal} {J. Stat. Mech.: Theory Exp.}\ }\textbf {\bibinfo {volume} {2022}}\bibinfo  {number} { (11)},\ \bibinfo {pages} {113101}}\BibitemShut {NoStop}%
\bibitem [{\citenamefont {Zhang}\ and\ \citenamefont {Barthel}(2022)}]{zhang2022criticality}%
  \BibitemOpen
\bibfield  {number} {  }\bibfield  {author} {\bibinfo {author} {\bibfnamefont {Y.}~\bibnamefont {Zhang}}\ and\ \bibinfo {author} {\bibfnamefont {T.}~\bibnamefont {Barthel}},\ }\href {https://doi.org/10.1103/PhysRevLett.129.120401} {\bibfield  {journal} {\bibinfo  {journal} {Phys. Rev. Lett.}\ }\textbf {\bibinfo {volume} {129}},\ \bibinfo {pages} {120401} (\bibinfo {year} {2022})}\BibitemShut {NoStop}%
\bibitem [{\citenamefont {Teretenkov}(2020{\natexlab{a}})}]{teretenkov2020dynamics}%
  \BibitemOpen
  \bibfield  {author} {\bibinfo {author} {\bibfnamefont {A.~E.}\ \bibnamefont {Teretenkov}},\ }\href {https://doi.org/10.1134/S0001434620030372} {\bibfield  {journal} {\bibinfo  {journal} {Math. Notes}\ }\textbf {\bibinfo {volume} {107}},\ \bibinfo {pages} {695} (\bibinfo {year} {2020}{\natexlab{a}})}\BibitemShut {NoStop}%
\bibitem [{\citenamefont {Ivanov}\ and\ \citenamefont {Teretenkov}(2022)}]{ivanov2022dynamics}%
  \BibitemOpen
  \bibfield  {author} {\bibinfo {author} {\bibfnamefont {D.~D.}\ \bibnamefont {Ivanov}}\ and\ \bibinfo {author} {\bibfnamefont {A.~E.}\ \bibnamefont {Teretenkov}},\ }\href {https://doi.org/10.1134/S0001434622070367} {\bibfield  {journal} {\bibinfo  {journal} {Math. Notes}\ }\textbf {\bibinfo {volume} {112}},\ \bibinfo {pages} {318} (\bibinfo {year} {2022})}\BibitemShut {NoStop}%
\bibitem [{\citenamefont {Nosal’}\ and\ \citenamefont {Teretenkov}(2020)}]{nosal2020exact}%
  \BibitemOpen
  \bibfield  {author} {\bibinfo {author} {\bibfnamefont {I.~A.}\ \bibnamefont {Nosal’}}\ and\ \bibinfo {author} {\bibfnamefont {A.~E.}\ \bibnamefont {Teretenkov}},\ }\href {https://doi.org/10.1134/S0001434620110358} {\bibfield  {journal} {\bibinfo  {journal} {Math. Notes}\ }\textbf {\bibinfo {volume} {108}},\ \bibinfo {pages} {911} (\bibinfo {year} {2020})}\BibitemShut {NoStop}%
\bibitem [{\citenamefont {Linowski}\ \emph {et~al.}(2022)\citenamefont {Linowski}, \citenamefont {Teretenkov},\ and\ \citenamefont {Rudnicki}}]{linowski2022dissipative}%
  \BibitemOpen
  \bibfield  {author} {\bibinfo {author} {\bibfnamefont {T.}~\bibnamefont {Linowski}}, \bibinfo {author} {\bibfnamefont {A.}~\bibnamefont {Teretenkov}},\ and\ \bibinfo {author} {\bibfnamefont {{\L}.}~\bibnamefont {Rudnicki}},\ }\href {https://doi.org/10.1103/PhysRevA.106.052206} {\bibfield  {journal} {\bibinfo  {journal} {Phys. Rev. A}\ }\textbf {\bibinfo {volume} {106}},\ \bibinfo {pages} {052206} (\bibinfo {year} {2022})}\BibitemShut {NoStop}%
\bibitem [{\citenamefont {Nosal}\ and\ \citenamefont {Teretenkov}(2022)}]{nosal2022higher}%
  \BibitemOpen
  \bibfield  {author} {\bibinfo {author} {\bibfnamefont {I.~A.}\ \bibnamefont {Nosal}}\ and\ \bibinfo {author} {\bibfnamefont {A.~E.}\ \bibnamefont {Teretenkov}},\ }\href {https://doi.org/10.1134/S1995080222100316} {\bibfield  {journal} {\bibinfo  {journal} {Lobachevskii J. Math.}\ }\textbf {\bibinfo {volume} {43}},\ \bibinfo {pages} {1726} (\bibinfo {year} {2022})}\BibitemShut {NoStop}%
\bibitem [{\citenamefont {Penc}\ and\ \citenamefont {Essler}(2026)}]{penc2026linear}%
  \BibitemOpen
  \bibfield  {author} {\bibinfo {author} {\bibfnamefont {P.}~\bibnamefont {Penc}}\ and\ \bibinfo {author} {\bibfnamefont {F.~H.~L.}\ \bibnamefont {Essler}},\ }\href {https://doi.org/10.21468/SciPostPhys.20.2.058} {\bibfield  {journal} {\bibinfo  {journal} {SciPost Phys.}\ }\textbf {\bibinfo {volume} {20}},\ \bibinfo {pages} {058} (\bibinfo {year} {2026})}\BibitemShut {NoStop}%
\bibitem [{\citenamefont {Teretenkov}\ and\ \citenamefont {Lychkovskiy}(2024)}]{teretenkov2024exact}%
  \BibitemOpen
  \bibfield  {author} {\bibinfo {author} {\bibfnamefont {A.}~\bibnamefont {Teretenkov}}\ and\ \bibinfo {author} {\bibfnamefont {O.}~\bibnamefont {Lychkovskiy}},\ }\href {https://doi.org/10.1103/PhysRevB.109.L140302} {\bibfield  {journal} {\bibinfo  {journal} {Phys. Rev. B}\ }\textbf {\bibinfo {volume} {109}},\ \bibinfo {pages} {L140302} (\bibinfo {year} {2024})}\BibitemShut {NoStop}%
\bibitem [{\citenamefont {Essler}\ and\ \citenamefont {Piroli}(2020)}]{essler2020integrability}%
  \BibitemOpen
  \bibfield  {author} {\bibinfo {author} {\bibfnamefont {F.~H.~L.}\ \bibnamefont {Essler}}\ and\ \bibinfo {author} {\bibfnamefont {L.}~\bibnamefont {Piroli}},\ }\href {https://doi.org/10.1103/PhysRevE.102.062210} {\bibfield  {journal} {\bibinfo  {journal} {Phys. Rev. E}\ }\textbf {\bibinfo {volume} {102}},\ \bibinfo {pages} {062210} (\bibinfo {year} {2020})}\BibitemShut {NoStop}%
\bibitem [{\citenamefont {Paszko}\ \emph {et~al.}()\citenamefont {Paszko}, \citenamefont {Turner}, \citenamefont {Rose},\ and\ \citenamefont {Pal}}]{paszko2025operator}%
  \BibitemOpen
  \bibfield  {author} {\bibinfo {author} {\bibfnamefont {D.}~\bibnamefont {Paszko}}, \bibinfo {author} {\bibfnamefont {C.~J.}\ \bibnamefont {Turner}}, \bibinfo {author} {\bibfnamefont {D.~C.}\ \bibnamefont {Rose}},\ and\ \bibinfo {author} {\bibfnamefont {A.}~\bibnamefont {Pal}},\ }\href@noop {} {}\Eprint {https://arxiv.org/abs/2506.16518} {arXiv:2506.16518} \BibitemShut {NoStop}%
\bibitem [{\citenamefont {Yashin}\ \emph {et~al.}(2020)\citenamefont {Yashin}, \citenamefont {Kiktenko}, \citenamefont {Mastiukova},\ and\ \citenamefont {Fedorov}}]{yashin2020minimal}%
  \BibitemOpen
  \bibfield  {author} {\bibinfo {author} {\bibfnamefont {V.}~\bibnamefont {Yashin}}, \bibinfo {author} {\bibfnamefont {E.}~\bibnamefont {Kiktenko}}, \bibinfo {author} {\bibfnamefont {A.}~\bibnamefont {Mastiukova}},\ and\ \bibinfo {author} {\bibfnamefont {A.}~\bibnamefont {Fedorov}},\ }\href@noop {} {\bibfield  {journal} {\bibinfo  {journal} {New J. Phys.}\ }\textbf {\bibinfo {volume} {22}},\ \bibinfo {pages} {103026} (\bibinfo {year} {2020})}\BibitemShut {NoStop}%
\bibitem [{\citenamefont {Kulikov}\ \emph {et~al.}(2024)\citenamefont {Kulikov}, \citenamefont {Yashin}, \citenamefont {Fedorov},\ and\ \citenamefont {Kiktenko}}]{kulikov2024minimizing}%
  \BibitemOpen
  \bibfield  {author} {\bibinfo {author} {\bibfnamefont {D.~A.}\ \bibnamefont {Kulikov}}, \bibinfo {author} {\bibfnamefont {V.~I.}\ \bibnamefont {Yashin}}, \bibinfo {author} {\bibfnamefont {A.~K.}\ \bibnamefont {Fedorov}},\ and\ \bibinfo {author} {\bibfnamefont {E.~O.}\ \bibnamefont {Kiktenko}},\ }\href@noop {} {\bibfield  {journal} {\bibinfo  {journal} {Phys. Rev. A}\ }\textbf {\bibinfo {volume} {109}},\ \bibinfo {pages} {012219} (\bibinfo {year} {2024})}\BibitemShut {NoStop}%
\bibitem [{\citenamefont {Hasen\"{o}hrl}\ and\ \citenamefont {Caro}(2023)}]{hasenohrl2023generators}%
  \BibitemOpen
  \bibfield  {author} {\bibinfo {author} {\bibfnamefont {M.}~\bibnamefont {Hasen\"{o}hrl}}\ and\ \bibinfo {author} {\bibfnamefont {M.~C.}\ \bibnamefont {Caro}},\ }\href {https://doi.org/10.1142/S1230161223500014} {\bibfield  {journal} {\bibinfo  {journal} {Open Syst. Inf. Dyn.}\ }\textbf {\bibinfo {volume} {30}},\ \bibinfo {pages} {2350001} (\bibinfo {year} {2023})}\BibitemShut {NoStop}%
\bibitem [{\citenamefont {Dhahri}\ \emph {et~al.}(2010)\citenamefont {Dhahri}, \citenamefont {Fagnola},\ and\ \citenamefont {Rebolledo}}]{dhahri2010decoherence}%
  \BibitemOpen
  \bibfield  {author} {\bibinfo {author} {\bibfnamefont {A.}~\bibnamefont {Dhahri}}, \bibinfo {author} {\bibfnamefont {F.}~\bibnamefont {Fagnola}},\ and\ \bibinfo {author} {\bibfnamefont {R.}~\bibnamefont {Rebolledo}},\ }\href {https://doi.org/10.1142/S0219025710004176} {\bibfield  {journal} {\bibinfo  {journal} {Inf. Dimens. Anal. Quantum Probab. Relat. Top.}\ }\textbf {\bibinfo {volume} {13}},\ \bibinfo {pages} {413} (\bibinfo {year} {2010})}\BibitemShut {NoStop}%
\bibitem [{\citenamefont {Dhahri}\ \emph {et~al.}(2011)\citenamefont {Dhahri}, \citenamefont {Fagnola},\ and\ \citenamefont {Rebolledo}}]{dhahri2011decoherence}%
  \BibitemOpen
  \bibfield  {author} {\bibinfo {author} {\bibfnamefont {A.}~\bibnamefont {Dhahri}}, \bibinfo {author} {\bibfnamefont {F.}~\bibnamefont {Fagnola}},\ and\ \bibinfo {author} {\bibfnamefont {R.}~\bibnamefont {Rebolledo}},\ }in\ \href@noop {} {\emph {\bibinfo {booktitle} {Quantum Probability And Related Topics}}}\ (\bibinfo  {publisher} {World Scientific},\ \bibinfo {year} {2011})\ pp.\ \bibinfo {pages} {131--147}\BibitemShut {NoStop}%
\bibitem [{\citenamefont {Deschamps}\ \emph {et~al.}(2016)\citenamefont {Deschamps}, \citenamefont {Fagnola}, \citenamefont {Sasso},\ and\ \citenamefont {Umanit\`{a}}}]{deschamps2016structure}%
  \BibitemOpen
  \bibfield  {author} {\bibinfo {author} {\bibfnamefont {J.}~\bibnamefont {Deschamps}}, \bibinfo {author} {\bibfnamefont {F.}~\bibnamefont {Fagnola}}, \bibinfo {author} {\bibfnamefont {E.}~\bibnamefont {Sasso}},\ and\ \bibinfo {author} {\bibfnamefont {V.}~\bibnamefont {Umanit\`{a}}},\ }\href {https://doi.org/10.1142/S0129055X16500033} {\bibfield  {journal} {\bibinfo  {journal} {Rev. Math. Phys.}\ }\textbf {\bibinfo {volume} {28}},\ \bibinfo {pages} {1650003} (\bibinfo {year} {2016})}\BibitemShut {NoStop}%
\bibitem [{\citenamefont {Agredo}\ \emph {et~al.}(2022)\citenamefont {Agredo}, \citenamefont {Fagnola},\ and\ \citenamefont {Poletti}}]{agredo2022decoherence}%
  \BibitemOpen
  \bibfield  {author} {\bibinfo {author} {\bibfnamefont {J.}~\bibnamefont {Agredo}}, \bibinfo {author} {\bibfnamefont {F.}~\bibnamefont {Fagnola}},\ and\ \bibinfo {author} {\bibfnamefont {D.}~\bibnamefont {Poletti}},\ }\href {https://doi.org/10.1007/s00032-022-00355-0} {\bibfield  {journal} {\bibinfo  {journal} {Milan J. Math.}\ }\textbf {\bibinfo {volume} {90}},\ \bibinfo {pages} {257} (\bibinfo {year} {2022})}\BibitemShut {NoStop}%
\bibitem [{\citenamefont {Choi}\ and\ \citenamefont {Kribs}(2006)}]{choi2006method}%
  \BibitemOpen
  \bibfield  {author} {\bibinfo {author} {\bibfnamefont {M.-D.}\ \bibnamefont {Choi}}\ and\ \bibinfo {author} {\bibfnamefont {D.~W.}\ \bibnamefont {Kribs}},\ }\href {https://doi.org/10.1103/PhysRevLett.96.050501} {\bibfield  {journal} {\bibinfo  {journal} {Phys. Rev. Lett.}\ }\textbf {\bibinfo {volume} {96}},\ \bibinfo {pages} {050501} (\bibinfo {year} {2006})}\BibitemShut {NoStop}%
\bibitem [{\citenamefont {Guan}\ \emph {et~al.}()\citenamefont {Guan}, \citenamefont {Feng},\ and\ \citenamefont {Ying}}]{guan2018structure}%
  \BibitemOpen
  \bibfield  {author} {\bibinfo {author} {\bibfnamefont {J.}~\bibnamefont {Guan}}, \bibinfo {author} {\bibfnamefont {Y.}~\bibnamefont {Feng}},\ and\ \bibinfo {author} {\bibfnamefont {M.}~\bibnamefont {Ying}},\ }\href@noop {} {}\Eprint {https://arxiv.org/abs/1802.04904} {arXiv:1802.04904} \BibitemShut {NoStop}%
\bibitem [{\citenamefont {Mahler}\ \emph {et~al.}(2012)\citenamefont {Mahler}, \citenamefont {Rozema}, \citenamefont {Darabi},\ and\ \citenamefont {Steinberg}}]{mahler2012identification}%
  \BibitemOpen
  \bibfield  {author} {\bibinfo {author} {\bibfnamefont {D.~H.}\ \bibnamefont {Mahler}}, \bibinfo {author} {\bibfnamefont {L.}~\bibnamefont {Rozema}}, \bibinfo {author} {\bibfnamefont {A.}~\bibnamefont {Darabi}},\ and\ \bibinfo {author} {\bibfnamefont {A.~M.}\ \bibnamefont {Steinberg}},\ }\href {https://doi.org/10.1103/PhysRevA.86.052101} {\bibfield  {journal} {\bibinfo  {journal} {Phys. Rev. A}\ }\textbf {\bibinfo {volume} {86}},\ \bibinfo {pages} {052101} (\bibinfo {year} {2012})}\BibitemShut {NoStop}%
\bibitem [{\citenamefont {Wang}\ \emph {et~al.}(2013)\citenamefont {Wang}, \citenamefont {Byrd},\ and\ \citenamefont {Jacobs}}]{wang2013numerical}%
  \BibitemOpen
  \bibfield  {author} {\bibinfo {author} {\bibfnamefont {X.}~\bibnamefont {Wang}}, \bibinfo {author} {\bibfnamefont {M.}~\bibnamefont {Byrd}},\ and\ \bibinfo {author} {\bibfnamefont {K.}~\bibnamefont {Jacobs}},\ }\href {https://doi.org/10.1103/PhysRevA.87.012338} {\bibfield  {journal} {\bibinfo  {journal} {Phys. Rev. A}\ }\textbf {\bibinfo {volume} {87}},\ \bibinfo {pages} {012338} (\bibinfo {year} {2013})}\BibitemShut {NoStop}%
\bibitem [{\citenamefont {Garraway}\ and\ \citenamefont {Knight}(1996)}]{garraway1996cavity}%
  \BibitemOpen
  \bibfield  {author} {\bibinfo {author} {\bibfnamefont {B.~M.}\ \bibnamefont {Garraway}}\ and\ \bibinfo {author} {\bibfnamefont {P.~L.}\ \bibnamefont {Knight}},\ }\href {https://doi.org/10.1103/PhysRevA.54.3592} {\bibfield  {journal} {\bibinfo  {journal} {Phys. Rev. A}\ }\textbf {\bibinfo {volume} {54}},\ \bibinfo {pages} {3592} (\bibinfo {year} {1996})}\BibitemShut {NoStop}%
\bibitem [{\citenamefont {Garraway}(1997)}]{garraway1997nonperturbative}%
  \BibitemOpen
  \bibfield  {author} {\bibinfo {author} {\bibfnamefont {B.~M.}\ \bibnamefont {Garraway}},\ }\href {https://doi.org/10.1103/PhysRevA.55.2290} {\bibfield  {journal} {\bibinfo  {journal} {Phys. Rev. A}\ }\textbf {\bibinfo {volume} {55}},\ \bibinfo {pages} {2290} (\bibinfo {year} {1997})}\BibitemShut {NoStop}%
\bibitem [{\citenamefont {Dalton}\ \emph {et~al.}(2001)\citenamefont {Dalton}, \citenamefont {Barnett},\ and\ \citenamefont {Garraway}}]{dalton2001theory}%
  \BibitemOpen
  \bibfield  {author} {\bibinfo {author} {\bibfnamefont {B.~J.}\ \bibnamefont {Dalton}}, \bibinfo {author} {\bibfnamefont {S.~M.}\ \bibnamefont {Barnett}},\ and\ \bibinfo {author} {\bibfnamefont {B.~M.}\ \bibnamefont {Garraway}},\ }\href {https://doi.org/10.1103/PhysRevA.64.053813} {\bibfield  {journal} {\bibinfo  {journal} {Phys. Rev. A}\ }\textbf {\bibinfo {volume} {64}},\ \bibinfo {pages} {053813} (\bibinfo {year} {2001})}\BibitemShut {NoStop}%
\bibitem [{\citenamefont {Tamascelli}\ \emph {et~al.}(2018)\citenamefont {Tamascelli}, \citenamefont {Smirne}, \citenamefont {Huelga},\ and\ \citenamefont {Plenio}}]{tamascelli2018nonperturbative}%
  \BibitemOpen
  \bibfield  {author} {\bibinfo {author} {\bibfnamefont {D.}~\bibnamefont {Tamascelli}}, \bibinfo {author} {\bibfnamefont {A.}~\bibnamefont {Smirne}}, \bibinfo {author} {\bibfnamefont {S.~F.}\ \bibnamefont {Huelga}},\ and\ \bibinfo {author} {\bibfnamefont {M.~B.}\ \bibnamefont {Plenio}},\ }\href {https://doi.org/10.1103/PhysRevLett.120.030402} {\bibfield  {journal} {\bibinfo  {journal} {Phys. Rev. Lett.}\ }\textbf {\bibinfo {volume} {120}},\ \bibinfo {pages} {030402} (\bibinfo {year} {2018})}\BibitemShut {NoStop}%
\bibitem [{\citenamefont {Teretenkov}(2019)}]{teretenkov2019pseudomode}%
  \BibitemOpen
  \bibfield  {author} {\bibinfo {author} {\bibfnamefont {A.~E.}\ \bibnamefont {Teretenkov}},\ }\href {https://doi.org/10.1134/S0081543819050201} {\bibfield  {journal} {\bibinfo  {journal} {Proc. Steklov Inst. Math.}\ }\textbf {\bibinfo {volume} {306}},\ \bibinfo {pages} {242} (\bibinfo {year} {2019})}\BibitemShut {NoStop}%
\bibitem [{\citenamefont {Tamascelli}\ \emph {et~al.}(2019)\citenamefont {Tamascelli}, \citenamefont {Smirne}, \citenamefont {Lim}, \citenamefont {Huelga},\ and\ \citenamefont {Plenio}}]{tamascelli2019efficient}%
  \BibitemOpen
  \bibfield  {author} {\bibinfo {author} {\bibfnamefont {D.}~\bibnamefont {Tamascelli}}, \bibinfo {author} {\bibfnamefont {A.}~\bibnamefont {Smirne}}, \bibinfo {author} {\bibfnamefont {J.}~\bibnamefont {Lim}}, \bibinfo {author} {\bibfnamefont {S.~F.}\ \bibnamefont {Huelga}},\ and\ \bibinfo {author} {\bibfnamefont {M.~B.}\ \bibnamefont {Plenio}},\ }\href {https://doi.org/10.1103/PhysRevLett.123.090402} {\bibfield  {journal} {\bibinfo  {journal} {Phys. Rev. Lett.}\ }\textbf {\bibinfo {volume} {123}},\ \bibinfo {pages} {090402} (\bibinfo {year} {2019})}\BibitemShut {NoStop}%
\bibitem [{\citenamefont {Pleasance}\ \emph {et~al.}(2020)\citenamefont {Pleasance}, \citenamefont {Garraway},\ and\ \citenamefont {Petruccione}}]{pleasance2020generalized}%
  \BibitemOpen
  \bibfield  {author} {\bibinfo {author} {\bibfnamefont {G.}~\bibnamefont {Pleasance}}, \bibinfo {author} {\bibfnamefont {B.~M.}\ \bibnamefont {Garraway}},\ and\ \bibinfo {author} {\bibfnamefont {F.}~\bibnamefont {Petruccione}},\ }\href {https://doi.org/10.1103/PhysRevResearch.2.043058} {\bibfield  {journal} {\bibinfo  {journal} {Phys. Rev. Res.}\ }\textbf {\bibinfo {volume} {2}},\ \bibinfo {pages} {043058} (\bibinfo {year} {2020})}\BibitemShut {NoStop}%
\bibitem [{\citenamefont {Kanazawa}\ and\ \citenamefont {Sornette}(2024)}]{kanazawa2024standard}%
  \BibitemOpen
  \bibfield  {author} {\bibinfo {author} {\bibfnamefont {K.}~\bibnamefont {Kanazawa}}\ and\ \bibinfo {author} {\bibfnamefont {D.}~\bibnamefont {Sornette}},\ }\href {https://doi.org/10.1103/PhysRevResearch.6.023270} {\bibfield  {journal} {\bibinfo  {journal} {Phys. Rev. Res.}\ }\textbf {\bibinfo {volume} {6}},\ \bibinfo {pages} {023270} (\bibinfo {year} {2024})}\BibitemShut {NoStop}%
\bibitem [{\citenamefont {Teretenkov}\ \emph {et~al.}(2025)\citenamefont {Teretenkov}, \citenamefont {Uskov},\ and\ \citenamefont {Lychkovskiy}}]{teretenkov2025pseudomode}%
  \BibitemOpen
  \bibfield  {author} {\bibinfo {author} {\bibfnamefont {A.}~\bibnamefont {Teretenkov}}, \bibinfo {author} {\bibfnamefont {F.}~\bibnamefont {Uskov}},\ and\ \bibinfo {author} {\bibfnamefont {O.}~\bibnamefont {Lychkovskiy}},\ }\href {https://doi.org/10.1103/PhysRevB.111.174308} {\bibfield  {journal} {\bibinfo  {journal} {Phys. Rev. B}\ }\textbf {\bibinfo {volume} {111}},\ \bibinfo {pages} {174308} (\bibinfo {year} {2025})}\BibitemShut {NoStop}%
\bibitem [{\citenamefont {Luchnikov}\ \emph {et~al.}(2020)\citenamefont {Luchnikov}, \citenamefont {Vintskevich}, \citenamefont {Grigoriev},\ and\ \citenamefont {Filippov}}]{luchnikov2020machine}%
  \BibitemOpen
  \bibfield  {author} {\bibinfo {author} {\bibfnamefont {I.~A.}\ \bibnamefont {Luchnikov}}, \bibinfo {author} {\bibfnamefont {S.~V.}\ \bibnamefont {Vintskevich}}, \bibinfo {author} {\bibfnamefont {D.~A.}\ \bibnamefont {Grigoriev}},\ and\ \bibinfo {author} {\bibfnamefont {S.~N.}\ \bibnamefont {Filippov}},\ }\href {https://doi.org/10.1103/PhysRevLett.124.140502} {\bibfield  {journal} {\bibinfo  {journal} {Phys. Rev. Lett.}\ }\textbf {\bibinfo {volume} {124}},\ \bibinfo {pages} {140502} (\bibinfo {year} {2020})}\BibitemShut {NoStop}%
\bibitem [{\citenamefont {Mukamel}(1995)}]{mukamel1995principles}%
  \BibitemOpen
  \bibfield  {author} {\bibinfo {author} {\bibfnamefont {S.}~\bibnamefont {Mukamel}},\ }\href@noop {} {\emph {\bibinfo {title} {Principles of Nonlinear Optical Spectroscopy}}}\ (\bibinfo  {publisher} {Oxford University Press},\ \bibinfo {address} {New York},\ \bibinfo {year} {1995})\BibitemShut {NoStop}%
\bibitem [{\citenamefont {Cho}(2009)}]{cho2009two}%
  \BibitemOpen
  \bibfield  {author} {\bibinfo {author} {\bibfnamefont {M.}~\bibnamefont {Cho}},\ }\href@noop {} {\emph {\bibinfo {title} {Two-Dimensional Optical Spectroscopy}}}\ (\bibinfo  {publisher} {CRC Press},\ \bibinfo {address} {Boca Raton},\ \bibinfo {year} {2009})\BibitemShut {NoStop}%
\bibitem [{\citenamefont {Yang}\ and\ \citenamefont {Fleming}(2002)}]{yang2002influence}%
  \BibitemOpen
  \bibfield  {author} {\bibinfo {author} {\bibfnamefont {M.}~\bibnamefont {Yang}}\ and\ \bibinfo {author} {\bibfnamefont {G.~R.}\ \bibnamefont {Fleming}},\ }\href {https://doi.org/10.1016/S0301-0104(01)00540-7} {\bibfield  {journal} {\bibinfo  {journal} {Chem. Phys.}\ }\textbf {\bibinfo {volume} {275}},\ \bibinfo {pages} {355} (\bibinfo {year} {2002})}\BibitemShut {NoStop}%
\bibitem [{\citenamefont {Seibt}\ and\ \citenamefont {Man{\v{c}}al}(2017)}]{seibt2017ultrafast}%
  \BibitemOpen
  \bibfield  {author} {\bibinfo {author} {\bibfnamefont {J.}~\bibnamefont {Seibt}}\ and\ \bibinfo {author} {\bibfnamefont {T.}~\bibnamefont {Man{\v{c}}al}},\ }\href {https://doi.org/10.1063/1.4981523} {\bibfield  {journal} {\bibinfo  {journal} {J. Chem. Phys.}\ }\textbf {\bibinfo {volume} {146}},\ \bibinfo {pages} {174109} (\bibinfo {year} {2017})}\BibitemShut {NoStop}%
\bibitem [{\citenamefont {Trushechkin}(2019)}]{trushechkin2019calculation}%
  \BibitemOpen
  \bibfield  {author} {\bibinfo {author} {\bibfnamefont {A.}~\bibnamefont {Trushechkin}},\ }\href {https://doi.org/10.1063/1.5100967} {\bibfield  {journal} {\bibinfo  {journal} {J. Chem. Phys.}\ }\textbf {\bibinfo {volume} {151}},\ \bibinfo {pages} {074101} (\bibinfo {year} {2019})}\BibitemShut {NoStop}%
\bibitem [{\citenamefont {Accardi}\ and\ \citenamefont {Kozyrev}(2002)}]{accardi2002lectures}%
  \BibitemOpen
  \bibfield  {author} {\bibinfo {author} {\bibfnamefont {L.}~\bibnamefont {Accardi}}\ and\ \bibinfo {author} {\bibfnamefont {S.}~\bibnamefont {Kozyrev}},\ }in\ \href@noop {} {\emph {\bibinfo {booktitle} {Quantum Interacting Particle Systems}}}\ (\bibinfo  {publisher} {World Scientific},\ \bibinfo {year} {2002})\ pp.\ \bibinfo {pages} {1--195}\BibitemShut {NoStop}%
\bibitem [{\citenamefont {Accardi}\ \emph {et~al.}(2016)\citenamefont {Accardi}, \citenamefont {Fagnola},\ and\ \citenamefont {Quezada}}]{accardi2016three}%
  \BibitemOpen
  \bibfield  {author} {\bibinfo {author} {\bibfnamefont {L.}~\bibnamefont {Accardi}}, \bibinfo {author} {\bibfnamefont {F.}~\bibnamefont {Fagnola}},\ and\ \bibinfo {author} {\bibfnamefont {R.}~\bibnamefont {Quezada}},\ }\href {https://doi.org/10.1142/S0219025716500090} {\bibfield  {journal} {\bibinfo  {journal} {Inf. Dimens. Anal. Quantum Probab. Relat. Top.}\ }\textbf {\bibinfo {volume} {19}},\ \bibinfo {pages} {1650009} (\bibinfo {year} {2016})}\BibitemShut {NoStop}%
\bibitem [{\citenamefont {Fagnola}\ and\ \citenamefont {Quezada}(2019)}]{fagnola2019characterization}%
  \BibitemOpen
  \bibfield  {author} {\bibinfo {author} {\bibfnamefont {F.}~\bibnamefont {Fagnola}}\ and\ \bibinfo {author} {\bibfnamefont {R.}~\bibnamefont {Quezada}},\ }\href {https://doi.org/10.1142/S0219025719500085} {\bibfield  {journal} {\bibinfo  {journal} {Infinite Dimens. Anal. Quantum Probab. Relat. Top.}\ }\textbf {\bibinfo {volume} {22}},\ \bibinfo {pages} {1950008} (\bibinfo {year} {2019})}\BibitemShut {NoStop}%
\bibitem [{\citenamefont {Hernandez-Cervantes}\ and\ \citenamefont {Quezada}(2020)}]{hernandez2020stationary}%
  \BibitemOpen
  \bibfield  {author} {\bibinfo {author} {\bibfnamefont {A.}~\bibnamefont {Hernandez-Cervantes}}\ and\ \bibinfo {author} {\bibfnamefont {R.}~\bibnamefont {Quezada}},\ }\href {https://doi.org/10.1142/S0219025720500034} {\bibfield  {journal} {\bibinfo  {journal} {Infinite Dimens. Anal. Quantum Probab. Relat. Top.}\ }\textbf {\bibinfo {volume} {23}},\ \bibinfo {pages} {2050003} (\bibinfo {year} {2020})}\BibitemShut {NoStop}%
\bibitem [{\citenamefont {Trushechkin}(2021)}]{trushechkin2021unified}%
  \BibitemOpen
  \bibfield  {author} {\bibinfo {author} {\bibfnamefont {A.}~\bibnamefont {Trushechkin}},\ }\href {https://doi.org/10.1103/PhysRevA.103.062226} {\bibfield  {journal} {\bibinfo  {journal} {Phys. Rev. A}\ }\textbf {\bibinfo {volume} {103}},\ \bibinfo {pages} {062226} (\bibinfo {year} {2021})}\BibitemShut {NoStop}%
\bibitem [{\citenamefont {Bola{\~n}os-Servin}\ \emph {et~al.}(2025)\citenamefont {Bola{\~n}os-Servin}, \citenamefont {Fagnola},\ and\ \citenamefont {Quezada}}]{bolanos2025gaussian}%
  \BibitemOpen
  \bibfield  {author} {\bibinfo {author} {\bibfnamefont {J.~R.}\ \bibnamefont {Bola{\~n}os-Servin}}, \bibinfo {author} {\bibfnamefont {F.}~\bibnamefont {Fagnola}},\ and\ \bibinfo {author} {\bibfnamefont {R.}~\bibnamefont {Quezada}},\ }\href {https://doi.org/10.1142/S0219025725500122} {\bibfield  {journal} {\bibinfo  {journal} {Infinite Dimens. Anal. Quantum Probab. Relat. Top.}\ ,\ \bibinfo {pages} {2550012}} (\bibinfo {year} {2025})}\BibitemShut {NoStop}%
\bibitem [{\citenamefont {Wightman}(1995)}]{wightman1995superselection}%
  \BibitemOpen
  \bibfield  {author} {\bibinfo {author} {\bibfnamefont {A.~S.}\ \bibnamefont {Wightman}},\ }\href {https://doi.org/10.1007/BF02741478} {\bibfield  {journal} {\bibinfo  {journal} {Il Nuovo Cimento B}\ }\textbf {\bibinfo {volume} {110}},\ \bibinfo {pages} {751} (\bibinfo {year} {1995})}\BibitemShut {NoStop}%
\bibitem [{\citenamefont {Cisneros}\ \emph {et~al.}(1998)\citenamefont {Cisneros}, \citenamefont {Mart{\'\i}nez-y Romero}, \citenamefont {N{\'u}{\~n}ez-Y{\'e}pez},\ and\ \citenamefont {Salas-Brito}}]{cisneros1998limitations}%
  \BibitemOpen
  \bibfield  {author} {\bibinfo {author} {\bibfnamefont {C.}~\bibnamefont {Cisneros}}, \bibinfo {author} {\bibfnamefont {R.~P.}\ \bibnamefont {Mart{\'\i}nez-y Romero}}, \bibinfo {author} {\bibfnamefont {H.~N.}\ \bibnamefont {N{\'u}{\~n}ez-Y{\'e}pez}},\ and\ \bibinfo {author} {\bibfnamefont {A.~L.}\ \bibnamefont {Salas-Brito}},\ }\href {https://doi.org/10.1088/0143-0807/19/3/005} {\bibfield  {journal} {\bibinfo  {journal} {Eur. J. Phys.}\ }\textbf {\bibinfo {volume} {19}},\ \bibinfo {pages} {237} (\bibinfo {year} {1998})}\BibitemShut {NoStop}%
\bibitem [{\citenamefont {Bartlett}\ \emph {et~al.}(2007)\citenamefont {Bartlett}, \citenamefont {Rudolph},\ and\ \citenamefont {Spekkens}}]{bartlett2007reference}%
  \BibitemOpen
  \bibfield  {author} {\bibinfo {author} {\bibfnamefont {S.~D.}\ \bibnamefont {Bartlett}}, \bibinfo {author} {\bibfnamefont {T.}~\bibnamefont {Rudolph}},\ and\ \bibinfo {author} {\bibfnamefont {R.~W.}\ \bibnamefont {Spekkens}},\ }\href {https://doi.org/10.1103/RevModPhys.79.555} {\bibfield  {journal} {\bibinfo  {journal} {Rev. Mod. Phys.}\ }\textbf {\bibinfo {volume} {79}},\ \bibinfo {pages} {555} (\bibinfo {year} {2007})}\BibitemShut {NoStop}%
\bibitem [{\citenamefont {Amosov}\ and\ \citenamefont {Filippov}(2017)}]{amosov2017spectral}%
  \BibitemOpen
  \bibfield  {author} {\bibinfo {author} {\bibfnamefont {G.~G.}\ \bibnamefont {Amosov}}\ and\ \bibinfo {author} {\bibfnamefont {S.~N.}\ \bibnamefont {Filippov}},\ }\href {https://doi.org/10.1007/s11128-016-1467-9} {\bibfield  {journal} {\bibinfo  {journal} {Quantum Inf. Process.}\ }\textbf {\bibinfo {volume} {16}},\ \bibinfo {pages} {2} (\bibinfo {year} {2017})}\BibitemShut {NoStop}%
\bibitem [{\citenamefont {Bratteli}\ and\ \citenamefont {Robinson}(2012)}]{bratteli2012operator}%
  \BibitemOpen
  \bibfield  {author} {\bibinfo {author} {\bibfnamefont {O.}~\bibnamefont {Bratteli}}\ and\ \bibinfo {author} {\bibfnamefont {D.~W.}\ \bibnamefont {Robinson}},\ }\href@noop {} {\emph {\bibinfo {title} {Operator Algebras and Quantum Statistical Mechanics 1: C*-and W*-Algebras. Symmetry Groups. Decomposition of States}}}\ (\bibinfo  {publisher} {Springer},\ \bibinfo {address} {Berlin},\ \bibinfo {year} {2012})\BibitemShut {NoStop}%
\bibitem [{\citenamefont {Moudgalya}\ and\ \citenamefont {Motrunich}(2022)}]{moudgalya2022hilbert}%
  \BibitemOpen
  \bibfield  {author} {\bibinfo {author} {\bibfnamefont {S.}~\bibnamefont {Moudgalya}}\ and\ \bibinfo {author} {\bibfnamefont {O.~I.}\ \bibnamefont {Motrunich}},\ }\href {https://doi.org/10.1103/PhysRevX.12.011050} {\bibfield  {journal} {\bibinfo  {journal} {Phys. Rev. X}\ }\textbf {\bibinfo {volume} {12}},\ \bibinfo {pages} {011050} (\bibinfo {year} {2022})}\BibitemShut {NoStop}%
\bibitem [{\citenamefont {Grigoletto}\ \emph {et~al.}()\citenamefont {Grigoletto}, \citenamefont {Viola},\ and\ \citenamefont {Ticozzi}}]{grigoletto2025model}%
  \BibitemOpen
  \bibfield  {author} {\bibinfo {author} {\bibfnamefont {T.}~\bibnamefont {Grigoletto}}, \bibinfo {author} {\bibfnamefont {L.}~\bibnamefont {Viola}},\ and\ \bibinfo {author} {\bibfnamefont {F.}~\bibnamefont {Ticozzi}},\ }\href@noop {} {}\Eprint {https://arxiv.org/abs/2510.25546} {arXiv:2510.25546} \BibitemShut {NoStop}%
\bibitem [{\citenamefont {Grigoletto}\ \emph {et~al.}(2025)\citenamefont {Grigoletto}, \citenamefont {Tao}, \citenamefont {Ticozzi},\ and\ \citenamefont {Viola}}]{grigoletto2025exact}%
  \BibitemOpen
  \bibfield  {author} {\bibinfo {author} {\bibfnamefont {T.}~\bibnamefont {Grigoletto}}, \bibinfo {author} {\bibfnamefont {Y.}~\bibnamefont {Tao}}, \bibinfo {author} {\bibfnamefont {F.}~\bibnamefont {Ticozzi}},\ and\ \bibinfo {author} {\bibfnamefont {L.}~\bibnamefont {Viola}},\ }\href@noop {} {\bibfield  {journal} {\bibinfo  {journal} {Quantum}\ }\textbf {\bibinfo {volume} {9}},\ \bibinfo {pages} {1814} (\bibinfo {year} {2025})}\BibitemShut {NoStop}%
\bibitem [{\citenamefont {de~Klerk}\ \emph {et~al.}(2011)\citenamefont {de~Klerk}, \citenamefont {Dobre},\ and\ \citenamefont {\.{P}asechnik}}]{de2011numerical}%
  \BibitemOpen
  \bibfield  {author} {\bibinfo {author} {\bibfnamefont {E.}~\bibnamefont {de~Klerk}}, \bibinfo {author} {\bibfnamefont {C.}~\bibnamefont {Dobre}},\ and\ \bibinfo {author} {\bibfnamefont {D.~V.}\ \bibnamefont {\.{P}asechnik}},\ }\href@noop {} {\bibfield  {journal} {\bibinfo  {journal} {Math. Program.}\ }\textbf {\bibinfo {volume} {129}},\ \bibinfo {pages} {91} (\bibinfo {year} {2011})}\BibitemShut {NoStop}%
\bibitem [{\citenamefont {Breuer}\ and\ \citenamefont {Petruccione}(2002)}]{breuer2002theory}%
  \BibitemOpen
  \bibfield  {author} {\bibinfo {author} {\bibfnamefont {H.-P.}\ \bibnamefont {Breuer}}\ and\ \bibinfo {author} {\bibfnamefont {F.}~\bibnamefont {Petruccione}},\ }\href@noop {} {\emph {\bibinfo {title} {The Theory of Open Quantum Systems}}}\ (\bibinfo  {publisher} {Oxford University Press},\ \bibinfo {address} {Oxford},\ \bibinfo {year} {2002})\BibitemShut {NoStop}%
\bibitem [{\citenamefont {Teretenkov}(2022)}]{teretenkov2022effective}%
  \BibitemOpen
  \bibfield  {author} {\bibinfo {author} {\bibfnamefont {A.~E.}\ \bibnamefont {Teretenkov}},\ }\href {https://doi.org/10.3390/e24081144} {\bibfield  {journal} {\bibinfo  {journal} {Entropy}\ }\textbf {\bibinfo {volume} {24}},\ \bibinfo {pages} {1144} (\bibinfo {year} {2022})}\BibitemShut {NoStop}%
\bibitem [{\citenamefont {Torres}(2014)}]{torres2014closed}%
  \BibitemOpen
  \bibfield  {author} {\bibinfo {author} {\bibfnamefont {J.~M.}\ \bibnamefont {Torres}},\ }\href {https://doi.org/10.1103/PhysRevA.89.052133} {\bibfield  {journal} {\bibinfo  {journal} {Phys. Rev. A}\ }\textbf {\bibinfo {volume} {89}},\ \bibinfo {pages} {052133} (\bibinfo {year} {2014})}\BibitemShut {NoStop}%
\bibitem [{\citenamefont {Teretenkov}(2020{\natexlab{b}})}]{teretenkov2020one}%
  \BibitemOpen
  \bibfield  {author} {\bibinfo {author} {\bibfnamefont {A.~E.}\ \bibnamefont {Teretenkov}},\ }\href {https://doi.org/https://doi.org/10.5890/DNC.2020.12.010} {\bibfield  {journal} {\bibinfo  {journal} {Discontinuity Nonlinearity Complex.}\ }\textbf {\bibinfo {volume} {9}},\ \bibinfo {pages} {567} (\bibinfo {year} {2020}{\natexlab{b}})}\BibitemShut {NoStop}%
\bibitem [{\citenamefont {Gorini}\ \emph {et~al.}(1976)\citenamefont {Gorini}, \citenamefont {Kossakowski},\ and\ \citenamefont {Sudarshan}}]{gorini1976completely}%
  \BibitemOpen
  \bibfield  {author} {\bibinfo {author} {\bibfnamefont {V.}~\bibnamefont {Gorini}}, \bibinfo {author} {\bibfnamefont {A.}~\bibnamefont {Kossakowski}},\ and\ \bibinfo {author} {\bibfnamefont {E.~C.~G.}\ \bibnamefont {Sudarshan}},\ }\href {https://doi.org/10.1063/1.522979} {\bibfield  {journal} {\bibinfo  {journal} {J. Math. Phys.}\ }\textbf {\bibinfo {volume} {17}},\ \bibinfo {pages} {821} (\bibinfo {year} {1976})}\BibitemShut {NoStop}%
\bibitem [{\citenamefont {Lindblad}(1976)}]{lindblad1976generators}%
  \BibitemOpen
  \bibfield  {author} {\bibinfo {author} {\bibfnamefont {G.}~\bibnamefont {Lindblad}},\ }\href {https://doi.org/10.1007/BF01608499} {\bibfield  {journal} {\bibinfo  {journal} {Commun. Math. Phys.}\ }\textbf {\bibinfo {volume} {48}},\ \bibinfo {pages} {119} (\bibinfo {year} {1976})}\BibitemShut {NoStop}%
\bibitem [{\citenamefont {Teretenkov}(2023)}]{Teretenkov2023MemoryTensor}%
  \BibitemOpen
  \bibfield  {author} {\bibinfo {author} {\bibfnamefont {A.~E.}\ \bibnamefont {Teretenkov}},\ }\href {https://doi.org/10.3390/math11183854} {\bibfield  {journal} {\bibinfo  {journal} {Mathematics}\ }\textbf {\bibinfo {volume} {11}},\ \bibinfo {pages} {3854} (\bibinfo {year} {2023})}\BibitemShut {NoStop}%
\bibitem [{\citenamefont {Milz}\ and\ \citenamefont {Modi}(2021)}]{milzQuantumStochasticProcesses2020}%
  \BibitemOpen
  \bibfield  {author} {\bibinfo {author} {\bibfnamefont {S.}~\bibnamefont {Milz}}\ and\ \bibinfo {author} {\bibfnamefont {K.}~\bibnamefont {Modi}},\ }\href {https://doi.org/10.1103/PRXQuantum.2.030201} {\bibfield  {journal} {\bibinfo  {journal} {PRX Quantum}\ }\textbf {\bibinfo {volume} {2}},\ \bibinfo {pages} {030201} (\bibinfo {year} {2021})}\BibitemShut {NoStop}%
\bibitem [{\citenamefont {Li}\ \emph {et~al.}(2018)\citenamefont {Li}, \citenamefont {Hall},\ and\ \citenamefont {Wiseman}}]{li2018concepts}%
  \BibitemOpen
  \bibfield  {author} {\bibinfo {author} {\bibfnamefont {L.}~\bibnamefont {Li}}, \bibinfo {author} {\bibfnamefont {M.~J.~W.}\ \bibnamefont {Hall}},\ and\ \bibinfo {author} {\bibfnamefont {H.~M.}\ \bibnamefont {Wiseman}},\ }\href {https://doi.org/https://doi.org/10.1016/j.physrep.2018.07.001} {\bibfield  {journal} {\bibinfo  {journal} {Phys. Rep.}\ }\textbf {\bibinfo {volume} {759}},\ \bibinfo {pages} {1} (\bibinfo {year} {2018})}\BibitemShut {NoStop}%
\bibitem [{\citenamefont {Evans}(1977)}]{evans1977irreducible}%
  \BibitemOpen
  \bibfield  {author} {\bibinfo {author} {\bibfnamefont {D.~E.}\ \bibnamefont {Evans}},\ }\href {https://doi.org/10.1007/BF01614091} {\bibfield  {journal} {\bibinfo  {journal} {Commun. Math. Phys.}\ }\textbf {\bibinfo {volume} {54}},\ \bibinfo {pages} {293} (\bibinfo {year} {1977})}\BibitemShut {NoStop}%
\bibitem [{\citenamefont {Ticozzi}\ and\ \citenamefont {Viola}(2008)}]{ticozzi2008quantum}%
  \BibitemOpen
  \bibfield  {author} {\bibinfo {author} {\bibfnamefont {F.}~\bibnamefont {Ticozzi}}\ and\ \bibinfo {author} {\bibfnamefont {L.}~\bibnamefont {Viola}},\ }\href {https://doi.org/10.1109/TAC.2008.929399} {\bibfield  {journal} {\bibinfo  {journal} {IEEE Trans. Autom. Control}\ }\textbf {\bibinfo {volume} {53}},\ \bibinfo {pages} {2048} (\bibinfo {year} {2008})}\BibitemShut {NoStop}%
\bibitem [{\citenamefont {Lidar}\ \emph {et~al.}(1998)\citenamefont {Lidar}, \citenamefont {Chuang},\ and\ \citenamefont {Whaley}}]{lidar1998decoherence}%
  \BibitemOpen
  \bibfield  {author} {\bibinfo {author} {\bibfnamefont {D.~A.}\ \bibnamefont {Lidar}}, \bibinfo {author} {\bibfnamefont {I.~L.}\ \bibnamefont {Chuang}},\ and\ \bibinfo {author} {\bibfnamefont {K.~B.}\ \bibnamefont {Whaley}},\ }\href {https://doi.org/10.1103/PhysRevLett.81.2594} {\bibfield  {journal} {\bibinfo  {journal} {Phys. Rev. Lett.}\ }\textbf {\bibinfo {volume} {81}},\ \bibinfo {pages} {2594} (\bibinfo {year} {1998})}\BibitemShut {NoStop}%
\bibitem [{\citenamefont {Lidar}\ and\ \citenamefont {Birgitta~Whaley}(2003)}]{lidar2003decoherence}%
  \BibitemOpen
  \bibfield  {author} {\bibinfo {author} {\bibfnamefont {D.~A.}\ \bibnamefont {Lidar}}\ and\ \bibinfo {author} {\bibfnamefont {K.}~\bibnamefont {Birgitta~Whaley}},\ }in\ \href {https://doi.org/10.1007/3-540-44874-8_5} {\emph {\bibinfo {booktitle} {Irreversible Quantum Dynamics}}},\ \bibinfo {editor} {edited by\ \bibinfo {editor} {\bibfnamefont {F.}~\bibnamefont {Benatti}}\ and\ \bibinfo {editor} {\bibfnamefont {R.}~\bibnamefont {Floreanini}}}\ (\bibinfo  {publisher} {Springer},\ \bibinfo {year} {2003})\ pp.\ \bibinfo {pages} {83--120}\BibitemShut {NoStop}%
\bibitem [{\citenamefont {Lidar}(2014)}]{lidar2014review}%
  \BibitemOpen
  \bibfield  {author} {\bibinfo {author} {\bibfnamefont {D.~A.}\ \bibnamefont {Lidar}},\ }in\ \href@noop {} {\emph {\bibinfo {booktitle} {Quantum Information and Computation for Chemistry}}},\ \bibinfo {editor} {edited by\ \bibinfo {editor} {\bibfnamefont {S.}~\bibnamefont {Kais}}}\ (\bibinfo  {publisher} {John Wiley \& Sons},\ \bibinfo {year} {2014})\ pp.\ \bibinfo {pages} {295--354}\BibitemShut {NoStop}%
\bibitem [{\citenamefont {Agredo}\ \emph {et~al.}(2014)\citenamefont {Agredo}, \citenamefont {Fagnola},\ and\ \citenamefont {Rebolledo}}]{agredo2014decoherence}%
  \BibitemOpen
  \bibfield  {author} {\bibinfo {author} {\bibfnamefont {J.}~\bibnamefont {Agredo}}, \bibinfo {author} {\bibfnamefont {F.}~\bibnamefont {Fagnola}},\ and\ \bibinfo {author} {\bibfnamefont {R.}~\bibnamefont {Rebolledo}},\ }\bibfield  {journal} {\bibinfo  {journal} {J. Math. Phys.}\ }\textbf {\bibinfo {volume} {55}},\ \href {https://doi.org/10.1063/1.4901009} {10.1063/1.4901009} (\bibinfo {year} {2014})\BibitemShut {NoStop}%
\bibitem [{\citenamefont {Am-Shallem}\ \emph {et~al.}()\citenamefont {Am-Shallem}, \citenamefont {Levy}, \citenamefont {Schaefer},\ and\ \citenamefont {Kosloff}}]{threeapproaches}%
  \BibitemOpen
  \bibfield  {author} {\bibinfo {author} {\bibfnamefont {M.}~\bibnamefont {Am-Shallem}}, \bibinfo {author} {\bibfnamefont {A.}~\bibnamefont {Levy}}, \bibinfo {author} {\bibfnamefont {I.}~\bibnamefont {Schaefer}},\ and\ \bibinfo {author} {\bibfnamefont {R.}~\bibnamefont {Kosloff}},\ }\href@noop {} {}\Eprint {https://arxiv.org/abs/1510.08634} {arXiv:1510.08634} \BibitemShut {NoStop}%
\bibitem [{\citenamefont {Paszke}\ \emph {et~al.}()\citenamefont {Paszke}, \citenamefont {Gross}, \citenamefont {Massa}, \citenamefont {Lerer}, \citenamefont {Bradbury}, \citenamefont {Chanan}, \citenamefont {Killeen}, \citenamefont {Lin}, \citenamefont {Gimelshein}, \citenamefont {Antiga}, \citenamefont {Desmaison}, \citenamefont {Köpf}, \citenamefont {Yang}, \citenamefont {DeVito}, \citenamefont {Raison}, \citenamefont {Tejani}, \citenamefont {Chilamkurthy}, \citenamefont {Steiner}, \citenamefont {Fang}, \citenamefont {Bai},\ and\ \citenamefont {Chintala}}]{PyTorch}%
  \BibitemOpen
  \bibfield  {author} {\bibinfo {author} {\bibfnamefont {A.}~\bibnamefont {Paszke}}, \bibinfo {author} {\bibfnamefont {S.}~\bibnamefont {Gross}}, \bibinfo {author} {\bibfnamefont {F.}~\bibnamefont {Massa}}, \bibinfo {author} {\bibfnamefont {A.}~\bibnamefont {Lerer}}, \bibinfo {author} {\bibfnamefont {J.}~\bibnamefont {Bradbury}}, \bibinfo {author} {\bibfnamefont {G.}~\bibnamefont {Chanan}}, \bibinfo {author} {\bibfnamefont {T.}~\bibnamefont {Killeen}}, \bibinfo {author} {\bibfnamefont {Z.}~\bibnamefont {Lin}}, \bibinfo {author} {\bibfnamefont {N.}~\bibnamefont {Gimelshein}}, \bibinfo {author} {\bibfnamefont {L.}~\bibnamefont {Antiga}}, \bibinfo {author} {\bibfnamefont {A.}~\bibnamefont {Desmaison}}, \bibinfo {author} {\bibfnamefont {A.}~\bibnamefont {Köpf}}, \bibinfo {author} {\bibfnamefont {E.}~\bibnamefont {Yang}}, \bibinfo {author} {\bibfnamefont {Z.}~\bibnamefont {DeVito}}, \bibinfo {author} {\bibfnamefont {M.}~\bibnamefont {Raison}}, \bibinfo {author} {\bibfnamefont {A.}~\bibnamefont {Tejani}}, \bibinfo
  {author} {\bibfnamefont {S.}~\bibnamefont {Chilamkurthy}}, \bibinfo {author} {\bibfnamefont {B.}~\bibnamefont {Steiner}}, \bibinfo {author} {\bibfnamefont {L.}~\bibnamefont {Fang}}, \bibinfo {author} {\bibfnamefont {J.}~\bibnamefont {Bai}},\ and\ \bibinfo {author} {\bibfnamefont {S.}~\bibnamefont {Chintala}},\ }\href@noop {} {}\Eprint {https://arxiv.org/abs/1912.01703} {arXiv:1912.01703} \BibitemShut {NoStop}%
\bibitem [{\citenamefont {Pechen}\ and\ \citenamefont {Tannor}(2012)}]{Pechen_Tannor_2012}%
  \BibitemOpen
  \bibfield  {author} {\bibinfo {author} {\bibfnamefont {A.~N.}\ \bibnamefont {Pechen}}\ and\ \bibinfo {author} {\bibfnamefont {D.~J.}\ \bibnamefont {Tannor}},\ }\href {https://doi.org/10.1002/ijch.201100165} {\bibfield  {journal} {\bibinfo  {journal} {Isr. J. Chem.}\ }\textbf {\bibinfo {volume} {52}},\ \bibinfo {pages} {467–472} (\bibinfo {year} {2012})}\BibitemShut {NoStop}%
\bibitem [{\citenamefont {Karnieli}\ \emph {et~al.}(2025)\citenamefont {Karnieli}, \citenamefont {Tziperman}, \citenamefont {Roques-Carmes},\ and\ \citenamefont {Fan}}]{karnieli2025decoherence}%
  \BibitemOpen
  \bibfield  {author} {\bibinfo {author} {\bibfnamefont {A.}~\bibnamefont {Karnieli}}, \bibinfo {author} {\bibfnamefont {O.}~\bibnamefont {Tziperman}}, \bibinfo {author} {\bibfnamefont {C.}~\bibnamefont {Roques-Carmes}},\ and\ \bibinfo {author} {\bibfnamefont {S.}~\bibnamefont {Fan}},\ }\href {https://doi.org/10.1103/PhysRevResearch.7.L012014} {\bibfield  {journal} {\bibinfo  {journal} {Phys. Rev. Res.}\ }\textbf {\bibinfo {volume} {7}},\ \bibinfo {pages} {L012014} (\bibinfo {year} {2025})}\BibitemShut {NoStop}%
\bibitem [{\citenamefont {Agarwal}\ and\ \citenamefont {Mullen}(1988)}]{agarwal1988partitions}%
  \BibitemOpen
  \bibfield  {author} {\bibinfo {author} {\bibfnamefont {A.~K.}\ \bibnamefont {Agarwal}}\ and\ \bibinfo {author} {\bibfnamefont {G.~L.}\ \bibnamefont {Mullen}},\ }\href {https://doi.org/10.1016/0097-3165(88)90079-9} {\bibfield  {journal} {\bibinfo  {journal} {J. Comb. Theory Ser. A}\ }\textbf {\bibinfo {volume} {48}},\ \bibinfo {pages} {120} (\bibinfo {year} {1988})}\BibitemShut {NoStop}%
\bibitem [{\citenamefont {Erickson}(2025)}]{erickson2025demazure}%
  \BibitemOpen
  \bibfield  {author} {\bibinfo {author} {\bibfnamefont {W.~Q.}\ \bibnamefont {Erickson}},\ }\href@noop {} {\bibfield  {journal} {\bibinfo  {journal} {Enumer. Comb. Appl.}\ }\textbf {\bibinfo {volume} {5}},\ \bibinfo {pages} {S2R11} (\bibinfo {year} {2025})}\BibitemShut {NoStop}%
\bibitem [{\citenamefont {Brigham}(1950)}]{brigham1950general}%
  \BibitemOpen
  \bibfield  {author} {\bibinfo {author} {\bibfnamefont {N.~A.}\ \bibnamefont {Brigham}},\ }\href@noop {} {\bibfield  {journal} {\bibinfo  {journal} {Proc. Am. Math. Soc.}\ }\textbf {\bibinfo {volume} {1}},\ \bibinfo {pages} {182} (\bibinfo {year} {1950})}\BibitemShut {NoStop}%
\end{thebibliography}%
\end{document}